\documentclass[twocolumn]{aastex6}
\usepackage{subfigure}
\usepackage{morefloats}
\usepackage{amsmath}
\usepackage{bm}
\usepackage[flushleft]{threeparttable}
\usepackage{booktabs}

\newcommand{\tprime}{\hbox{$t^{\prime}$}}

\newcommand{\ism}{\textsc{ism}}
\newcommand{\bc}{\rm BC03}
\newcommand{\cb}{\rm CB07}
\newcommand{\ma}{\rm M05}
\newcommand{\galev}{\rm GALEV}
\newcommand{\ynII}{\rm Yunnan-II}
\newcommand{\bpass}{\rm BPASS V2.0}
\newcommand{\Msun}{\rm M_{\odot}}
\newcommand{\burst}{\rm Burst}
\newcommand{\const}{\rm Constant}
\newcommand{\expd}{\rm Exp-dec}
\newcommand{\expi}{\rm Exp-inc}
\newcommand{\delayed}{\rm Delayed-$\tau$}
\newcommand{\Calzetti}{\rm Calzetti law}
\newcommand{\mw}{\rm MW}
\newcommand{\smc}{\rm SMC}
\newcommand{\lmc}{\rm LMC}
\newcommand{\tmass}{\rm 2MASS}
\newcommand{\sdss}{\rm SDSS}
\newcommand{\cosmos}{\rm COSMOS}
\newcommand{\vista}{\rm UltraVISTA}
\newcommand{\candels}{\rm CANDELS}
\newcommand{\sdhst}{\rm 3D-HST}
\begin{document}

\title{A comprehensive Bayesian discrimination of the simple stellar population model, star formation history and dust attenuation law in the spectral energy distribution modeling of galaxies}
\shorttitle{Bayesian discrimination of the SED modelings of galaxies}

\author{Yunkun Han\altaffilmark{1,2,3}, Zhanwen Han\altaffilmark{1,2,3}}
\shortauthors{Han \& Han}
\altaffiltext{1}{Yunnan Observatories, Chinese Academy of Sciences, 396 Yangfangwang, Guandu District, Kunming, 650216, P. R. China}
\altaffiltext{2}{Center for Astronomical Mega-Science, Chinese Academy of Sciences, 20A Datun Road, Chaoyang District, Beijing, 100012, P. R. China}
\altaffiltext{3}{Key Laboratory for the Structure and Evolution of Celestial Objects, Chinese Academy of Sciences, 396 Yangfangwang, Guandu District, Kunming, 650216, P. R. China}
\email{hanyk@ynao.ac.cn}
\email{zhanwenhan@ynao.ac.cn}

\begin{abstract}
When modeling and interpreting the spectral energy distributions (SEDs) of galaxies, the simple stellar population (SSP) model, star formation history (SFH) and dust attenuation law (DAL) are three of the most important components.
However, each of them carries significant uncertainties which have seriously limited our ability to reliably recover the physical properties of galaxies from the analysis of their SEDs.
In this paper, we present a Bayesian framework to deal with these uncertain components simultaneously.
Based on the Bayesian evidence, a quantitative implement of the principle of Occam's razor, the method allows a more objective and quantitative discrimination among the different assumptions about these uncertain components.
With a Ks-selected sample of 5467 low-redshift (mostly with $z\lesssim 1$) galaxies in the COSMOS/UltraVISTA field and classified into passively evolving galaxies (PEGs) and star-forming galaxies (SFGs) with UVJ diagram, we present a Bayesian discrimination of a set of $16$ SSP models from five research groups (\bc\ and \cb, \ma, \galev, \ynII, \bpass), five forms of SFH (\burst, \const, \expd, \expi, \delayed), and four kinds of DAL (\Calzetti, \mw, \lmc, \smc).
We show that the results obtained with the method are either obvious or understandable in the context of galaxy physics.
We conclude that the Bayesian model comparison method, especially that for a sample of galaxies, is very useful for discriminating the different assumptions in the SED modeling of galaxies.
The new version of the BayeSED code, which is used in this work, is publicly available at \url{https://bitbucket.org/hanyk/bayesed/}.

\end{abstract}


\section{Introduction}\label{s:intro}
Understanding the formation and evolution of galaxies is one of the biggest challenges in modern astrophysics \citep{Mo2010a,DeLuciaG2014b,SomervilleR2015a,NaabT2017a}.
Various complex and not well understood baryonic processes, such as the formation and evolution of stars \citep{Kennicutt1998a,McKee2007a,Heber2009a,Kennicutt2012a,Duchene2013a}, the accretion and feedback of super-massive black holes \citep{Melia2001a,Merloni04,Kormendy2013a,Fabian2012b} and the chemical enrichment of interstellar medium (ISM) \citep{McKeeC1977a,SpitzerL1978a,LiA1997a,DraineB2003a,DeLucia2004a,Scannapieco2005a,DraineB2010a,NomotoK2013a}, are involved.
What make the problem even more challenging is the fact that all of these complex baryonic processes are also tightly entangled \citep{HamannF1993a,TimmesF1995a,Ferrarese00,Hopkins2008a,Hopkins2008b,Marulli2008a,Bonoli2009a,HeckmanT2014a}.
It is often not trivial to decouple any one of them from the others to allow a complete independent study.
To disentangle these complex and highly related baryonic processes involved in the formation and evolution of galaxies, we need to make use of all available sources of information \citep{BartosI2017a}.

Despite the recent progress in the detection of the cosmic-rays \citep{MuraseK2008a,AdrianiO2009a}, neutrinos \citep{AhmadQ2002a,BeckerJ2008a}, and gravitational-waves \citep{AbbottB2016a,Abbott2017a}, electromagnetic emissions are still the main source of information for our understanding of galaxies.
All of those complex baryonic processes involved in the formation and evolution of galaxies leave their imprint on the spectral energy distributions (SEDs) of the electromagnetic emissions from galaxies.
In the last decades, large photometric and spectroscopic surveys, such as \tmass\ \citep{SkrutskieM2006a}, \sdss\ \citep{York2000a}, \cosmos\ \citep{ScovilleN2007a}, \vista\ \citep{McCrackenH2012a,Muzzin2013a}, \candels\ \citep{Koekemoer2011a,GroginN2011a}, and \sdhst\ \citep{Brammer2012a,SkeltonR2014a}, have provided us with rich multi-wavelength observational data for millions of galaxies covering a large range of redshift. 
These massive data sets present a tremendous opportunity and challenge for us to understand the formation and evolution of galaxies from the analysis of their SEDs.

The process of solving the inverse problem of deriving the physical properties of galaxies from their observational SEDs is known as SED-fitting \citep{BolzonellaM2000a,MassarottiM2001a,IlbertO2006a,Salim2007a,Walcher2011a}.
In principle, a SED-fitting method which is capable of effectively extracting all the information encoded in these SEDs of galaxies would allow us to fully understand their physical properties. 
Traditionally, SED-fitting is considered as an optimization problem, where some $\chi^2$ minimization techniques are employed to find the best-fit model and corresponding value of parameters \citep{ArnoutsS1999a,BolzonellaM2000a,CidFernandes2005a,Kriek2009a,Koleva2009a,Sawicki2012a,GomesJ2017a}.
However, due to the large number of often degenerated free parameters, it should be more reasonable to consider the problem of SED-fitting as a Bayesian inference problem \citep{Benitez2000a,Kauffmann2003a}.
Recently, it has becoming quite popular to employ the Markov Chain Monte Carlo (MCMC) sampling method to efficiently obtain not only the best-fit results but also the detailed posterior probability distribution of all parameters \citep{Benitez2000a,Kauffmann2003a,Serra2011a,AcquavivaV2011a,Pirzkal2012a,Johnson2013a,CalistroRiveraG2016a,LejaJ2017a}.

Despite the popularity of Bayesian parameter estimation method, the Bayesian model comparison/selection method, which is based on the computation of the Bayesian evidences of different models, has not yet been widely used in the field of SED-fitting of galaxies.
The Bayesian evidence quantitatively implements the principle of Occam's razor, according to which a simpler model with compact parameter space should be preferred over a more complicated one with a large fraction of useless parameter space, unless the latter can provide a significantly better explanation to the data \citep{MacKayD1992a,MacKayD2003a}. 
Based on the Bayesian framework initially introduced by \cite{SuyuS2006a} for solving the gravitational lensing problem, \cite{DyeS2008a} presented an approach to determine the star formation history of galaxies from multiband photometry, where the most probable model of star formation history is obtained by the maximization of the Bayesian evidence.
In \cite{Han2012a}, we have presented a Bayesian model comparison for the SED modeling of hyperluminous infrared galaxies (HLIRGs), where the multimodal-nested sampling (MultiNest) techniques \citep{Feroz2008a,Feroz2009a,Feroz2013a} has been employed to allow a more efficient calculation of the Bayesian evidence of different SED models. 
\cite{SalmonB2016a} presented a Bayesian approach based on Bayesian evidence to check the universality of the dust attenuation law.
For a sample of $z\sim1.5-3$ galaxies from \candels with rest-frame UV to near-IR photometric data, they found that  some galaxies show strong Bayesian evidence in favor of one particular dust attenuation law over another, and this preference is consistent with their observed distribution on the infrared excess (IRX) and UV slope ($\beta$) plane.
\cite{DriesM2016a,DriesM2018a} presented a hierarchical Bayesian approach to reconstructing the initial mass function (IMF) in single and composite stellar populations (SSPs and CSPs), where the Bayesian evidence is employed to compare different choices of the IMF prior parameters, and to determine the number of SSPs required in CSPs by the maximization of the Bayesian evidence.

In \cite{HanY2014a}, with the first publicly available version of our BayeSED code,  we have presented a Bayesian model comparison between two of the most widely used stellar population synthesis (SPS) model \citep[][hereafter BC03 and M05]{Bruzual2003a,Maraston2005a} for the first time.
With the distribution of Bayes factor (the ratio of Bayesian evidence) for a Ks-selected sample of galaxies in the COSMOS/UltraVISTA field \citep{Muzzin2013a}, we found that the BC03 model statistically has larger Bayesian evidence than the M05 model.
In \cite{HanY2014a}, the reliability of the BayeSED code for physical parameter estimation has also been systematically tested.
The internal consistency of the code has been tested with a  mock sample of galaxies, while its external consistency has been tested by the comparison with the results of the widely used FAST code \citep{Kriek2009a}.
However, the work still has many limitations.
For example, a fixed exponentially declining SFH and the \cite{Calzetti2000a} dust attenuation law have been assumed to be universal for all galaxies.
However, from either an observational or a theoretical point of view, the form of star formation history and dust attenuation law of different galaxies are not likely to be the same \citep{WittA2000a,Maraston2010a,WuytsS2011a,Simha2014a}.
Besides, the numerous uncertainties carried by almost all the components involved in the process of stellar population synthesis \citep{Conroy2009a,Conroy2010c,Conroy2010d,Conroy2013e} have resulted in the diversity of SPS models.
Except for the BC03 and M05 model, there are numerous SPS models from many other groups, which have employed different stellar evolution tracks, stellar spectral libraries, IMFs and/or synthesis methods \citep{Buzzoni1989a,Fioc1997a,Leitherer1999a,Zhang2005a,KotullaR2009a,Eldridge2009a,Conroy2009a,Vazdekis2010a}.

As three of the most important components in modeling and interpreting the SEDs of galaxies, the simple stellar population  model, star formation history and dust attenuation law all carry significant uncertainties.
The existence of these uncertainties would seriously limit the possibility of reliably recovering the physical properties of galaxies from the analysis of their SEDs.
Besides, it is not easy to reasonably quantify the impact of any one of them without mention the other two.
So, it is very important to find an unitized method to quantify the propagation of these uncertainties into the estimation of the physical parameters of galaxies, and to quantitatively discriminate their different choices.
In this work, we present an unitized Bayesian framework to deal with all of these uncertain components simultaneously.

This paper is structured as follows.
In \S \ref{s:model}, we introduce the new SED modeling module of the BayeSED code, including the composite stellar population (CSP) synthesis method in \S \ref{ss:csp}, the SED modeling of a simple stellar population (SSP) in \S \ref{ss:ssp}, the form of star formation history (SFH) in \S \ref{ss:sfh}, and the dust attenuation law (DAL) in \S \ref{ss:dal}.
We then briefly review the Bayesian inference methods in \S \ref{s:methods}, including the Bayesian parameter estimation in \S \ref{ss:bayes_par} and the Bayesian model comparison in \S \ref{ss:bayes_model}.
In the next two sections, we introduce our new methods for calculating the Bayesian evidence and associated Occam factor for the SED modeling of an individual galaxy (\S \ref{s:ev_1SED}) and a sample of galaxies (\S \ref{s:ev_NSED}), respectively.
In \S  \ref{s:apply}, we present the results of applying our new methods to a Ks-selected sample in the COSMOS/UltraVISTA field for discriminating among the different choices of SSP model, SFH and DAL when modeling the SEDs of galaxies.
Some discussions about the different SSPs, SFHs and DALs are presented in \S \ref{sec:disc}.
Finally, a summary of our new methods and results is presented in \S \ref{sec:sum}.


\section{The spectral energy distribution modeling of galaxies in BayeSED} \label{s:model}
For a detailed Bayesian analysis of the observed multi-wavelength SED of a galaxy, the modeling of its SED is often the most computationally demanding.
So, the efficiency of the whole Bayesian analysis process is strongly depends on the efficiency of the SED modeling method.
In the previous version of BayeSED \citep{Han2012a,HanY2014a}, some machine learning methods, such as artificial neural network (ANN) and  K-nearest neighbor searching (KNN) algorithm, have been employed.
After the training with a pre-computed library of model SEDs, the machine learning methods allow a very efficient computation of a massive number of model SEDs during the sampling of an often high-dimensional parameter space of a SED model.
By using the machine learning methods, very different SED models can be easily integrated into the BayeSED code with the same procedure.
Therefore, the BayeSED code can be easily extended to solve the SED fitting problem in different fields.

\begin{figure*}
  \centering 
  \includegraphics[scale=0.4]{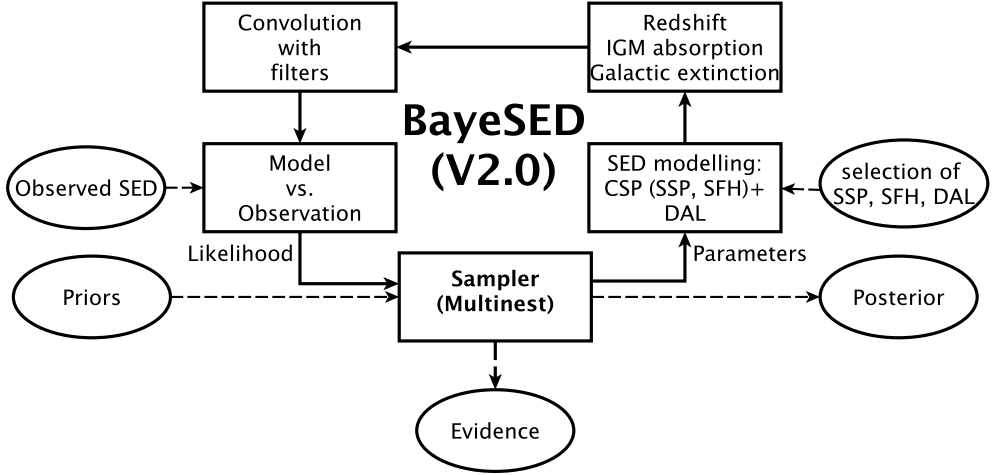} 
  \caption
  {
  The flowchart for modeling and interpreting the multi-wavelength photometric SED of a galaxy with BayeSED V2.0. 
  Most parts of V2.0 are similar to that of V1.0 \cite[see Figure 14 of][]{HanY2014a}.
  The major difference between them is the method used for SED modeling.
  In BayeSED V1.0, some machine learning (ML) techniques (e.g. PCA, ANN and KNN) have been used for interpolating the model SED grid pre-computed with the widely used FAST code.
  Instead, in BayeSED V2.0, we have built a module for modeling the SEDs of galaxies, which allow the free selection of SSP, SFH and DAL within a large set.
  The ML based methods are not used in this work, but have not been abandoned.
  See the text for a discussion about the advantages and disadvantages of the two methods.
  }
  \label{fig:flowchart}
\end{figure*}

Despite these interesting benefits, the machine learning based SED modeling methods are not so convenient during the development a SED model, since any modification to the model components requires a new and often time-consuming machine learning procedure.
To explore the effects of assuming different simple stellar population model, star formation history, and dust attenuation law in the SED modeling of galaxies, we have built a SED modeling module into the new version (V2.0) of our BayeSED code (see the flowchart in Figure \ref{fig:flowchart}).
Currently, we do not intend to build a very sophisticated SED modeling procedure into the BayeSED code.
To be consistent with the principle of Occam's razor, according to which ``Entities should not be multiplied unnecessarily'', we prefer to start with a simple but still useful SED modeling procedure, and gradually increase its complexity.

\subsection{Composite Stellar Population synthesis} \label{ss:csp}
The SED of a galaxy  as a complex stellar system can be obtained with  composite stellar population synthesis  as:
\begin{align}
	L_\lambda(t) &= \int_0^t {\rm d}\tprime \, \psi(t-\tprime) \, S_\lambda[\tprime,{Z}(t-\tprime)] \, T^\ism_\lambda(t,t^\prime) \label{eq:csp1}\\
	&= T^\ism_\lambda \,\int_0^t {\rm d}\tprime \, \psi(t-\tprime) \, S_\lambda[\tprime,Z_0] \label{eq:csp2}\,  ,
\end{align}
where $\psi(t-\tprime)$ is the star formation rate at time $t-\tprime$ (SFH: the star formation history), $S_\lambda[\tprime,{Z}(t-t^\prime)]$ the luminosity emitted per unit wavelength per unit mass by a simple stellar population (SSP) of age $t^\prime$ and chemical composition ${Z}(t-t^\prime)$, and $T^\ism_\lambda(t,t^\prime)$ the transmission function of the ISM.
We assume a time-independent metallicity $Z_0$ and dust attenuation law $T^\ism_\lambda$ for the entire composite population.

\subsection{The SED modeling of a simple stellar population} \label{ss:ssp}
According to the most widely used isochrone synthesis approach \citep{Charlot1991a,Bruzual1993a,Bruzual2003a}, the SED of an SSP is obtained as:
\begin{multline}
	\label{eq:ssp}
	S_\lambda(\tprime,{Z}) = \\
	\int_{m_\textnormal{low}}^{m_\textnormal{up}} {\rm d}m \, \phi(m) \, f_\lambda[L_\textnormal{bol}(m,{Z},\tprime),T_\textnormal{eff}(m,{Z},\tprime),{Z}] \, ,
\end{multline}
where $m$ is the stellar mass, $\phi(m)$ the stellar initial mass function (IMF)  with lower and upper mass cutoffs $m_\textnormal{low}$ and $m_\textnormal{up}$, and  $f_\lambda[L_\textnormal{bol}(m,{Z},\tprime),T_\textnormal{eff}(m,{Z},\tprime),{Z}]$ the SED of a star with bolometric luminosity $L_\textnormal{bol}(m,{Z},\tprime)$, effective temperature $T_\textnormal{eff}(m,{Z},\tprime)$, and metallicity ${Z}$.
So, different choices for any of the IMF, stellar isochrone and stellar spectral library will result in different SSP models.

Alternatively, the fuel consumption theorem \citep{RenziniA1986a,Maraston1998a,Maraston2005a} has been used to allow an easier calculation of the luminosity contribution of the short-lived and often less understood post-main sequence stellar evolution stages, such as the thermally-pulsing asymptotic giant branch (TP-AGB) phase.
According to the theorem, the luminosity contribution of each stellar evolutionary phase is proportional to the amount of hydrogen and/or helium (the fuel) burned by nuclear fusion within the stars.
It also provides analytical relations between the main sequence and post-main sequence stellar evolution, and the SEDs can be obtained using the relations between colors/spectra and bolometric luminosities.
There are other approaches to obtain the integrated SED of an SSP, such as the use of empirical spectra of star clusters as templates for SSPs \citep{BicaE1986a,BicaE1988a,CidFernandesR2001a,KongX2003a} and the employment of Monte Carlo technique \citep{Zhang2005a,Han2007a,daSilva2012a,Cervino2013b}.
 
There are many publicly available SSP models (See \url{http://www.sedfitting.org/Models.html}).
In this work, we have selected a set of $16$ different SSP models from five groups, including the \bc\ \citep{Bruzual2003a} and \cb\ \citep{BruzualA2007a}, \ma\ \citep{Maraston2005a}, \galev\ \citep{KotullaR2009a}, \ynII\ \citep{Zhang2005a}, and \bpass\ \citep{Eldridge2009a} models.
Many SSP models from other research groups \citep[e.g.][]{Buzzoni1989a,Fioc1997a,Leitherer1999a,Conroy2009a,Vazdekis2010a}, many of which have been widely used in many works, are not included in our list.
It is straightforward for us to add all of these SSP models to the new version of the BayeSED code.
However, the main purpose of this paper is to demonstrate the Bayesian model comparison method, and to evaluate its effectiveness.
So, we try to randomly select a small set of representative models that are as diverse as possible, although they could be biased to those that are either popular, easier to obtain, or more familiar to us.
The physical considerations about the effectiveness of the SSP models for the galaxy sample have not been used as the criterion for the selection of them.
Actually, they are considered to be equally likely a priori (i.e. before the comparison with data).
A summary of the $16$ SSP models used in this paper is presented in Table \ref{tab:SSPs}.
As shown clearly in the table, the SSP models which differ in any model component (Track/Spectral library/IMF/Binary/Nebular) are treated as different SSP models.
In the following of this section, we present a short description of each chosen SSP model, with a focus on their differences.
\begin{table*}
	\caption{Summary of SSP models}
		\begin{tabular}{lllllll}
			\toprule
			Short name&Model family &Track/Isochrone&Spectral library&IMF&Binary&Nebular\\
			\midrule
			bc03\_ch&\bc\tablenotemark{a}&Padova94+Charlot97&BaSeL 3.1&{Chabrier03}&No&No\\
			bc03\_kr&\bc&Padova94+Charlot97&BaSeL 3.1&{Kroupa01}&No&No\\
			bc03\_sa&\bc&Padova94+Charlot97&BaSeL 3.1&{Salpeter55}&No&No\\
			cb07\_ch&\cb\tablenotemark{b}&Padova94+Marigo07&BaSeL 3.1&{Chabrier03}&No&No\\
			cb07\_kr&\cb&Padova94+Marigo07&BaSeL 3.1&{Kroupa01}&No&No\\
			cb07\_sa&\cb&Padova94+Marigo07&BaSeL 3.1&{Salpeter55}&No&No\\
			m05\_sa&\ma\tablenotemark{c}&\cite{CassisiS1997a,CassisiS1997b,CassisiS2000a}&BaSeL 3.1&{Salpeter55}&No&No\\
			m05\_kr&\ma&\cite{CassisiS1997a,CassisiS1997b,CassisiS2000a}&BaSeL 3.1&{Kroupa01}&No&No\\
			galev0\_sa&\galev\tablenotemark{d}&Padova94&BaSeL 2.0&{Salpeter55}&No&No\\
			galev0\_kr&\galev&Padova94&BaSeL 2.0&{Kroupa01}&No&No\\
			galev\_sa&\galev&Padova94&BaSeL 2.0&{Salpeter55}&No&Yes\\
			galev\_kr&\galev&Padova94&BaSeL 2.0&{Kroupa01}&No&Yes\\
			ynII\_s &\ynII\tablenotemark{e}&\cite{PolsO1998a}\tablenotemark{g}&BaSeL 2.0&\cite{MillerG1979a}\tablenotemark{i}&No&No\\
			ynII\_b &\ynII&\cite{PolsO1998a}&BaSeL 2.0&\cite{MillerG1979a}&Yes&No\\
			bpass\_s&\bpass\tablenotemark{f}&\cite{EldridgeJ2008a}\tablenotemark{h}&BaSeL 3.1&Broken power law\tablenotemark{j}&No&No\\
			bpass\_b&\bpass&\cite{EldridgeJ2008a}&BaSeL 3.1&Broken power law&Yes&No\\
			\bottomrule
		\end{tabular}
		\tablecomments
		{
			\tablenotetext{a}{\url{http://www.bruzual.org/bc03/}}
			\tablenotetext{b}{\url{http://www.bruzual.org/cb07/}}
			\tablenotetext{c}{\url{http://www-astro.physics.ox.ac.uk/~maraston/SSPn/SED/}}
			\tablenotetext{d}{\url{http://model.galev.org/}}
			\tablenotetext{e}{\url{http://www1.ynao.ac.cn/~zhangfh/YN_SP.html}}
			\tablenotetext{f}{\url{http://www.bpass.org.uk/}}
			\tablenotetext{g}{Based on the Cambridge stellar evolutionary tracks as given by the rapid stellar evolution code of \cite{HurleyJ2000a,HurleyJ2002a}.}
			\tablenotetext{h}{Based on a detailed stellar evolution with a custom version of the Cambridge STARS stellar evolution code.}
			\tablenotetext{i}{This IMF is supported by the studies of \cite{KroupaP1993a} and \cite{ZoccaliM2000a}.}
			\tablenotetext{j}{A IMF with a slope of $-1.30$ from $0.1$ to $0.5\Msun$ and $-2.35$ from $0.5$ to $300\Msun$, which is similar to that of Kroupa01 and Chabrier03.}
		}
		\label{tab:SSPs}
\end{table*}
\subsubsection{BC03 and updated CB07} \label{ss:bc03}
The \bc\ \citep{Bruzual2003a} model is the one most widely used in the literature.
It is a good choice for a standard model that will be compared with.
Besides, the isochrone synthesis technique first introduced in this model have been employed by many other more recent models.
So, the \bc\ model is also a good representative of the set of models which have employed similar technique.
We have used the version built with the Padova 1994 evolutionary tracks, the BaSeL 3.1 spectral library, and the IMF of \cite{Chabrier2003a}, \cite{Kroupa2001a}, and \cite{Salpeter1955a}, respectively.
The model contains the SED of SSPs with ${\rm log}(age/\rm{yr})=[5\ 10.3]$ and ${\rm log}(Z/Z_{\odot})=[-2.30\ 0.70]$.
The \cb\ \citep{BruzualA2007a} model is very similar to the \bc\ model, with the former including an updated prescription \citep{Marigo2007a} for the TP-AGB evolution of low- and intermediate-mass stars, which produces much redder near-IR colors for young and intermediate-age stellar populations.
However, whether this represents a much better treatment of the TP-AGB phase remains an open issue \citep{Kriek2010a,Zibetti2013a,CapozziD2016a}.

\subsubsection{M05} \label{ss:m05}
The \ma\ \citep{Maraston2005a} model is also very widely used in many works and often used to be compared with the \bc\ model.
A main feature of this model lies on its treatment of the post-main sequence stellar evolution stages, such as TP-AGB, based on the fuel consumption theorem.
The contribution of TP-AGB stars is expected to be crucial for modelling the SEDs of young and intermediate age ($0.1-2\rm Gyr$) stellar populations, which predominate the $1.5 \lesssim z\lesssim 3$ redshift range \citep{Maraston2005a,MarastonC2006a,Henriques2011}.
Except for the different treatment of TP-AGB stars, \ma\ model has employed the input stellar evolution tracks/isochrones of \cite{CassisiS1997a,CassisiS1997b,CassisiS2000a}, which is different from that used in \bc\ and \cb\ model.
The public version  of \ma\ model contains the SED of SSPs with ${\rm log}(age/\rm{yr})=[3\ 10.2]$ and ${\rm log}(Z/Z_{\odot})=[-2.25\ 0.67]$.
In this work, we have used the version with a red horizontal branch morphology, and the IMF of \cite{Kroupa2001a} and \cite{Salpeter1955a}, respectively.

\subsubsection{GALEV} \label{ss:galev}
The GALEV (GALaxy EVolution) evolutionary synthesis model \citep{KotullaR2009a} has many properties that are in common with the \bc\ model.
What makes the GALEV model special is its consistent treatment of the chemical evolution of the gas and the spectral evolution of the stellar content.
However, to be more easily compared with the SSPs from other groups, we prefer to use the version with metallicity fixed to some specific values, instead of that obtained with a chemically consistent treatment.
Actually, we just want to select an SSP model that has nebular emission included, while the \galev\ model is the only one which we found to be publicly available and much easier to obtain.
Although the treatment of nebular emission in \galev\ model is relatively simple, it is still useful to test the importance of including nebular emission in the SED model of galaxies.
We have used the web interface at \url{http://model.galev.org/model_1.php} to generate the SED of SSPs with ${\rm log}(age/\rm{yr})=[6.6\ 10.2]$ and ${\rm log}(Z/Z_{\odot})=[-1.7\ 0.4]$.
Both the version with and without the contribution of nebular emission have been used in this work.

\subsubsection{Yunnan-II} \label{ss:ynII}
The Yunnan models have been built at our binary population synthesis (BPS) team of Yunnan observatory \citep{ZhangF2004a,Zhang2005a,Zhang2005b,Han2007a}.
The main feature of these models is the consideration of various binary interactions, which is implemented with the help of a Monte Carlo technique.
The Yunnan models have employed the Cambridge stellar evolutionary tracks in the form given in the rapid stellar evolution code of \cite{HurleyJ2000a,HurleyJ2002a} as a set of analytic evolution functions fitted to the model tracks of \cite{PolsO1998a}.
In this work,we have used the Yunnan-II version \citep{Zhang2005a} with the BaSeL 2.0 spectral library, and the IMF of \cite{MillerG1979a}.
The model contains the SED of SSPs with ${\rm log}(age/\rm{yr})=[5.0\ 10.2]$ and ${\rm log}(Z/Z_{\odot})=[-2.3\ 0.18]$.
To test the importance of considering the effects of binary interactions, both the version with and without binary interactions have been used in this work.

\subsubsection{BPASS} \label{ss:bpass}
The Binary Population and Spectral Synthesis (BPASS) code is another publicly available population synthesis model which has considered the effects of binary evolution in the SED modelling of stellar populations.
Instead of an approximate rapid population synthesis method, detailed stellar evolution models, which are obtained with a custom version of the long-established Cambridge STARS stellar evolution code, have been used in the code.
The authors of the model also try to only use theoretical model spectra with as few empirical inputs as possible in the population syntheses to create synthetic models as pure as possible to compare with observations.
In this work, we have used the V2.0 fiducial models which have assumed a broken power law IMF with the slope to be $-1.30$ from $0.1$ to $0.5\Msun$, and $-2.35$ from $0.5$ to $300\Msun$.
The model contains the SED of SSPs with ${\rm log}(age/\rm{yr})=[6.0\ 10.0]$ and ${\rm log}(Z/Z_{\odot})=[-1.3\ 0.30]$.

The BPASS model is undergoing a rather rapid development.
During the writing of this paper, the BPASS team have released their V2.1 \citep{EldridgeJ2017a} and V2.2 \citep{StanwayE2018a}  models.
The BPASS V2.0 model, which is used in this paper, was released in 2015 and has been widely used in many stellar and extragalactic works.
However, it was not formally described in detail until the V2.1 data release paper of \cite{EldridgeJ2017a}.
There are a few refinements in the V2.1 models, but no major changes to the V2.0 results.
In \cite{EldridgeJ2017a}, the authors also discussed some key caveats and uncertainties in  their current models.
Especially, they identified several aspects of the old stellar populations ($>1 \rm Gyr$), such as the binary fraction in lower mass stars, as problematic in their current model set.
In \cite{StanwayE2018a}, the authors stated  that some of these issues have been partly addressed in their recently released V2.2 models.

Given the limitations of the BPASS V2.0 model and the improved V2.1 and V2.2 of the same model, it may seem nonsensical to still use the older one.
However, in addition to those regarding binary evolution, there are still many other uncertainties involved in the SSP model.
Given this, the model would be undergoing an intensive development for a long time, during which the older version of the same model will be rapidly replaced by the newer ones.
Actually, many of the models from other groups also have their more updated version \citep[e.g.,][]{BruzualA2011a,MarastonC2011a,Zhang2013a}.
Here, we need to point out that it is by no means the aim of this paper to find out the most cutting-edge SSP model.
In this paper, we aim at evaluating the effectiveness of applying the Bayesian model comparison method to the SSP models.
So, we prefer to use the more stable version of those models that have been used for a relatively long time, and the performance of them have been known to some extent.
Certainly, in the future, we would like to compare these more updated models using the Bayesian methods developed in this paper.

\subsection{The form of star formation history} \label{ss:sfh}
Due to its complex formation and evolution history, the detailed star formation history (SFH) of a real galaxy could be arbitrarily complex.
However, to derive the physical parameters, such as star formation rate (SFR) and  stellar mass, from the  multi-wavelength photometric SED from a very limited number of filter bands, we need to make a priori simple assumptions about its SFH. 

The exponentially declining (Exp-dec for short) SFH of the form $\psi(t)\propto e^{-t/\tau}$, the so-called $\tau$ model, is the most widely used assumption.
However, some works suggest that it leads to biased estimation of the stellar mass of individual galaxies and the stellar mass functions \citep{WuytsS2011a,Simha2014a}.
The exponentially increasing (Exp-inc for short) SFH of the form $\psi(t)\propto e^{t/\tau}$, the so-called inverted-$\tau$ model \citep{Maraston2010a,Pforr2012a}, and the delayed-$\tau$ (Delayed for short) model of form $\psi(t)\propto te^{-t/\tau}$ \citep{LeeS2010a} has been suggested to explain the SEDs of high-redshift star-forming galaxies. 
Besides, we also considered the simpler single-burst (Burst for short) and constant SFH for reference.
So, in total, we have considered five analytical forms of SFHs.

Recently, some authors have suggested some more complicated parameterizations of SFH \citep{GladdersM2013a,AbramsonL2016a,DiemerB2017a,CieslaL2017a,CarnallA2018a}, and physically motivated prescriptions of SFHs drawn from either the hydrodynamic or the semi-analytic models of galaxy formation \citep{FinlatorK2007a,Pacifici2012a,IyerK2017a}.
Besides, it is also possible to directly employ the non-parametric form of SFH, an approach that has been employed by many works \citep{Heavens2000a,CidFernandes2005a,Ocvirk2006a,Tojeiro2007a,Koleva2009a,Diaz-GarciaL2015a,MagrisC.G2015a,LejaJ2017a,ZhangH2017a}.
However, the aim of this paper is to evaluate the effectiveness of the Bayesian model comparison method and  build a reference for future study, it is better to start with much simpler forms of SFH.
We leave the exploration of these more complicated forms of SFH for future study.

\subsection{Dust attenuation curve} \label{ss:dal}
The existence of interstellar medium \citep{DraineB2010a} can significantly change the SED of the stellar populations.
For example, the UV-Optical starlight can be absorbed by the interstellar dust and re-emitted in the infrared.
Besides, the UV and ionizing photon flux from the stellar populations can be reduced by the interstellar nebular gas, and re-emitted as a nebular continuum component and strong emission lines in the optical and infrared.
In this paper, we only consider the effect of dust attenuation as a uniform dust screen with different dust attenuation laws, while leaving the consideration of dust emission for future study.

The dust attenuation law of different galaxies  are likely to be different due to different star-dust geometry and/or composition \citep{WittA2000a,ReddyN2015a,CullenF2017a,CullenF2017b}.
In this work, we have selected four widely used attenuation curves, including the \cite{Calzetti2000a} dust attenuation law for starburst galaxies (SB for short), the MW, LMC, and SMC attenuation laws\footnote{We have used the version of these attenuation curves as implemented in the HyperZ code \citep {BolzonellaM2000a}.}.
As for the nebular emission, we have selected the SSP models from \galev, which has included a self-consistent treatment of this, to test the importance of including nebular emission in the SED modeling of galaxies.
We leave the consideration of the physically motivated time-dependent attenuation model \citep{Charlot2000a} and more complicated parameterizations \citep{WittA2000a}, the more sophisticated modeling of the nebular emission with the photoionization codes, such as CLOUDY \citep{Ferland98,Ferland2013a,FerlandG2017a} and MAPPINGS \citep{SutherlandR1993a,GrovesB2004a}, for future study.

\section{Bayesian inference methods} \label{s:methods}
In BayeSED, the Bayesian inference methods are employed to interpret the SEDs of galaxies.
The base for all these methods is Bayes' theorem.
It can be used to solve both the parameter estimation problem and model comparison/selection problems.
\subsection{Bayesian parameter estimation} \label{ss:bayes_par}
With the application of Bayes' theorem in the parameter space, the posterior probability of the parameters $\bm \theta$ of a model $\bm M$ given a set of observational data $\bm d$, the model $\bm M$ itself, and all the other background assumptions $\bm I$ is related to the prior probability $p(\bm \theta|\bm M,\bm I)$ and the likelihood function $p(\bm d|\bm \theta, \bm M,\bm I)\equiv\mathcal{L}(\bm \theta )$ such that:
\begin{equation}
 p(\bm \theta|\bm d, \bm M,\bm I) = \frac{p(\bm d|\bm \theta, \bm M,\bm I) p(\bm \theta|\bm M,\bm I)}{p(\bm d|\bm M,\bm I)},
 \label{eq:bayes_theorem_par}
\end{equation}
where $p(\bm d|\bm M,\bm I)$ is a normalization factor called Bayesian evidence, or model likelihood.
With the joint posterior parameter probability distribution in Equation \ref{eq:bayes_theorem_par}, the marginalized posterior probability distribution for each parameter $\theta_X$ can be obtained as:
\begin{multline}
	p(\theta_X|\bm d, \bm M, \bm I) =\\
	\int p(\bm \theta|\bm d, \bm M,\bm I)
	\mathrm{d}\theta_1 \cdots \mathrm{d}\theta_{X-1} \mathrm{d}\theta_{X+1} \cdots \mathrm{d}\theta_{N} .
\end{multline}
The mean, median, or maximum of the marginalized posterior probability distribution can be used as an estimation of the value of a parameter, while the typical width of the distribution can be used as an estimation of the associated uncertainty.

Assuming a Gaussian form of noise, we define the likelihood function for $n$ independent wavelength band as:
\begin{align}
	\mathcal{L}(\bm \theta )&\equiv p(\bm d|\bm \theta, \bm M,\bm I)\notag\\
	&= \prod\limits_{i = 1}^{n} {\frac{1}{{\sqrt {2\pi } {\sigma_i}}}\exp ( - \frac{1}{2}\frac{{{{({F_{\rm obs,i}} - {F_{M(\bm \theta ),i}})}^2}}}{\sigma_i^2})},
 \label{eq:likelihood}
\end{align}
where $F_{\rm obs,i}$ and  $\sigma_i$ represent the observational flux and associated uncertainty in each band, and $F_{M(\bm \theta ),i}$ represents the value of flux for the $i$-th band predicted by the model $M$ which has a set of free parameters (as indicated by the vector $\bm \theta$).
The uncertainty $\sigma_i$ for the $i$-th band is not just the observational error, which is often an underestimation.
It is a common practice to additionally consider the potential systematic uncertainties in the observed fluxes and the systematic uncertainties in the employed model itself.
So, $\sigma_i$ should contain three terms such that:
\begin{equation}
	\sigma_i^2=\sigma_{\rm obs,i}^2 + {\sigma_{\rm sys}^{\rm obs,i}}^2 + {\sigma_{\rm sys}^{\rm model,i}}^2,
\end{equation}
where $\sigma_{\rm obs,i}$ is the purely observational error, $\sigma_{\rm sys}^{\rm obs,i}$  represents the systematic uncertainties regarding the observational procedure, and $\sigma_{\rm sys}^{\rm model,i}$ represents the systematic uncertainties regarding the modeling procedure.

In principle, $\sigma_{\rm obs,i}$ should be considered as a function of the observer-frame wavelength, while  $\sigma_{\rm sys}^{\rm model,i}$ should be considered as a function of the rest-frame wavelength.
For example, \cite{Brammer2008a} have introduced a rest-frame template error function to account for the systematic uncertainties in the SED model.
However, the form of the rest-frame template error function, which is likely to be model-dependent, must be determined in advance, instead of during the SED fitting.
Besides, the definition of a flexible form of wavelength-dependent $\sigma_{\rm obs,i}$ and $\sigma_{\rm sys}^{\rm model,i}$ would require too much free parameters, which cannot be well constrained by the limited number of photometric data.
So, in BayeSED V2.0, the two additional terms are simply defined as:
\begin{equation}
	\sigma_{\rm sys}^{\rm obs,i}=err_{\rm sys}^{\rm obs} * {F_{\rm obs,i}}
	\label{eq:sys_err_obs}
\end{equation}
and
\begin{equation}
	\sigma_{\rm sys}^{\rm model,i}=err_{\rm sys}^{\rm model} * {F_{\rm model,i}},
	\label{eq:sys_err_model}
\end{equation}
where $err_{\rm sys}^{\rm obs}$ and $err_{\rm sys}^{\rm model}$ are two wavelength-independent free parameters. 

In the literature \cite[e.g.][]{Dale2012a,Dahlen2013a}, only one of the $err_{\rm sys}^{\rm obs}$ and $err_{\rm sys}^{\rm model}$ is usually used and fixed to a pre-determined value (typically, $0.02-0.2$).
So, to start from a simpler assumption and not go beyond too much from the common practice, in this work, only $err_{\rm sys}^{\rm obs}$ is considered as a free parameter spanning $(0,1)$, while $err_{\rm sys}^{\rm model}$  is fixed to be zero.
Due to the simple definition in Equation \ref{eq:sys_err_obs} and \ref{eq:sys_err_model}, the two free parameters $err_{\rm sys}^{\rm obs}$ and $err_{\rm sys}^{\rm model}$ are likely to be degenerated with each other to some extent.
In practice, we found that the reduced $\chi^2$ tend to be closer to $1$ in most cases if only $err_{\rm sys}^{\rm obs}$ is considered as a free parameter.
Besides, We found that the free parameter $err_{\rm sys}^{\rm obs}$ can be well constrained by the data, and very close to the typical value (See Table \ref{tab:twoG} and Figures \ref{fig:pdftreex}, \ref{fig:pdftreey}).
On the other hand, if $err_{\rm sys}^{\rm model}$ is left to vary as a free parameter, the model deficiencies would be deposited in this free parameter, and it is potentially possible to use the value of $err_{\rm sys}^{\rm model}$ as an indicator of the quality of a certain model.
However, if $err_{\rm sys}^{\rm model}$ is also considered as a free parameter, the difference between different SED model as shown in the Bayesian evidence, which is the focus of this paper, would likely be diluted.
We leave the exploration of the effects of $err_{\rm sys}^{\rm model}$ and more complicated form of both $err_{\rm sys}^{\rm model}$ and $err_{\rm sys}^{\rm obs}$ for future study.

\subsection{Bayesian model comparison} \label{ss:bayes_model}
Bayesian model comparison try to compare competing models, which may have similar or different parameters, by calculating the probability of each model as a whole.
Similar to Bayesian parameter estimation, Bayesian model comparison can be achieved by the application of Bayes' theorem in the model space:
\begin{equation}
 p(\bm M|\bm d, \bm I) = \frac{p(\bm d|\bm M,\bm I) p(\bm M|\bm I)}{p(\bm d|\bm I)}.
 \label{eq:bayes_theorem_model}
\end{equation}
Here, the a priori probability distribution of models in the model space, $p(\bm M|\bm I)$, can be used to represent our aesthetic and/or empirical motivated like or dislike of a model, although it is often assumed to be uniform in practice.
The Bayesian evidence, or model likelihood of the model $\bm M$, $p(\bm d|\bm M,\bm I)$, which is just a normalization factor in Equation \ref{eq:bayes_theorem_par} and not relevant to parameter estimation, is crucial for Bayesian model comparison.
The Bayesian evidence of a model $p(\bm d |\bm M, \bm I)$ can be obtained by the marginalization (integration) over the entire parameter space:
\begin{equation}
 p(\bm d |\bm M, \bm I) \equiv {\int p(\bm d | \bm \theta, \bm M, \bm I)
 p(\bm \theta| \bm M, \bm I){\rm d}^N \bm \theta}.
\label{eq:Bayes_evidence}
\end{equation}
In Equation \ref{eq:bayes_theorem_model}, ${p(\bm d|\bm I)}$ is a normalization factor, which is not relevant to the Bayesian comparison of different models $\bm M$, but could be crucial for the Bayesian comparison of different background assumptions $\bm I$ in an even higher level of Bayesian inference.

Two models, $\bm M_2$ and $\bm M_1$, can be formally compared with the ratio of their posterior probabilities given the same observational dataset $\bm d$ and the background knowledge and assumptions $\bm I$:
\begin{equation}
\frac{{p({\bm M_2}|\bm d, \bm I)}}{{p(\bm {M_1}|\bm d, \bm I)}} = \frac{{p(\bm d|{\bm M_2}, \bm I)p({\bm M_2}| \bm I)}}{{p(\bm d|{\bm M_1}, \bm I)p({\bm M_1}| \bm I)}},
  \label{eq:bayes_model}
\end{equation}
where $p({\bm M_2}| \bm I)/p({\bm M_1}| \bm I)$ is the prior odds ratio of the two models.
If none of the two models is more favored a priori, the  Bayes factor, which is defined as
\begin{equation}
  {B_{2,1}} \equiv \frac{{p(\bm d|{\bm M_2, \bm I})}}{{p(\bm d|{\bm M_1, \bm I})}},
  \label{eq:bayes_factor}
\end{equation}
can be directly used for the Bayesian model comparison.
In practice, the empirically calibrated Jeffrey's scale \citep{Jeffreys1998a,Trotta2008a} suggests that $\rm ln(B_{2,1}) > 0, 1, 1.5$ and $5$ (corresponding to the odds of about 1:1, 3:1, 12:1 and 150:1) can be used to indicate inconclusive, weak, moderate and strong evidence in favor of $M_2$, respectively \cite[See also][]{JenkinsC2014a}.
If more than two models need to be compared, it would be convenient to define a standard model $M_0$ and compute the Bayes factors ${B_{i,0}}$ of all models with respect to the standard model.
When comparing models with their Bayes factors, it is important to make sure that the data $\bm d$ and all of the background knowledge/assumptions $\bm I$ used in all models are the same, or the results of comparison would be meaningless.

\subsection{Occam factor} \label{ss:Occam_factor}
As the prior-weighted average of likelihood over the entire parameter space, the Bayesian evidence obtained with Equation \ref{eq:Bayes_evidence} automatically implements the principle of Occam's razor.
Actually, the Bayesian evidence is determined by the trade-off between the complexity of a model and its goodness-of-fit to the data.
The Occam factor \citep[see e.g.][]{MacKayD2003a, Gregory2005a}, which represents a penalty to a model for having to finely tune its free parameters to match the observations, is related to the variety of the predictions that a model makes in the data space.
By adopting the suggestion of \cite{Gregory2005a}, we define the Occam factor of a model as:
\begin{flalign}
	{\Omega _{\bm \theta}}\triangleq \frac{p(\bm d |\bm M, \bm I)}{\mathcal{L}_{\rm max}(\hat{\bm \theta})}\label{eq:Occam_factor},
\end{flalign}
where ${\mathcal{L}_{\rm max}(\hat{\bm \theta})}$ is the maximum of the likelihood function at $\hat{\bm \theta}$.
So, the Occam factor defined here is just the ratio of average-likelihood and maximum-likelihood which is never greater than one.
It ensures that:
\begin{flalign}
	{p(\bm d |\bm M, \bm I)}\equiv{\mathcal{L}_{\rm max}(\hat{\bm \theta})}{\Omega _{\bm \theta}}.
\end{flalign}

A complex model would require a fine-tuning of its parameters to give a better fit to the observational data.
Then, a large fraction of its parameter space would be useless and consequently wasted.
In that case, its average-likelihood will be much smaller than its maximum-likelihood, which lead to  smaller Occam factor.
The Occam factor defined in Equation \ref{eq:Occam_factor} is not directly related to the algorithmic complexity of a model, and cannot be obtained independently of the observational data.
So, it would be interesting to find out whether this simple quantification of the complexity of a model is consistent with our intuition about the complexity of the model.
Some examples about this will be presented in \S \ref{s:apply}.

\section{The Bayesian evidence for the SED modeling of an individual galaxy}\label{s:ev_1SED}
When modeling the SED of a galaxy, it is clear from \S \ref{s:model} that we need to make assumptions about the SSP model, the form of SFH, and the properties of the interstellar medium given by the DAL.
Since our understandings of these physical processes are far from complete, we have many possible choices for each one of them.
Apparently, different choices of these components would result in very different SED modelings.
In this section, we introduce the methods of compute the Bayesian evidence for the different SED modelings.

In practice, it is meaningful to distinguish between two kinds of SED modelings: the SED modelings with the SSP, SFH and DAL all being fixed and the SED modelings with one of the SSP, SFH and DAL being fixed while the other two being free to vary.
The Bayesian model comparison of the first kind of SED modelings can be used to ask the question like: Which specific combination of SSP, SFH and DAL is the best?
Differently and more interestingly, the Bayesian model comparison of the second kind of SED modelings can be used to ask the question like: Which SSP/SFH/DAL is the best regardless of the choices of the other two?
In \S \ref{ss:ev_1SED_0} and  \ref{ss:ev_1SED_1}, we will introduce our method to compute the Bayesian evidence for the two different kinds of SED modelings, respectively.

\subsection{The SED modeling of a galaxy with SSP, SFH and DAL all being fixed} \label{ss:ev_1SED_0}
\begin{figure}[]
  \begin{center}
	\includegraphics[scale=0.45]{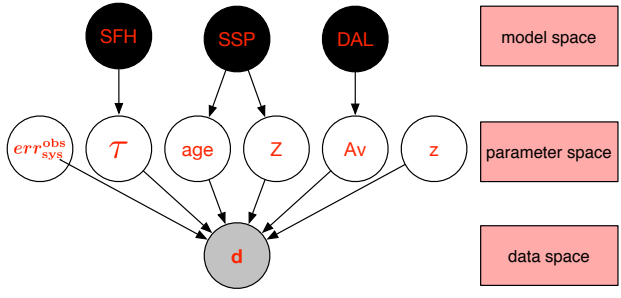}
  \end{center}
  \caption{The hierarchical structure for the ${{\bm M}({ssp}_{0},{sfh}_{0},{dal}_{0})}$ like SED modeling of a galaxy, where SSP, SFH and DAL are fixed to ${ssp}_{0}$, ${sfh}_{0}$, and ${dal}_{0}$, respectively. The black nodes indicate certain quantities (or fixed parameters), while the empty nodes indicate uncertain quantities (or free parameters). The gray nodes indicate observational data with errors. In the language of Bayesian hierarchical modeling, SSP, SFH, and DAL are called hyperparameters. They are just used to indicate the different selections of the three uncertain components. For the SED modeling of galaxies, they define a three-dimensional model space. Finally, the conditional dependence between nodes are specified with arrow lines. Hereafter, we set the ${{\bm M}({ssp}_{0},{sfh}_{0},{dal}_{0})}$ like SED modeling of a galaxy with ${ssp}_{0}=$bc03\_ch, ${sfh}_{0}=$Exp-dec, and ${dal}_{0}=$SB-like as the standard model $M_0^1$.}
  \label{fig:HBM000}
\end{figure}

Since we have many possible choices for the SSP, SFH and DAL when modeling the SED of galaxies, it would be interesting to ask:
within all the possible choices, which combination of the SSP, SFH and DAL is the best for the interpretation of a given observational data?
This question can be answered by the Bayesian model comparison of the ${{\bm M}({ssp}_{0},{sfh}_{0},{dal}_{0})}$ like SED model  which has assumed a specific SSP model ${ssp}_{0}$, SFH ${sfh}_{0}$, and  DAL ${dal}_{0}$, respectively.
The hierarchical (or multilevel) structure of this kind of SED modeling of a galaxy is shown in Figure \ref{fig:HBM000}.

As mentioned above, the computation of Bayesian evidence is crucial for the Bayesian model comparison.
The Bayesian evidence for a ${{\bm M}({ssp}_{0},{sfh}_{0},{dal}_{0})}$ like SED model can be obtained as:
\begin{flalign}
	&p(\bm d_1|{{\bm M}({ssp}_{0},{sfh}_{0},{dal}_{0})},\bm I)=\notag\\
	&\int p(\bm d_1|\bm \theta_1,{{\bm M}({ssp}_{0},{sfh}_{0},{dal}_{0})},\bm I)\notag\\
	&p(\bm \theta_1|{{\bm M}({ssp}_{0},{sfh}_{0},{dal}_{0})},\bm I) \mathrm{d}\bm \theta_1\label{eq:bayes_ev_1SED_000}\\
	&\equiv\mathcal{L}_{\rm max}(\hat{\bm \theta_1}){\Omega _{\bm \theta_1}}\label{eq:Omega_1SED_000},
\end{flalign}
where
\begin{multline}
	\mathcal{L}_{\rm max}(\hat{\bm \theta_1}) \equiv {\mathop {\rm max} \limits_ {\bm \theta_1}}[p(\bm d_1|{\bm \theta_1}, {{\bm M}({ssp}_{0},{sfh}_{0},{dal}_{0})},\bm I)]
	\label{eq:MaxL_1SED_000}
\end{multline}
is the maximum of the likelihood function at $\hat{\bm \theta_1}$, and ${\Omega _{\bm \theta_1}}$ is  the defined Occam factor associated with the free parameters $\bm \theta_1$ of the ${{\bm M}({ssp}_{0},{sfh}_{0},{dal}_{0})}$ like SED model.
If we use the shorthand ``$||\,{{\bm M}({ssp}_{0},{sfh}_{0},{dal}_{0})}$'' to indicate that ${{\bm M}({ssp}_{0},{sfh}_{0},{dal}_{0})}$ is the conditioning information common to all displayed probabilities in the equation, then Equation \ref{eq:bayes_ev_1SED_000} can be significantly shortened as:
\begin{flalign}
	&p(\bm d_1|\bm I)=\notag\\
	&\int p(\bm d_1|\bm \theta_1,\bm I) p(\bm \theta_1|\bm I) \mathrm{d}\bm \theta_1\qquad||\,{{\bm M}({ssp}_{0},{sfh}_{0},{dal}_{0})}.\label{eq:bayes_ev_1SED_000_s}
\end{flalign}
Similar shorthand will be used throughout this paper.


\subsection{The SED modeling of a galaxy with one of the SSP, SFH and DAL being fixed while the other two being free to vary} \label{ss:ev_1SED_1}
\subsubsection{The case for a fixed SSP but free SFH and DAL} \label{ss:ev_1SED_011}
\begin{figure}[]
  \begin{center}
	\includegraphics[scale=0.45]{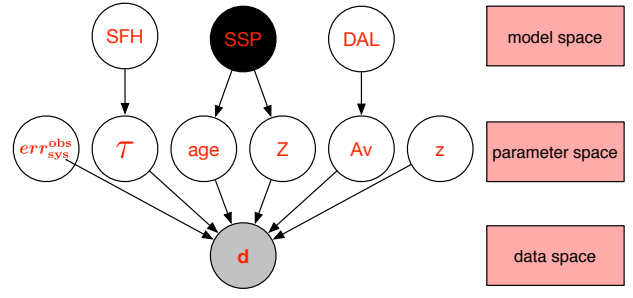}
  \end{center}
  \caption{Similar to Figure \ref{fig:HBM000}, but for the ${{\bm M}({ssp}_{0},{sfh},{dal})}$ like SED modeling, where a fixed SSP (${ssp}_{0}$), free SFH ($sfh$) and DAL ($dal$)  are assumed. Here, the selection of the form of SFH and DAL are considered as two additional free parameters, which will be marginalized out when comparing the different SSP models.}
  \label{fig:HBM011}
\end{figure}

Given the observational data of a galaxy, it is even more interesting to ask a question like: Which SSP model is the best regardless of the choices of the SFH and DAL?
To answer this question, it is useful to define a SED model ${\bm M}({ssp}_{0},{sfh},{dal})$, where the SSP model is fixed to the specific choice ${ssp}_{0}$, while the star formation history and the dust attenuation law are free to vary.
The hierarchical  structure of this kind of SED modeling of a galaxy is shown in Figure \ref{fig:HBM011}.
So, ${sfh}$ and ${dal}$ are considered as two free parameters in addition to $\bm \theta_1$, while ${ssp}_{0}$ represents a given SSP model.

The Bayesian evidence for the ${{\bm M}({ssp}_{0},{sfh},{dal})}$ like SED model can be obtained as:
\begin{flalign}
&p(\bm d_1|\bm I)=\notag&\\
&\sum\limits_{j,k} \int p(\bm d_1|\bm \theta_1, {sfh}_{j},{dal}_{k},\bm I)p(\bm \theta_1,{sfh}_{j},{dal}_{k}|\bm I)d{\bm \theta_1}\notag& \\
& \qquad||\,{{\bm M}({ssp}_{0},{sfh},{dal})}\label{eq:bayes_ev_1SED_011_a}&\\
&\equiv\mathcal{L}_{\rm max}(\hat{\bm \theta_1}, \hat{sfh},\hat{dal}){\Omega _{\bm \theta_1}}{\Omega _{sfh}}{\Omega _{dal}}\label{eq:Omega_1SED_011}&\\
&\equiv\mathcal{L}_{\rm max}(\hat{\bm \theta_1}, \hat{sfh},\hat{dal}){\Omega _{Total}}\label{eq:Omega_total_1SED_011},&
\end{flalign}
where
\begin{flalign}
	&\mathcal{L}_{\rm max}(\hat{\bm \theta_1}, \hat{sfh},\hat{dal})\notag\\
	&\equiv {\mathop {\rm max} \limits_ {\bm \theta_1,j,k}}[p(\bm d_1|{\bm \theta_1},{sfh}_{j},{dal}_{k}, {{\bm M}({ssp}_{0},{sfh},{dal})},\bm I)]
	\label{eq:MaxL_1SED_011}
\end{flalign}
 is the maximum of the likelihood function at $(\hat{\bm \theta_1}, \hat{sfh}, \hat{dal})$, and ${\Omega _{\bm \theta_1}}$, ${\Omega _{sfh}}$, and ${\Omega _{dal}}$ is  the defined Occam factor associated with the free parameters of this SED model.
 The additional Occam factors ${\Omega _{sfh}}$ and ${\Omega _{dal}}$ imply that the ${\bm M}({ssp}_{0},{sfh},{dal})$ like SED model  will be further punished for having to freely select the SFH and DAL to match the observations.

Using the product rule of probability, we can obtain the identity equation:
\begin{flalign}
&p(\bm \theta_1,{sfh}_j,{dal}_k|\bm I)=p(\bm \theta_1|{sfh}_j,{dal}_k,\bm I)p({sfh}_j,{dal}_k|\bm I)&\notag\\
&\qquad||\,{{\bm M}({ssp}_{0},{sfh},{dal})}.&
\label{eq:identity1}
\end{flalign}
So, Equation \ref{eq:bayes_ev_1SED_011_a}  can be rewritten as:
\begin{flalign}
&p(\bm d_1|\bm I)=&\notag\\
&\sum\limits_{j,k} \int p(\bm d_1|\bm \theta_1, {sfh}_{j},{dal}_{k},\bm I)p(\bm \theta_1|{sfh}_{j},{dal}_{k},\bm I)&\notag\\
&p({sfh}_{j},{dal}_{k}|\bm I) d{\bm \theta_1}\qquad||\,{{\bm M}({ssp}_{0},{sfh},{dal})}&\notag\\
&=\sum\limits_{j,k} p({sfh}_{j},{dal}_{k}|\bm I)\int p(\bm d_1|\bm \theta_1, {sfh}_{j},{dal}_{k},\bm I)&\notag\\
&p(\bm \theta_1|{sfh}_{j},{dal}_{k},\bm I) d{\bm \theta_1}\qquad||\,{{\bm M}({ssp}_{0},{sfh},{dal})}&\notag\\
&=\sum\limits_{j,k} p({sfh}_{j},{dal}_{k}|{{\bm M}({ssp}_{0},{sfh},{dal})},\bm I)&\notag\\
&p(\bm d_1|{{\bm M}({ssp}_{0},{sfh}_{j},{dal}_{k})},\bm I)\label{eq:bayes_ev_1SED_011_b}
\end{flalign}


With the assumptions that the SSP , SFH and DAL are independent of each other, and the $N_{ssp}$ of SSP , the $N_{sfh}$ of SFH , and the $N_{dal}$ of DAL are equally likely a priori, respectively, Equations \ref{eq:bayes_ev_1SED_011_b} can be further simplified as:
\begin{flalign}
&p(\bm d_1|\bm I)=&\notag\\
&\sum\limits_{j,k} p({sfh}_{j}|\bm I)p({dal}_{k}|\bm I)p(\bm d_1|\bm I)\qquad||\,{{\bm M}({ssp}_{0},{sfh},{dal})}&\notag\\
&=\frac{1}{N_{sfh}N_{dal}}\sum\limits_{j,k} p(\bm d_1|{{\bm M}({ssp}_{0},{sfh}_{j},{dal}_{k})},\bm I)\label{eq:bayes_ev_1SED_011}.&
\end{flalign}

\begin{figure}[]
  \begin{center}
	\includegraphics[scale=0.45]{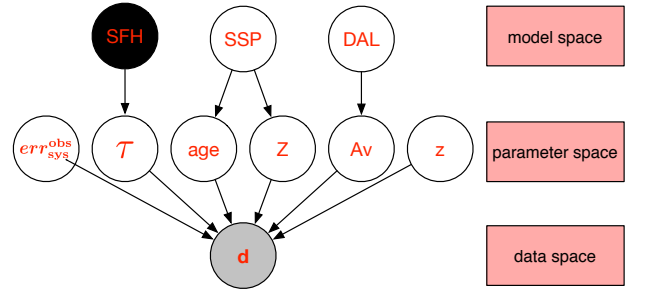}
  \end{center}
  \caption{Similar to Figure \ref{fig:HBM011}, but for the ${{\bm M}({ssp},{sfh}_{0},{dal})}$ like SED modeling, where a fixed SFH (${sfh}_{0}$), free SSP ($ssp$) and DAL ($dal$)  are assumed. Similarly, the uncertain selection of SSP model and the form of DAL will be marginalized out when comparing the different forms of SFH.}
  \label{fig:HBM101}
\end{figure}

\begin{figure}[]
  \begin{center}
	\includegraphics[scale=0.45]{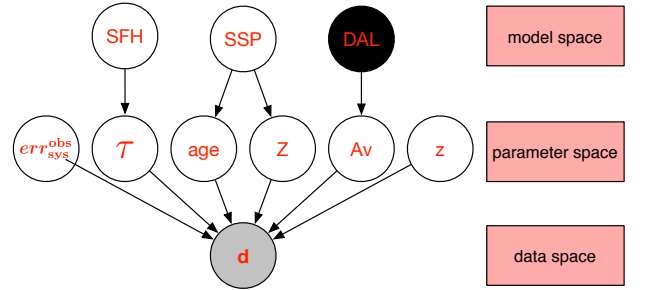}
  \end{center}
  \caption{Similar to Figure \ref{fig:HBM011}, but for the ${{\bm M}({ssp},{sfh},{dal}_{0})}$ like SED modeling, where a fixed DAL (${dal}_{0}$), free SSP ($ssp$) and SFH ($sfh$)  are assumed. Similarly, the uncertain selection of SSP model and the form of SFH will be marginalized out when comparing the different forms of DAL.}
  \label{fig:HBM110}
\end{figure}

The method of calculating the Bayesian evidence for the ${\bm M}({ssp}_{0},{sfh},{dal})$ like SED modeling presented above can be similarly applied to the ${\bm M}({ssp},{sfh}_{0},{dal})$ and ${\bm M}({ssp},{sfh},{dal}_{0})$ like SED modelings.
The hierarchical  structure of the latter two kinds of SED modelings of a galaxy are shown in Figures \ref{fig:HBM101} and \ref{fig:HBM110}, respectively.
The Bayesian evidence of ${\bm M}({ssp},{sfh}_{0},{dal})$ like SED can be obtained as:
\begin{flalign}
&p(\bm d_1|{{\bm M}({ssp},{sfh}_{0},{dal})},\bm I)&\notag\\
&=\frac{1}{N_{ssp}N_{dal}}\sum\limits_{i,k} p(\bm d_1|{{\bm M}({ssp}_{i},{sfh}_{0},{dal}_{k})},\bm I)\label{eq:bayes_ev_1SED_101}.&
\end{flalign}
It can be used to answer the question: Given the observational data of a galaxy, which SFH model is the best regardless of the choices of SSP and DAL?
Similarly, the Bayesian evidence of the ${\bm M}({ssp},{sfh},{dal}_{0})$ like SED modeling can be obtained as:
\begin{flalign}
&p(\bm d_1|{{\bm M}({ssp},{sfh},{dal}_{0})},\bm I)&\notag\\
&=\frac{1}{N_{ssp}N_{sfh}}\sum\limits_{i,j} p(\bm d_1|{{\bm M}({ssp}_{i},{sfh}_{j},{dal}_{0})},\bm I)\label{eq:bayes_ev_1SED_110}.&
\end{flalign}
It can be used to answer the question: Given the observational data of a galaxy, which DAL is the best regardless of the choices of SSP and SFH?


\section{The Bayesian evidence for the SED modeling of a sample of galaxies} \label{s:ev_NSED}
When modeling and interpreting the SEDs of a sample of galaxies, we need to make assumptions about the SSP, the form of SFH and DAL for all galaxies in the sample.
In many works in the literature, a common SSP, SFH and DAL (e.g. the BC03 SSP with a Chabrier03 IMF, exponentially declining SFH, and Calzetti law) are often assumed for all galaxies in their sample.
However, we cannot make sure that the SFH and DAL for different galaxies must be the same.
Generally, the different assumptions about the universality of SSP, SFH and DAL result in different SED modelings of a sample of galaxies, and the correctness of them need to be properly justified.
This can be achieved by the Bayesian model comparison of the SED modelings of a sample of galaxies with different assumptions about the universality of SSP, SFH and DAL.
The foundation for this kind of study is the computation of the Bayesian evidences for the different cases.
In this paper, we limit ourselves to two kinds of SED modelings of a sample of galaxies: the one with SSP, SFH and DAL all being assumed to be universal, and the one with one of the SSP, SFH and DAL being assumed to be universal while the other two object-dependent.
We introduce our method for computing the Bayesian evidence for them in \S \ref{ss:ev_NSED_UUU}, \S \ref{ss:ev_NSED_U}, respectively.

\subsection{The SED modeling of a sample of galaxies with SSP, SFH and DAL all being assumed to be universal} \label{ss:ev_NSED_UUU}
As a widely used approach when modeling and interpreting the SEDs of a sample of galaxies, the same SSP, SFH and DAL are often assumed for all galaxies in a sample, especially when the size of the sample is very large.
This is a natural choice, since it would be much more computational demanding to use different SSP, SFH and/or DAL for different galaxies when we have a large sample.
In this subsection, we introduce the method to compute the Bayesian for this case.
Although the SSP, SFH and DAL are all assumed to be universal for all galaxies in a sample, we still have many possible choices for each one of them.
This is very similar to the case for an individual galaxy in \S \ref{s:ev_1SED}.
In \S \ref{ss:ev_NSED_UUU_000}, \S \ref{ss:ev_NSED_UUU_011}, we introduce our method for computing the Bayesian evidence for the different cases respectively.

\subsubsection{The case for a fixed SSP, SFH and DAL} \label{ss:ev_NSED_UUU_000}
\begin{figure*}[]
  \begin{center}
  \includegraphics[scale=0.32]{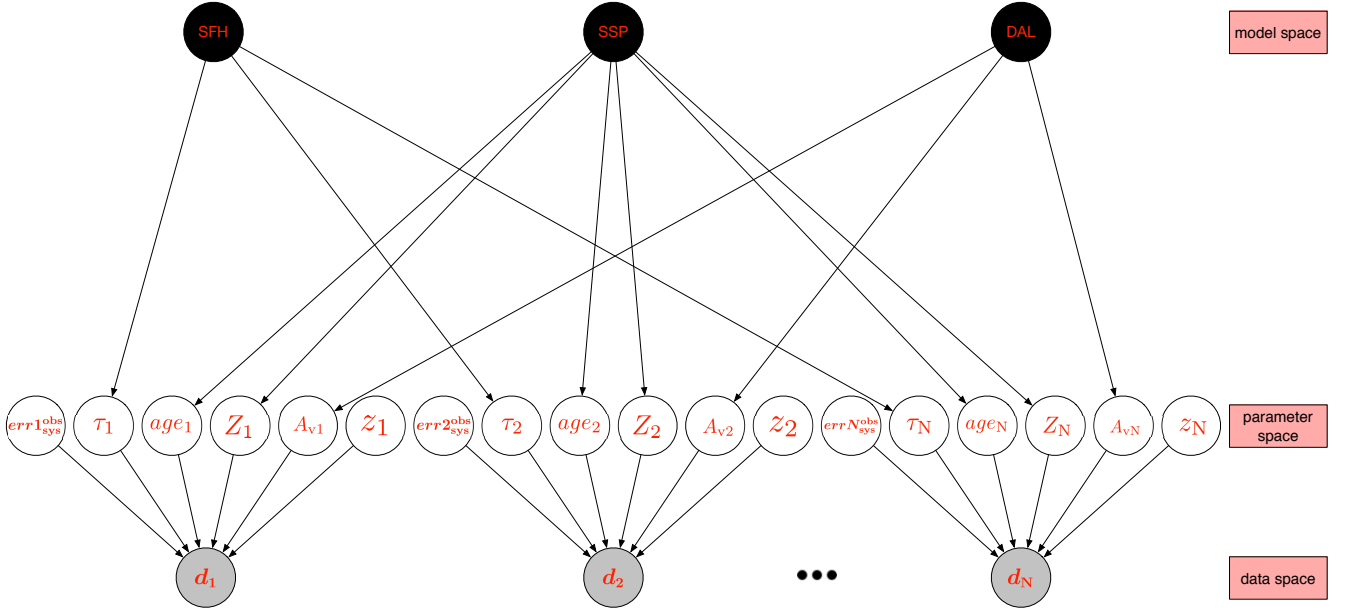}
  \end{center}
  \caption{The hierarchical structure for the ${{\bm M}({ssp}_{0},{sfh}_{0},{dal}_{0})}$ like SED modeling of a sample of $N$ galaxies, where SSP, SFH and DAL are assumed to be universal and fixed to ${ssp}_{0}$, ${sfh}_{0}$, and ${dal}_{0}$, respectively. Hereafter, we set the ${{\bm M}({ssp}_{0},{sfh}_{0},{dal}_{0})}$ like SED modeling of $N$ galaxy with ${ssp}_{0}=$bc03\_ch, ${sfh}_{0}=$Exp-dec, and ${dal}_{0}=$SB-like as the standard model $M_0^{\rm N}$.}
  \label{fig:HBM_N0}
\end{figure*}

As the most widely used approach for the SED modeling of a sample of galaxies, the ${{\bm M}({ssp}_{0},{sfh}_{0},{dal}_{0})}$ like SED modeling assumes a particular SSP, SFH, and DAL for all galaxies in a sample. 
The hierarchical  structure of this kind of SED modeling of a sample of $N$ galaxies is shown in Figure \ref{fig:HBM_N0}.
The Bayesian evidence of this kind of SED modeling for a sample of galaxies can be obtained as:
\begin{flalign}
&p(\bm d_1,\bm d_2,\dots,\bm d_N|\bm I)=\notag\\
&\int p(\bm d_1,\bm d_2,\dots,\bm d_N|{\bm \theta_1,\bm \theta_2,\dots,\bm \theta_N}, \bm I)\notag\\
& p({\bm \theta_1,\bm \theta_2,\dots,\bm \theta_N}|\bm I) {\mathrm{d}\bm \theta_1\mathrm{d}\bm \theta_2\dots\mathrm{d}\bm \theta_N}\ ||\,{{\bm M}({ssp}_{0},{sfh}_{0},{dal}_{0})}&\label{eq:bayes_ev_NSED_UUU_000_a}\\
&\equiv\mathcal{L}_{\rm max}(\hat{\bm \theta}_1,\hat{\bm \theta}_2,\dots,\hat{\bm \theta}_N){\Omega _{\bm \theta_1},\Omega _{\bm \theta_2},\dots,\Omega _{\bm \theta_N}}\label{eq:Omega_NSED_UUU_000}\\
&\equiv\mathcal{L}_{\rm max}(\hat{\bm \theta}_1,\hat{\bm \theta}_2,\dots,\hat{\bm \theta}_N){\Omega _{Total}}\label{eq:Omega_total_NSED_UUU_000},
\end{flalign}
where
\begin{multline}
	\mathcal{L}_{\rm max}(\hat{\bm \theta}_1,\hat{\bm \theta}_2,\dots,\hat{\bm \theta}_N) \equiv\\
	{\mathop {\rm max} \limits_ {\bm \theta_1,\bm \theta_2,\dots,\bm \theta_N}}[p(\bm d_1,\bm d_2,\dots,\bm d_N|{\bm \theta_1,\bm \theta_2,\dots,\bm \theta_N}\\
, {{\bm M}({ssp}_{0},{sfh}_{0},{dal}_{0})},\bm I)]
\label{eq:MaxL_NSED_UUU_000}
\end{multline}
is the maximum of the likelihood function at $(\hat{\bm \theta}_1,\hat{\bm \theta}_2,\dots,\hat{\bm \theta}_N)$, and ${\Omega _{\bm \theta_1},\Omega _{\bm \theta_2},\dots,\Omega _{\bm \theta_N}}$ is  the defined Occam factor associated with the free parameters of $N$ galaxies, respectively.

As shown in Figure \ref{fig:HBM_N0}, we assume that the observational data $\bm d_i$ of different galaxies are independent of each other, and that the parameters of a galaxy ${\bm \theta_i}$ tell nothing about the observational data $\bm d_j$ of any other galaxy.
With these assumptions, the Bayesian evidence of a ${{\bm M}({ssp}_{0},{sfh}_{0},{dal}_{0})}$ like SED model in Equation \ref{eq:bayes_ev_NSED_UUU_000_a} can be obtained as:
\begin{flalign}
&p(\bm d_1,\bm d_2,\dots,\bm d_N|{{\bm M}({ssp}_{0},{sfh}_{0},{dal}_{0})},\bm I)=&\notag\\
&\prod\limits_{g = 1}^N\int p(\bm d_g|{\bm \theta_g}, {{\bm M}({ssp}_{0},{sfh}_{0},{dal}_{0})},\bm I)&\notag\\
&p({\bm \theta_g}|{{\bm M}({ssp}_{0},{sfh}_{0},{dal}_{0})},\bm I) {\mathrm{d}\bm \theta_g}&\notag\\
&=\prod\limits_{g = 1}^N p(\bm d_g|{{\bm M}({ssp}_{0},{sfh}_{0},{dal}_{0})},\bm I).
\label{eq:bayes_ev_NSED_UUU_000}
\end{flalign}

\subsubsection{The case for a fixed SSP but free SFH and DAL} \label{ss:ev_NSED_UUU_011}
It is interesting to check the performance of a particular SSP model for a sample of galaxies and independently of the selection of SFH and DAL.
This can be achieved by defining a ${{\bm M}({ssp}_0,{sfh},{dal})}$ like SED modeling for a sample of $N$ galaxies, where a particular SSP model  ${ssp}_0$ and a free SFH and DAL are assumed for all galaxies in the sample.
The hierarchical  structure of this kind of SED modeling of a sample of $N$ galaxies is similar to Figure \ref{fig:HBM_N0}, but with the nodes for SFH and DAL being empty.
With the Bayesian evidence for this case, we can answer the question: Given the observational dataset of a sample of $N$ galaxies, which SSP model is the best regardless of the choices of the SFH and DAL?
The Bayesian evidence for this case can be obtained as:
\begin{flalign}
&p(\bm d_1,\bm d_2,\dots,\bm d_N|\bm I)=&\notag\\
&\sum\limits_{j,k} \int p(\bm d_1,\bm d_2,\dots,\bm d_N|\bm \theta_1,\bm \theta_2,\dots,\bm \theta_N, {sfh}_{j},{dal}_{k},\bm I)&\notag\\
&p(\bm \theta_1,\bm \theta_2,\dots,\bm \theta_N,{sfh}_{j},{dal}_{k}|\bm I) d{\bm \theta_1}{d{\bm \theta_2}\dots d{\bm \theta_N}}&\notag\\
&\qquad||\,{{\bm M}({ssp}_0,{sfh},{dal})}\label{eq:bayes_ev_NSED_UUU_011_a}&\\
&\equiv\mathcal{L}_{\rm max}({\hat{\bm \theta}_1,\hat{\bm \theta}_2,\dots,\hat{\bm \theta}_N}, \hat{sfh},\hat{dal})&\notag\\
&\qquad\qquad*{{\Omega _{\bm \theta_1}}{\Omega _{\bm \theta_2}}\dots{\Omega _{\bm \theta_N}}}{\Omega _{sfh}}{\Omega _{dal}}\label{eq:Omega_NSED_UUU_011}&\\
&\equiv\mathcal{L}_{\rm max}({\hat{\bm \theta}_1,\hat{\bm \theta}_2,\dots,\hat{\bm \theta}_N}, \hat{sfh},\hat{dal}){\Omega _{Total}}\label{eq:Omega_total_NSED_UUU_011},&
\end{flalign}
where
\begin{multline}
	\mathcal{L}_{\rm max}(\hat{\bm \theta}_1,\hat{\bm \theta}_2,\dots,\hat{\bm \theta}_N, \hat{sfh},\hat{dal}) \equiv\\
	{\mathop {\rm max} \limits_ {\bm \theta_1,\bm \theta_2,\dots,\bm \theta_N,j,k}}[p(\bm d_1,\bm d_2,\dots,\bm d_N|{\bm \theta_1,\bm \theta_2,\dots,\bm \theta_N,{sfh}_{j},{dal}_{k}}\\
, {{\bm M}({ssp}_{0},{sfh},{dal})},\bm I)]
\label{eq:MaxL_NSED_UUU_011}
\end{multline}
is the maximum of the likelihood function at $(\hat{\bm \theta}_1,\hat{\bm \theta}_2,\dots,\hat{\bm \theta}_N, \hat{sfh},\hat{dal})$, and ${\Omega _{\bm \theta_1},\Omega _{\bm \theta_2},\dots,\Omega _{\bm \theta_N}}$ is  the defined Occam factor associated with the free parameters of $N$ galaxies, respectively.
Since the SFH and DAL are assumed to be universal for all galaxies in the sample, we only need to add two free parameters ($sfh$ and $dal$) to represent the selection of the form of SFH and DAL.
 The associated two additional Occam factors ${\Omega _{sfh}}$ and ${\Omega _{dal}}$ imply that the ${\bm M}({ssp}_{0},{sfh},{dal})$ like SED modeling for a sample of galaxies  will be further punished for having to freely select the SFH and DAL to match the observations.

As in Equation \ref{eq:identity1},  we can obtain a similar identity equation for $N$ galaxies as:
\begin{flalign}
&p(\bm \theta_1,\bm \theta_2,\dots,\bm \theta_N,{sfh}_{j},{dal}_{k}|\bm I)=&\notag\\
&p(\bm \theta_1,\bm \theta_2,\dots,\bm \theta_N|{sfh}_{j},{dal}_{k},\bm I)p({sfh}_{j},{dal}_{k}|\bm I)&\notag\\
&\qquad||\,{{\bm M}({ssp}_0,{sfh},{dal})}&
\label{eq:identity_NSED_UUU_011}
\end{flalign}
So, the Bayesian evidence in Equation \ref{eq:bayes_ev_NSED_UUU_011_a} can be rewritten as:
\begin{flalign}
&p(\bm d_1,\bm d_2,\dots,\bm d_N|\bm I)=&\notag\\
&\sum\limits_{j,k} \int p(\bm d_1,\bm d_2,\dots,\bm d_N|\bm \theta_1,\bm \theta_2,\dots,\bm \theta_N, {sfh}_{j},{dal}_{k},\bm I)&\notag\\
&p(\bm \theta_1,\bm \theta_2,\dots,\bm \theta_N|{sfh}_{j},{dal}_{k},\bm I)p({sfh}_{j},{dal}_{k}|,\bm I)\notag&\\
&d{\bm \theta_1}{d{\bm \theta_2}\dots d{\bm \theta_N}}\qquad||\,{{\bm M}({ssp}_0,{sfh},{dal})}&\notag\\
&=\sum\limits_{j,k} p({sfh}_{j},{dal}_{k}|\bm I)&\notag\\
&\int p(\bm d_1,\bm d_2,\dots,\bm d_N|\bm \theta_1,\bm \theta_2,\dots,\bm \theta_N, {sfh}_{j},{dal}_{k},\bm I)&\notag\\
&p(\bm \theta_1,\bm \theta_2,\dots,\bm \theta_N|{sfh}_{j},{dal}_{k},\bm I)d{\bm \theta_1}{d{\bm \theta_2}\dots d{\bm \theta_N}}&\notag\\
&\qquad||\,{{\bm M}({ssp}_0,{sfh},{dal})}&\notag\\
&=\sum\limits_{j,k} p({sfh}_{j},{dal}_{k}|\bm I)p(\bm d_1,\bm d_2,\dots,\bm d_N|{sfh}_{j},{dal}_{k},\bm I)&\notag\\
&\qquad||\,{{\bm M}({ssp}_0,{sfh},{dal})}.&
\label{eq:bayes_ev_NSED_UUU_011_b}
\end{flalign}

As in Equation \ref{eq:bayes_ev_1SED_011}, we assume that the SSP, SFH and DAL are independent of each other, and the $N_{ssp}$ kinds of SSP model, the $N_{sfh}$ forms of SFH model, and the $N_{dal}$ kinds of DAL are equally likely a priori, respectively.
Besides, we assume that the physical properties of different galaxies are independent of each other.
With these assumptions, Equations \ref{eq:bayes_ev_NSED_UUU_011_b}  can be further simplified as:
\begin{flalign}
&p(\bm d_1,\bm d_2,\dots,\bm d_N|\bm I)=&\notag\\
&\sum\limits_{j,k} p({sfh}_{j}|\bm I)p({dal}_{k}|\bm I)p(\bm d_1,\bm d_2,\dots,\bm d_N|{sfh}_{j},{dal}_{k},\bm I)&\notag\\
&\qquad||\,{{\bm M}({ssp}_0,{sfh},{dal})}&\notag\\
&=\sum\limits_{j,k}p({sfh}_{j}|\bm I)p({dal}_{k}|\bm I)\prod\limits_{g = 1}^N p(\bm d_g|{{sfh}_{j}},{{dal}_{k}},\bm I)\notag&\\
&\qquad||\,{{\bm M}({ssp}_0,{sfh},{dal})}&\notag\\
&=\frac{1}{N_{sfh}N_{dal}}\sum\limits_{j,k}{\prod\limits_{g = 1}^N p(\bm d_g|{{\bm M}({ssp}_0,{sfh}_j,{dal}_k)},\bm I)}&\label{eq:bayes_ev_NSED_UUU_011}
\end{flalign}

The  above method of calculating the Bayesian evidence for the ${\bm M}({ssp}_{0},{sfh},{dal})$ like SED modeling for a sample of $N$ galaxies  can also be applied to the ${\bm M}({ssp},{sfh}_{0},{dal})$ like and ${\bm M}({ssp},{sfh},{dal}_{0})$ like SED modeling for a sample of $N$ galaxies.
The Bayesian evidence of the ${\bm M}({ssp},{sfh}_{0},{dal})$ like SED modeling for a sample of $N$ galaxies can be obtained as:
\begin{flalign}
&p(\bm d_1,\bm d_2,\dots,\bm d_N|{{\bm M}({ssp},{sfh}_0,{dal})},\bm I)&\notag\\
&=\frac{1}{N_{ssp}N_{dal}}\sum\limits_{i,k}\prod\limits_{g = 1}^N p(\bm d_g|{{\bm M}({ssp}_i,{sfh}_0,{dal}_k)},\bm I),&\label{eq:bayes_ev_NSED_UUU_101}
\end{flalign}
It can be used to answer the question: Given the observational dataset of a sample of $N$ galaxies, which SFH model is the best regardless of the choices of the SSP and DAL?
Similarly, the Bayesian evidence of the ${\bm M}({ssp},{sfh},{dal}_{0})$ like SED modeling for a sample of $N$ galaxies can be obtained as:
\begin{flalign}
&p(\bm d_1,\bm d_2,\dots,\bm d_N|{{\bm M}({ssp},{sfh},{dal}_0)},\bm I)&\notag\\
&=\frac{1}{N_{ssp}N_{sfh}}\sum\limits_{i,j}\prod\limits_{g = 1}^N p(\bm d_g|{{\bm M}({ssp}_i,{sfh}_j,{dal}_0)},\bm I),&\label{eq:bayes_ev_NSED_UUU_110}
\end{flalign}
It can be used to answer the question: Given the observational dataset of a sample of $N$ galaxies, which DAL is the best regardless of the choices of the SSP and SFH model?

\subsection{The SED modeling of a sample of galaxies with one of the SSP, SFH and DAL being assumed to be universal while the other two object-dependent} \label{ss:ev_NSED_U}
In \S \ref{ss:ev_NSED_UUU}, we have introduced the method of calculating the Bayesian evidence for the SED modeling of a sample of galaxies where the SSP, SFH and DAL are all assumed to be universal.
However, this could be too strong an assumption.
So, in this subsection we introduce the method of calculating the Bayesian evidence for the SED modelings with only one of the SSP, SFH and DAL being assumed to be universal while the other two object-dependent.

\subsubsection{The case for a universal SSP but object-dependent SFH and DAL} \label{sss:ev_NSED_UDD}
\begin{figure*}[]
  \begin{center}
  \includegraphics[scale=0.32]{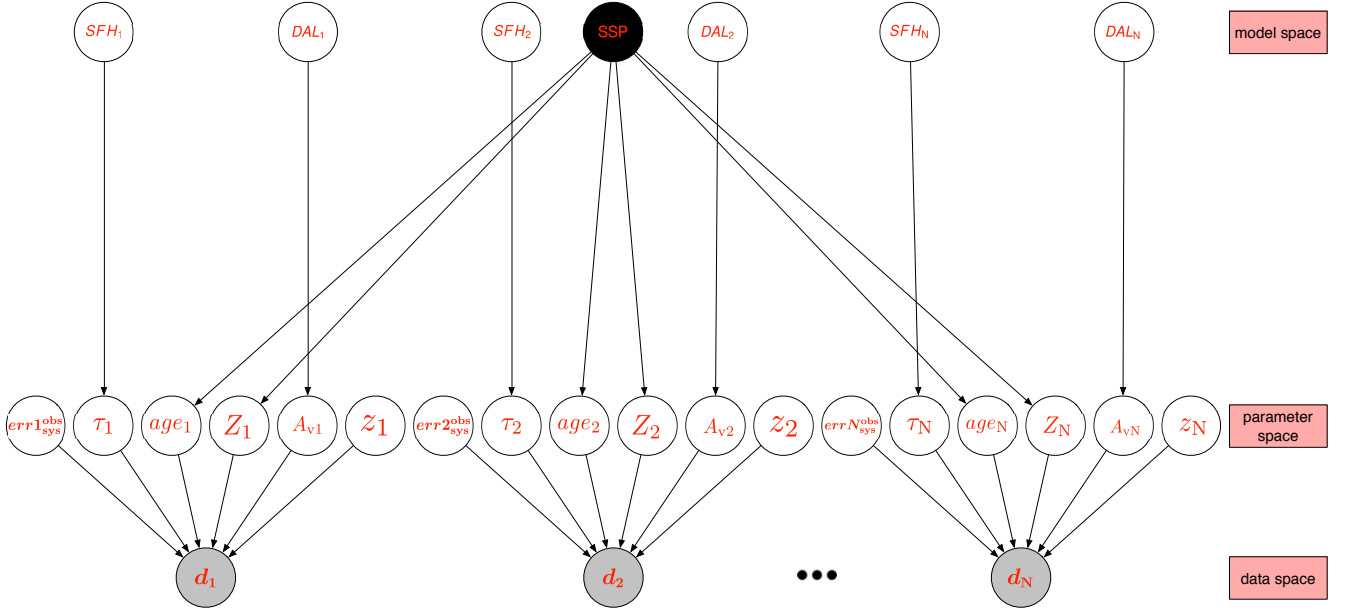}
  \end{center}
  \caption{The hierarchical structure for the ${{\bm M}({ssp}_{0},{sfh_1},{sfh_2},\dots,{sfh_N},{dal_1},{dal_2},\dots,{dal_N})}$ like SED modeling of a sample of $N$ galaxies, where a universal and fixed SSP, object-dependent and free SFH and DAL are assumed.}
  \label{fig:HBM_NUDD}
\end{figure*}

In practice, it is very interesting to ask: given the observational dataset of a sample of $N$ galaxies, which SSP model is the best regardless of the different choices of the SFH and DAL for different galaxies?
This question can be answered by calculating the Bayesian evidence for a ${{\bm M}({ssp}_{0},{sfh_1},{sfh_2},\dots,{sfh_N},{dal_1},{dal_2},\dots,{dal_N})}$ like SED modeling of a sample of $N$ galaxies where a particular SSP model ${ssp}_{0}$ is assumed for all galaxies in the sample but the form of SFH and DAL for different galaxies are allowed to be different.
The hierarchical structure of this kind of SED modeling of a sample of $N$ galaxies is shown in Figure \ref{fig:HBM_NUDD}.
The Bayesian evidence for this case can be defined as:
\begin{flalign}
&p(\bm d_1,\bm d_2,\dots,\bm d_N|\bm I)=&\notag\\
&\sum\limits_{{j_1,j_2,\dots,j_N},{k_1,k_2,\dots,k_N}} \int p(\bm d_1,\bm d_2,\dots,\bm d_N|\bm \theta_1,\bm \theta_2,\dots,\bm \theta_N&\notag\\
&, {{sfh}_{j_1},{sfh}_{j_2},\dots,{sfh}_{j_N}},{{dal}_{k_1},{dal}_{k_2},\dots,{dal}_{k_N}},\bm I)&\notag\\
&p(\bm \theta_1,\bm \theta_2,\dots,\bm \theta_N,{{sfh}_{j_1},{sfh}_{j_2},\dots,{sfh}_{j_N}}\notag&\\
&,{{dal}_{k_1},{dal}_{k_2},\dots,{dal}_{k_N}}|\bm I)d{\bm \theta_1}{d{\bm \theta_2}\dots d{\bm \theta_N}}\notag&\\
&\quad||\,{\bm M}({ssp}_{0},{sfh_1},{sfh_2},\dots,{sfh_N},{dal_1},{dal_2},\dots,{dal_N})\label{eq:bayes_ev_NSED_UDD_011_a}&\\
&\equiv\mathcal{L}_{\rm max}({\hat{\bm \theta}_1,\hat{\bm \theta}_2,\dots,\hat{\bm \theta}_N}, {\hat{sfh_1},\hat{sfh_2},\dots,\hat{sfh_N}}&\notag\\
&,{\hat{dal_1},\hat{dal_2},\dots,\hat{dal_N}}){{\Omega _{\bm \theta_1}},{\Omega _{\bm \theta_2}},\dots,{\Omega _{\bm \theta_N}}}&\notag\\
&{{\Omega _{sfh_1}},{\Omega _{sfh_2}},\dots,{\Omega _{sfh_N}}}{{\Omega _{dal_1}},{\Omega _{dal_2}},\dots,{\Omega _{dal_N}}}\label{eq:Omega_NSED_UDD_011}&\\
&\equiv\mathcal{L}_{\rm max}({\hat{\bm \theta}_1,\hat{\bm \theta}_2,\dots,\hat{\bm \theta}_N}, {\hat{sfh_1},\hat{sfh_2},\dots,\hat{sfh_N}}&\notag\\
&,{\hat{dal_1},\hat{dal_2},\dots,\hat{dal_N}}){\Omega _{Total}}\label{eq:Omega_total_NSED_UDD_011},&
\end{flalign}
where
\begin{multline}
	\mathcal{L}_{\rm max}(\hat{\bm \theta}_1,\hat{\bm \theta}_2,\dots,\hat{\bm \theta}_N, {\hat{sfh_1},\hat{sfh_2},\dots,\hat{sfh_N}}\\
	,{\hat{dal_1},\hat{dal_2},\dots,\hat{dal_N}}) \equiv\\
	{\mathop {\rm max} \limits_ {\bm \theta_1,\bm \theta_2,\dots,\bm \theta_N,{{j_1,j_2,\dots,j_N},{k_1,k_2,\dots,k_N}}}}[p(\bm d_1,\bm d_2,\dots,\bm d_N|{\bm \theta_1,\bm \theta_2}\\
	{,\dots,\bm \theta_N},{{sfh}_{j_1},{sfh}_{j_2},\dots,{sfh}_{j_N}},{{dal}_{k_1},{dal}_{k_2},\dots,{dal}_{k_N}}\\
, {{\bm M}({ssp}_{0},{sfh},{dal})},\bm I)]
\label{eq:MaxL_NSED_UDD_011}
\end{multline}
is the maximum of the likelihood function at $(\hat{\bm \theta}_1,\hat{\bm \theta}_2,\dots,\hat{\bm \theta}_N, {\hat{sfh_1},\hat{sfh_2},\dots,\hat{sfh_N}},{\hat{dal_1},\hat{dal_2},\dots,\hat{dal_N}})$, and ${{\Omega _{\bm \theta_1}},{\Omega _{\bm \theta_2}},\dots,{\Omega _{\bm \theta_N}}}$, ${{\Omega _{sfh_1}},{\Omega _{sfh_2}},\dots,{\Omega _{sfh_N}}}$, and ${{\Omega _{dal_1}},{\Omega _{dal_2}},\dots,{\Omega _{dal_N}}}$ are the defined Occam factors associated with the free parameters of the $N$ galaxies, respectively.
Since the SFH and DAL are not assumed to be universal for all galaxies in the sample, we need to add two free parameters to represent the selection of the form of SFH and DAL for each galaxy.
So, the associated $2*N$ additional Occam factors  ${{\Omega _{sfh_1}},{\Omega _{sfh_2}},\dots,{\Omega _{sfh_N}}}$ and ${{\Omega _{dal_1}},{\Omega _{dal_2}},\dots,{\Omega _{dal_N}}}$ imply that the ${\bm M}({ssp}_{0},{sfh_1},{sfh_2},\dots,{sfh_N},{dal_1},{dal_2},\dots,{dal_N})$ like SED modeling for a sample of $N$ galaxies will be further punished for having to freely select the SFH and DAL for each galaxy in the sample to match the observations.

With the identity equation as:
\begin{flalign}
&p(\bm \theta_1,\bm \theta_2,\dots,\bm \theta_N,{{sfh}_{j_1},{sfh}_{j_2},\dots,{sfh}_{j_N}}\notag&\\
&,{{dal}_{k_1},{dal}_{k_2},\dots,{dal}_{k_N}}|\bm I)=p(\bm \theta_1,\bm \theta_2,\dots,\bm \theta_N|{sfh}_{j_1}\notag&\\
&,{sfh}_{j_2},\dots,{sfh}_{j_N},{{dal}_{k_1},{dal}_{k_2},\dots,{dal}_{k_N}},\bm I)\notag&\\
&*p({{sfh}_{j_1},{sfh}_{j_2},\dots,{sfh}_{j_N}},{{dal}_{k_1},{dal}_{k_2},\dots,{dal}_{k_N}}|\bm I)\notag&\\
&\quad||\,{\bm M}({ssp}_{0},{sfh_1},{sfh_2},\dots,{sfh_N},{dal_1},{dal_2},\dots,{dal_N}),&
\label{eq:identity_NSED_UDD_011}
\end{flalign}
the Bayesian evidence in Equation \ref{eq:bayes_ev_NSED_UDD_011_a} can be rewritten as:
\begin{flalign}
&p(\bm d_1,\bm d_2,\dots,\bm d_N|\bm I)&\notag\\
&=\sum\limits_{{j_1,j_2,\dots,j_N},{k_1,k_2,\dots,k_N}} \int p(\bm d_1,\bm d_2,\dots,\bm d_N|\bm \theta_1,\bm \theta_2,\dots,\bm \theta_N&\notag\\
&, {{sfh}_{j_1},{sfh}_{j_2},\dots,{sfh}_{j_N}},{{dal}_{k_1},{dal}_{k_2},\dots,{dal}_{k_N}},\bm I)&\notag\\
&p(\bm \theta_1,\bm \theta_2,\dots,\bm \theta_N|{{sfh}_{j_1},{sfh}_{j_2},\dots,{sfh}_{j_N}}\notag&\\
&,{{dal}_{k_1},{dal}_{k_2},\dots,{dal}_{k_N}},\bm I)p({{sfh}_{j_1},{sfh}_{j_2},\dots,{sfh}_{j_N}}\notag&\\
&,{{dal}_{k_1},{dal}_{k_2},\dots,{dal}_{k_N}}|\bm I)d{\bm \theta_1}{d{\bm \theta_2}\dots d{\bm \theta_N}}\notag&\\
&||\,{\bm M}({ssp}_{0},{sfh_1},{sfh_2},\dots,{sfh_N},{dal_1},{dal_2},\dots,{dal_N})&\\
&=\sum\limits_{{j_1,j_2,\dots,j_N},{k_1,k_2,\dots,k_N}} p({{sfh}_{j_1},{sfh}_{j_2},\dots,{sfh}_{j_N}}\notag&\\
&,{{dal}_{k_1},{dal}_{k_2},\dots,{dal}_{k_N}}|\bm I)\int p(\bm d_1,\bm d_2,\dots,\bm d_N|\bm \theta_1,\bm \theta_2,\dots,\bm \theta_N&\notag\\
&, {{sfh}_{j_1},{sfh}_{j_2},\dots,{sfh}_{j_N}},{{dal}_{k_1},{dal}_{k_2},\dots,{dal}_{k_N}},\bm I)&\notag\\
&p(\bm \theta_1,\bm \theta_2,\dots,\bm \theta_N|{{sfh}_{j_1},{sfh}_{j_2},\dots,{sfh}_{j_N}}\notag&\\
&,{{dal}_{k_1},{dal}_{k_2},\dots,{dal}_{k_N}},\bm I)d{\bm \theta_1}{d{\bm \theta_2}\dots d{\bm \theta_N}}\notag&\\
&||\,{\bm M}({ssp}_{0},{sfh_1},{sfh_2},\dots,{sfh_N},{dal_1},{dal_2},\dots,{dal_N})&\\
&=\sum\limits_{{j_1,j_2,\dots,j_N},{k_1,k_2,\dots,k_N}} p({{sfh}_{j_1},{sfh}_{j_2},\dots,{sfh}_{j_N}}\notag&\\
&,{{dal}_{k_1},{dal}_{k_2},\dots,{dal}_{k_N}}|{\bm M}({ssp}_{0},{sfh_1},{sfh_2},\dots,{sfh_N}\notag&\\
&,{dal_1},{dal_2},\dots,{dal_N}),\bm I)&\notag\\
& p(\bm d_1,\bm d_2,\dots,\bm d_N|{\bm M}({ssp}_{0},{sfh}_{j_1},{sfh}_{j_2},\dots,{sfh}_{j_N}\notag&\\
&,{dal}_{k_1},{dal}_{k_2},\dots,{dal}_{k_N}),\bm I).\label{eq:bayes_ev_NSED_UDD_011_b}
\end{flalign}

With the assumption that the SSP, SFH and DAL are independent of each other, and the physical properties of different galaxies are independent of each other, Equations \ref{eq:bayes_ev_NSED_UDD_011_b}  can be further simplified as:
\begin{flalign}
&p(\bm d_1,\bm d_2,\dots,\bm d_N|{\bm M}({ssp}_{0},{sfh_1},{sfh_2},\dots,{sfh_N}&\notag\\
&,{dal_1},{dal_2},\dots,{dal_N}),\bm I)&\notag\\
&=\sum\limits_{{j_1,j_2,\dots,j_N},{k_1,k_2,\dots,k_N}} p({{sfh}_{j_1},{sfh}_{j_2},\dots,{sfh}_{j_N}}|{\bm M}({ssp}_{0}&\notag\\
&,{sfh_1},{sfh_2},\dots,{sfh_N},{dal_1},{dal_2},\dots,{dal_N}),\bm I)\notag&\\
&p({{dal}_{k_1},{dal}_{k_2},\dots,{dal}_{k_N}}|{\bm M}({ssp}_{0},{sfh_1},{sfh_2},\dots,{sfh_N}&\notag\\
&,{dal_1},{dal_2},\dots,{dal_N}),\bm I)&\notag\\
& p(\bm d_1,\bm d_2,\dots,\bm d_N|{\bm M}({ssp}_{0},{sfh}_{j_1},{sfh}_{j_2},\dots,{sfh}_{j_N}\notag&\\
&,{dal}_{k_1},{dal}_{k_2},\dots,{dal}_{k_N}),\bm I)\notag&\\
&=\sum\limits_{{j_1,j_2,\dots,j_N},{k_1,k_2,\dots,k_N}}\notag&\\
&\prod\limits_{g = 1}^N p({sfh}_{j_g}|{\bm M}({ssp}_{0},{sfh_1},{sfh_2},\dots,{sfh_N}&\notag\\
&,{dal_1},{dal_2},\dots,{dal_N}),\bm I)&\notag\\
&p({dal}_{j_g}|{\bm M}({ssp}_{0},{sfh_1},{sfh_2},\dots,{sfh_N}&\notag\\
&,{dal_1},{dal_2},\dots,{dal_N}),\bm I)p(\bm d_g|{\bm M}({ssp}_{0},{sfh}_{j_g},{dal}_{k_g}),\bm I).\label{eq:bayes_ev_NSED_UDD_011_c}
\end{flalign}

Then, we assume that the $N_{ssp}$ kinds of SSP, the $N_{sfh}$ forms of SFH, and the $N_{dal}$ kinds of DAL are equally likely a priori.
So,
\begin{flalign}
&p(\bm d_1,\bm d_2,\dots,\bm d_N|{\bm M}({ssp}_{0},{sfh_1},{sfh_2},\dots,{sfh_N}&\notag\\
&,{dal_1},{dal_2},\dots,{dal_N}),\bm I)&\notag\\
&=\sum\limits_{{j_1,j_2,\dots,j_N},{k_1,k_2,\dots,k_N}}\notag&\\
&\prod\limits_{g = 1}^N \frac{1}{N_{sfh}N_{dal}}p(\bm d_g|{ssp}_{0},{sfh}_{j_g},{dal}_{k_g},\bm I)\notag&\\
&=(\frac{1}{N_{sfh}N_{dal}})^N\sum\limits_{{j_1,j_2,\dots,j_N},{k_1,k_2,\dots,k_N}}\notag&\\
&\prod\limits_{g = 1}^N p(\bm d_g|{\bm M}({ssp}_{0},{sfh}_{j_g},{dal}_{k_g}),\bm I).\label{eq:bayes_ev_NSED_UDD_011}
\end{flalign}


The above method of calculating the Bayesian evidence for the ${{\bm M}({ssp}_{0},{sfh_1},{sfh_2},\dots,{sfh_N},{dal_1},{dal_2},\dots,{dal_N})}$ like SED modeling for $N$ galaxies can also be applied to the ${{\bm M}({ssp_1},{ssp_2},\dots,{ssp_N},{sfh}_{0},{dal_1},{dal_2},\dots,{dal_N})}$ and ${\bm M}({ssp_1},{ssp_2},\dots,{ssp_N},{sfh_1},{sfh_2},\dots,{sfh_N},{dal}_{0})$ like SED modelings.
The Bayesian evidence for the ${{\bm M}({ssp_1},{ssp_2},\dots,{ssp_N},{sfh}_{0},{dal_1},{dal_2},\dots,{dal_N})}$ like SED modeling of a sample fo $N$ galaxies can be obtained as:
\begin{flalign}
&p(\bm d_1,\bm d_2,\dots,\bm d_N|{\bm M}({ssp_1},{ssp_2},\dots,{ssp_N},{sfh}_{0}&\notag\\
&,{dal_1},{dal_2},\dots,{dal_N}),\bm I)&\notag\\
&=(\frac{1}{N_{ssp}N_{dal}})^N\sum\limits_{{i_1,i_2,\dots,i_N},{k_1,k_2,\dots,k_N}}\notag&\\
&\prod\limits_{g = 1}^N p(\bm d_g|{\bm M}({{ssp}_{i_g}},{sfh}_{0},{{dal}_{k_g}}),\bm I),\label{eq:bayes_ev_NSED_UDD_101}
\end{flalign}
It can be used to answer the question: Given the observational dataset of a sample of $N$ galaxies, which SFH model is the best regardless of the choices of the SSP and DAL for different galaxies?

Similarly, the Bayesian evidence for the ${\bm M}({ssp_1},{ssp_2},\dots,{ssp_N},{sfh_1},{sfh_2},\dots,{sfh_N},{dal}_{0})$ like SED modeling of a sample of $N$ galaxies can be obtained as:
\begin{flalign}
&p(\bm d_1,\bm d_2,\dots,\bm d_N|{\bm M}({ssp_1},{ssp_2},\dots,{ssp_N},{sfh}_{1}&\notag\\
&,{sfh_2},\dots,{sfh_N},{dal_0}),\bm I)&\notag\\
&=(\frac{1}{N_{ssp}N_{sfh}})^N\sum\limits_{{i_1,i_2,\dots,i_N},{j_1,j_2,\dots,j_N}}\notag&\\
&\prod\limits_{g = 1}^N p(\bm d_g|{\bm M}({{ssp}_{i_g}},{{sfh}_{j_g}},{dal}_{0}),\bm I),\label{eq:bayes_ev_NSED_UDD_110}
\end{flalign}
It can be used to answer the question: Given the observational dataset of a sample of $N$ galaxies, which DAL is the best regardless of the choices of the SSP and SFH for different galaxies?


\section{Application to a Ks-selected sample in the COSMOS/UltraVISTA field} \label{s:apply}
In this section, by using the new methods for calculating the Bayesian evidence, we present a Bayesian discrimination among the different choices of SSP model, SFH and DAL in the SED modeling of galaxies, with the multi-wavelength observational data of an individual galaxy (\S \ref{ss:1SED_par}, and \ref{ss:1SED_model}) and of a sample of galaxies (\S \ref{ss:NSED_model}), respectively.

\subsection{Sample selection and classification of galaxies}\label{sec:samples}
As in \cite{HanY2014a}, from the \cite{Muzzin2013a} Ks-selected catalog in the COSMOS/UltraVISTA field which provides reliable spectroscopic redshifts and photometries in 30 bands covering the wavelength range $0.15-24 \micron$, we have selected a sample of 5467 galaxies mostly with $z\lesssim 1$.
The galaxies in the sample are classified into star-forming galaxies (SFGs) and passively evolving galaxies (PEGs) according to the box regions defined in \cite{Muzzin2013b} which are similar but not identical to those defined in other works \citep{WilliamsR2009a,Whitaker2011a,BrammerG2011a}.
Specifically, PEGs are defined as:
\begin{flalign}
U - V > 1.3, V - J < 1.5\\
U - V > (V - J)\times0.88 + 0.69
\label{eq:classify}
\end{flalign} 

\begin{figure*}[]
  \begin{center}
	\includegraphics[scale=0.72]{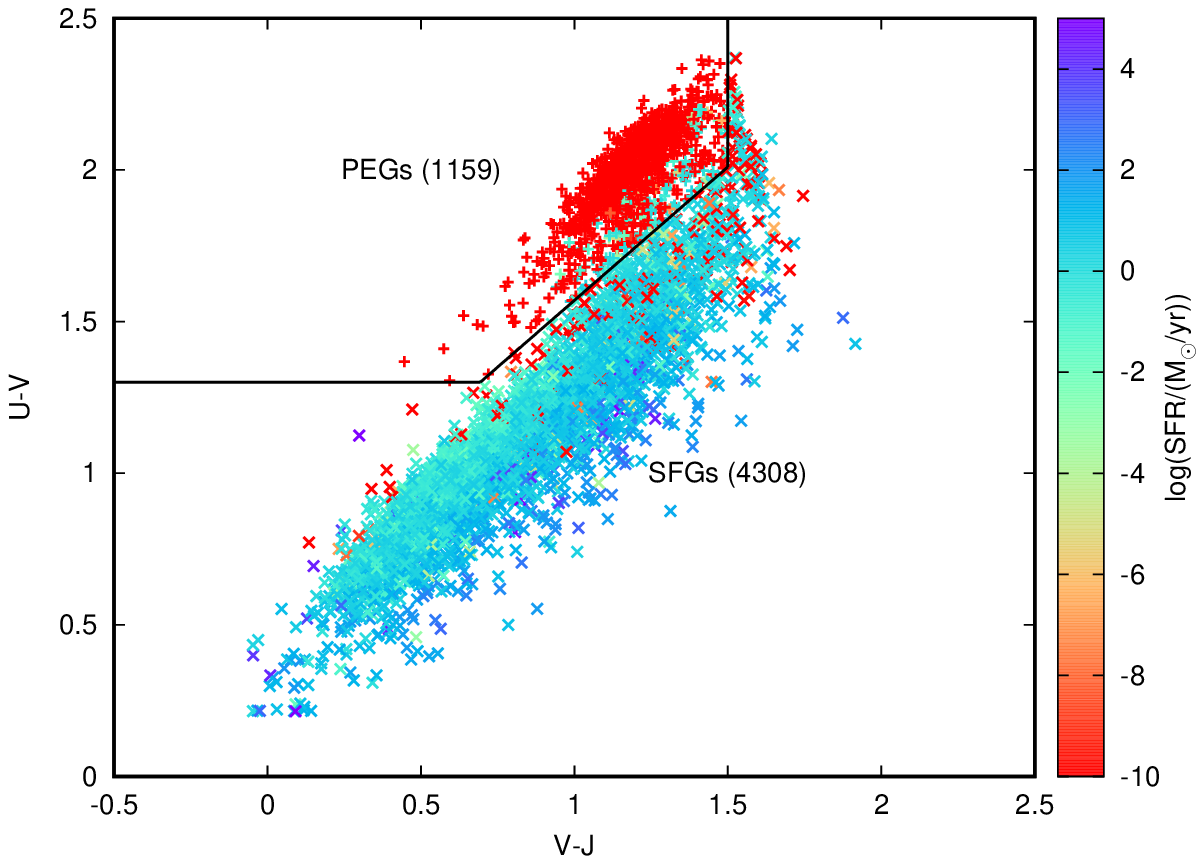}
	\includegraphics[scale=0.74]{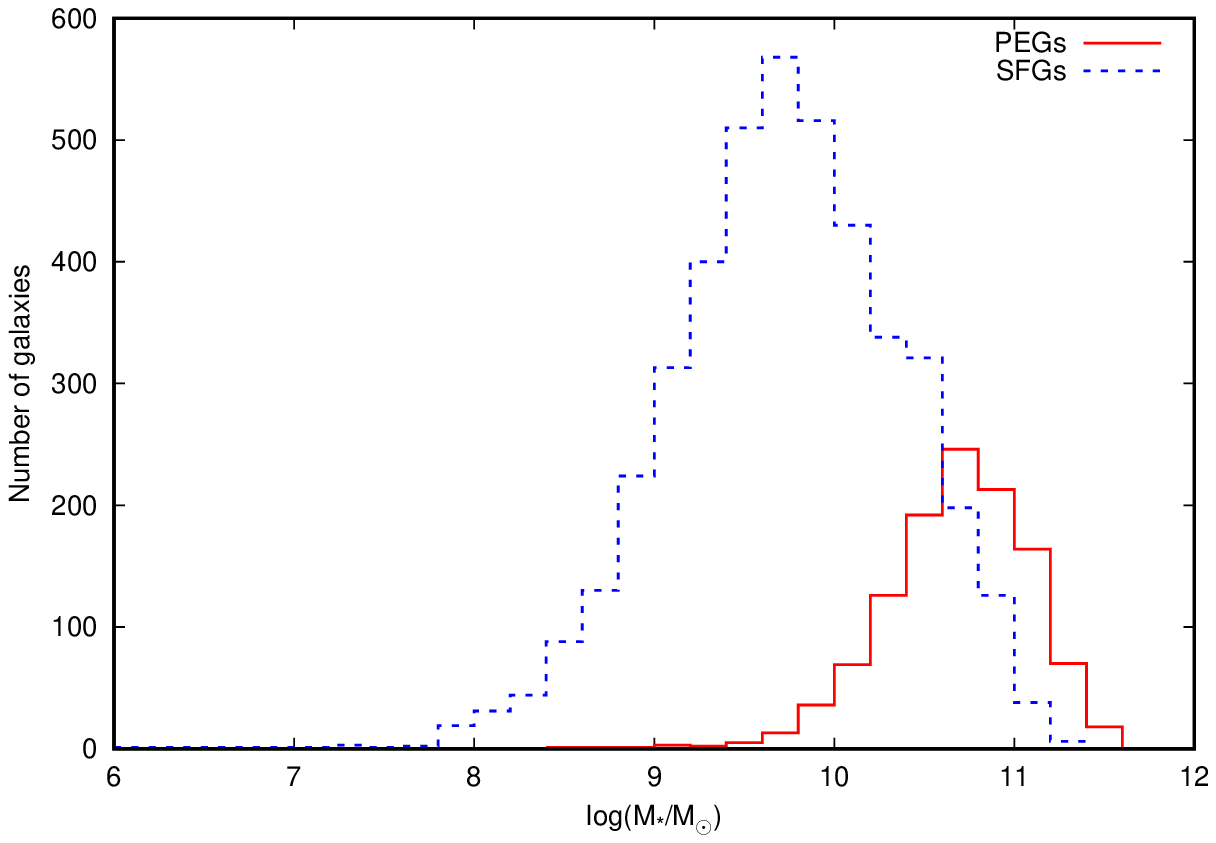}
  \end{center}
  \caption{Left: Classification of galaxies in our sample according to the definition of box regions in the UVJ color-color diagram as given by \cite{Muzzin2013b}, and color coded with SFR. Right: The distribution of stellar mass for galaxies in our sample. It is clear that the galaxies in our sample are distributed widely in both the color-color and stellar mass space.}
  \label{fig:UVJ_SFR}
\end{figure*}

Generally, there are $1159$ PEGs and $4308$ SFGs in our sample.
In the left panel of Figure \ref{fig:UVJ_SFR}, we show the distribution of galaxies in our sample in the UVJ color-color diagram.
The estimated SFRs of these galaxies with BC03 model as given in the catalog of \cite{Muzzin2013a} are shown color-coded.
It is clear that the classification of galaxies into SFGs and PEGs is consistent with the estimation of SFR.
In the right panel of Figure \ref{fig:UVJ_SFR}, we show the distribution of stellar mass for galaxies in the sample.
Most of PEGs in our sample are massive galaxies with stellar mass larger than $10^{10}\Msun$, while the SFGs spans a much wider range of stellar mass from $10^8\Msun$ to $10^{11}\Msun$.
As shown in the figure, the galaxies in our sample are distributed widely in both the color-color and stellar mass space.
The diversity of galaxies in the sample ensure that robust conclusions can be obtained with them.

\subsection{Bayesian parameter estimation for individual galaxies}\label{ss:1SED_par}
In this subsection, we demonstrate the methods of Bayesian parameter estimation with a PEG (ULTRAVISTA114558) and an SFG (ULTRAVISTA99938) by assuming the commonly used BC03 SSP model with a \cite{Chabrier2003a} IMF (bc03\_ch), an Exp-dec SFH, and the \cite{Calzetti2000a} dust attenuation law. 
The $6$ free parameters of the model and the priors for them are given in Table \ref{tab:priors}.
Generally, a uniform  prior truncated at a physically reasonable range is assumed for all free parameters.
Besides, the age of a galaxy is forced to be less than the age of the Universe at the given redshift $z$.
More physically reasonable and informative priors can be provided by assuming a model for the redshift-dependent distribution of physical parameters of galaxies.
However, in this work, we are only interested in the comparison of different SED models.
So, the truncated uniform prior only reflect the fact that the SED model itself tell us nothing about the detailed distribution of any physical parameter of galaxies, except for the allowed range.
\begin{table}
	\caption{Summary of the free parameters and priors}
	\begin{tabular}{llll}
		\toprule
		Parameter & Prior range & Prior type & \\
		z & [0 6] & Uniform & \\
		$err_{\rm sys}^{\rm obs}$ & [0 1] & Uniform & \\
		${\rm log}(age/\rm{yr})$ & [5 10.3] & Uniform and $age<age_{\rm U}(z)$& \\
		${\rm log}(Z/Z_{\odot})$ & [-2.30 0.70] & Uniform & \\
		${\rm log}(\tau/\rm{yr})$& [6 12] & Uniform & \\
		$A_{\rm v}$ & [0 4] & Uniform & \\
		\bottomrule
	\end{tabular}
	\label{tab:priors}
\end{table}

\begin{figure*}[]
  \begin{center}
	  \includegraphics[scale=0.145]{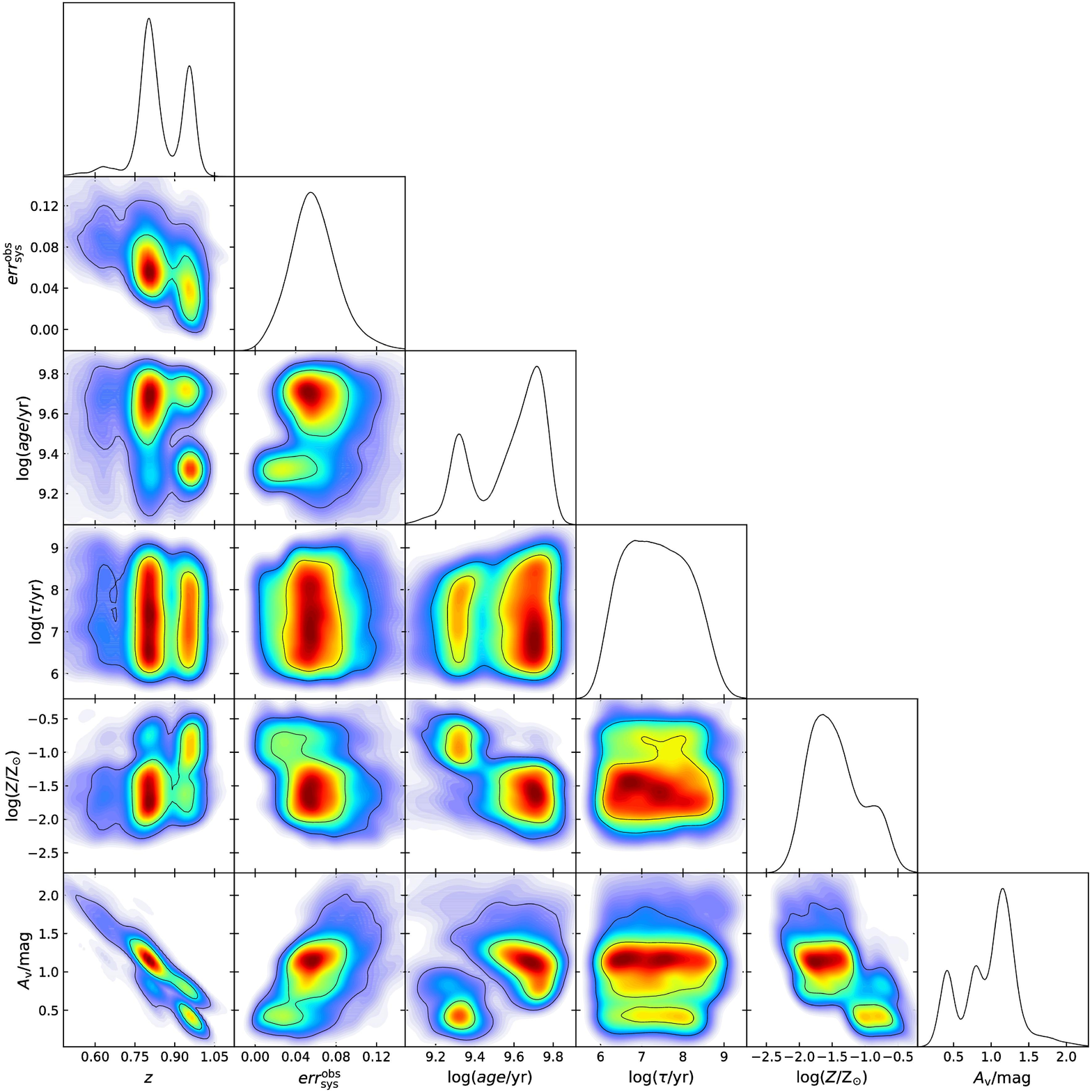}
  \end{center}
  \caption{The 1 and 2-D posterior PDFs of free parameters for the PEG ULTRAVISTA114558. They represent our state of knowledge about them. The presence of multiple peaks and/or strong correlations in the 2D PDFs indicate the degeneracies between the free parameters of the SED model.}
  \label{fig:pdftreex}
\end{figure*}

\begin{figure*}[]
  \begin{center}
	  \includegraphics[scale=0.145]{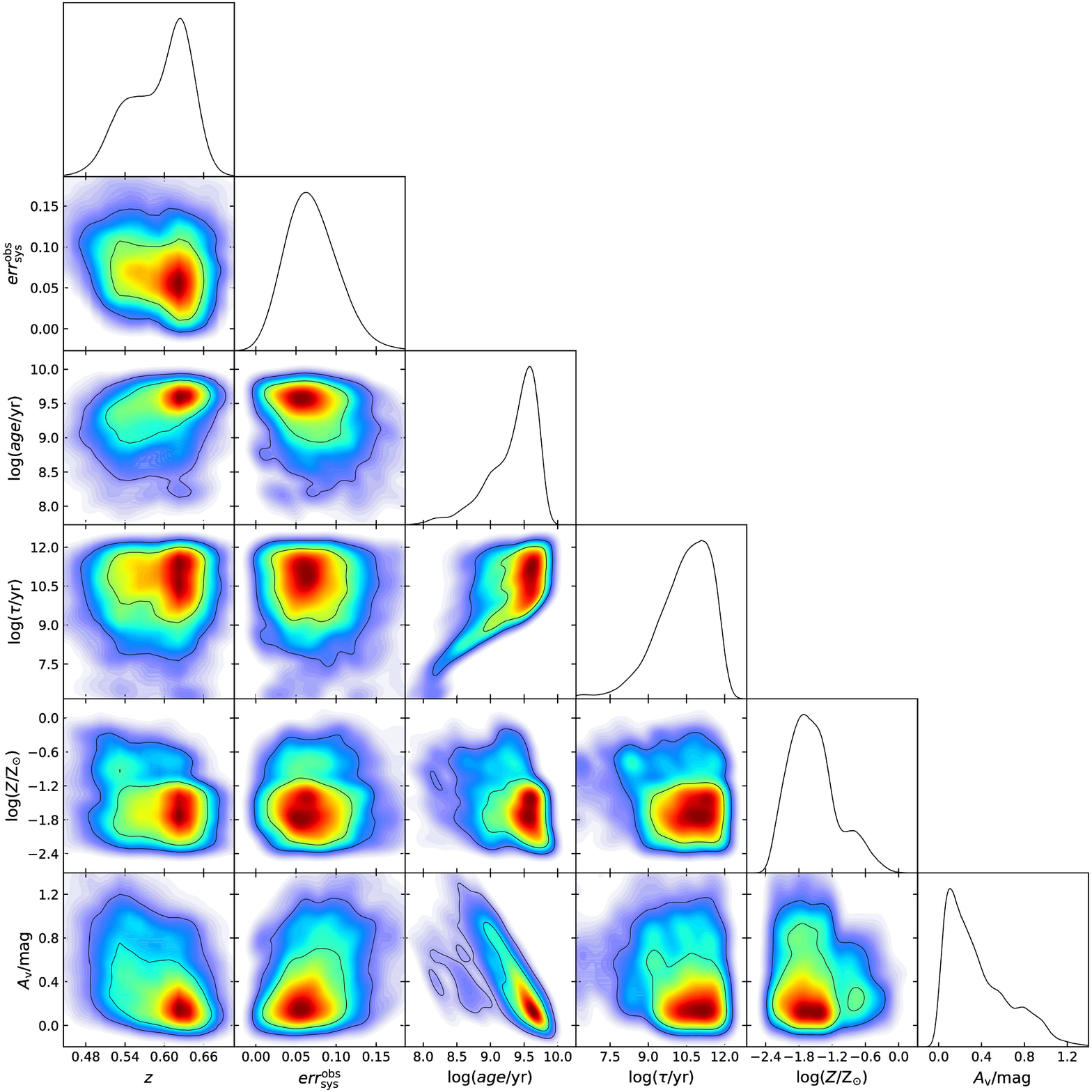}
  \end{center}
  \caption{Same as Figure \ref{fig:pdftreex}, but for the SFG ULTRAVISTA99938.}
  \label{fig:pdftreey}
\end{figure*}

As a benefit of the Bayesian parameter estimation, in addition to the best-fit results and associated estimation of parameters, the detailed posterior probability distribution functions (PDFs) for all of the free and derived parameters of a model can be obtained.
The posterior PDFs of parameters fully described our current state of knowledge about them.
In Figures \ref{fig:pdftreex} and \ref{fig:pdftreey}, we show the obtained posterior PDFs for all parameters of the PEG ULTRAVISTA114558 and the SFG ULTRAVISTA99938.
The degeneracies between free parameters can be recognized as multiple peaks and/or strong correlations in the 2D PDFs.
Besides, the peak and width of the 1D PDFs can be directly used to estimate the value and associated uncertainty of all parameters.
For example, the results of parameter estimation for the PEG ULTRAVISTA114558 and the SFG ULTRAVISTA99938 are shown in the Table \ref{tab:twoG}.
The results suggest that the PEG ULTRAVISTA114558 is only slightly older than the SFG ULTRAVISTA99938.
However, it is much more massive than the latter, and have experienced a much shorter duration of active star formation, which was started much earlier.

\begin{figure*}[]
  \begin{center}
	\includegraphics[scale=1.4]{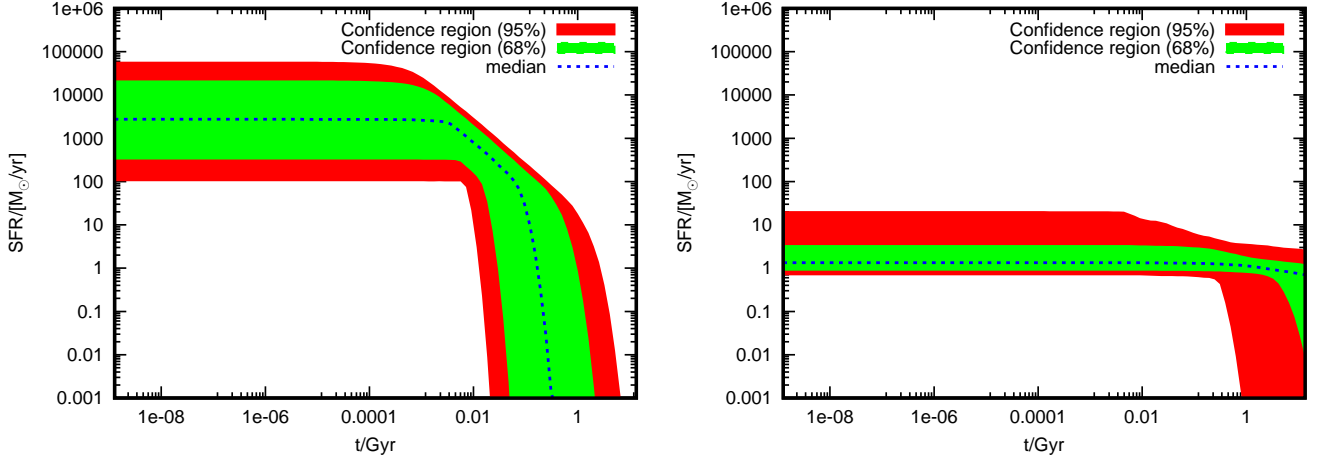}
  \end{center}
  \caption{The posterior PDF for the SFH of the PEG ULTRAVISTA114558 (left) and the SFG ULTRAVISTA99938 (right). Only the median, $68\%$ and $95\%$ credible region obtained from the posterior PDF of the SFH for the two galaxies are shown.}
  \label{fig:plot_sfh}
\end{figure*}

It is often very hard, if not impossible, to determine the SFH of a galaxy with only the photometric data.
However, with the Bayesian parameter estimation, we can at least obtain the posterior PDF for the SFH of a galaxy.
In Figure \ref{fig:plot_sfh}, we show the detailed posterior PDF for the SFHs of the PEG ULTRAVISTA114558 and SFG ULTRAVISTA99938, respectively.
It is clear from the figure that the obtained SFHs of the two galaxies are very uncertain, although the same Exp-dec SFH has been assumed for them.
However, the posterior PDF of their SFHs still allows us to recognize the very different nature of their SFHs.
The PEG ULTRAVISTA114558 has experienced a very intensive ($\gtrsim1000\rm \Msun/\rm{yr}$) star formation phase during the 1st $10$Myr, after which the star formation activity has been quenched very quickly.
Differently, the SFG ULTRAVISTA99938 has experienced a stable ($\gtrsim1\rm \Msun/\rm{yr}$) star formation phase during the 1st $1$Gyr, after which the star formation rate has only been slightly decreased.
These results are consistent with the merger-driven formation mechanism for the massive PEGs and the secular evolution of the disk dominated SFGs.

\begin{figure*}[]
  \begin{center}
	\includegraphics[scale=1.5]{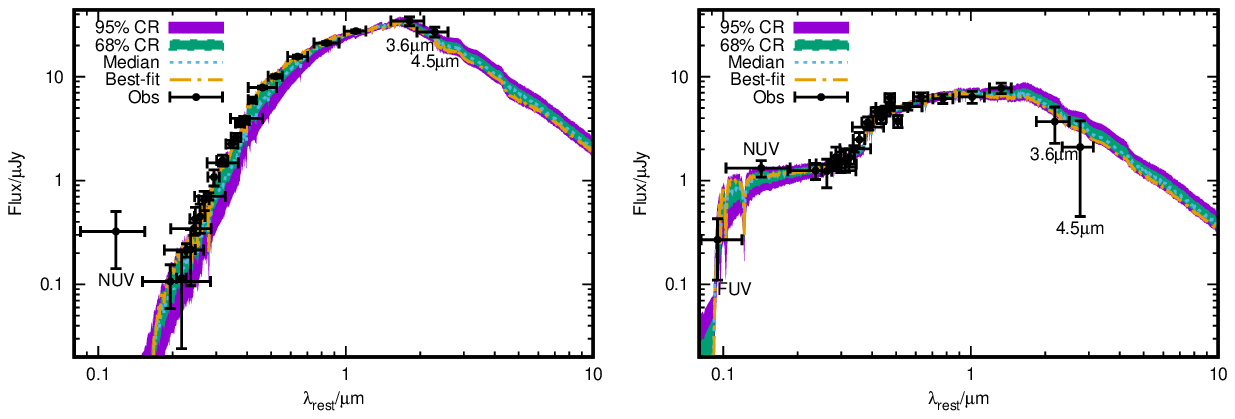}
  \end{center}
  \caption{The results of SED fitting  for the PEG ULTRAVISTA114558 (left) and the SFG ULTRAVISTA99938 (right). Except for the best-fit SED, the median, $68\%$ and $95\%$ credible region obtained from the posterior PDFs of the model SEDs are also shown. The GALEX FUV and NUV, Spitzer IRAC $3.6$ and $4.5\micron$ data have been labeled in the figure.}
  \label{fig:plot_fit}
\end{figure*}

Finally, in Figure \ref{fig:plot_fit}, we show the results of SED fitting for the PEG ULTRAVISTA114558 and the SFG ULTRAVISTA99938.
Except for the best-fit SED as can be given by the traditional SED-fitting methods, the Bayesian SED-fitting method allow us to obtain the detailed posterior PDF of the model SEDs.
From the compact credible regions  \footnote{See \url{https://en.wikipedia.org/wiki/Credible_interval} for the difference between the credible regions/intervals in Bayesian statistics and the confidence regions/intervals in frequentist statistics.} and the similarity between the median SED and the best-fit SED, it is clear that the SED model is well constrained by the data.
For the PEG ULTRAVISTA114558, the GALEX NUV data is far beyond the  $95\%$ credible region of the posterior model SEDs.
This indicates a failure of the employed SED model.
Except for the \bc\ model, we have also tested the \ynII\ and \bpass\ model, which includes UV contribution by hot stars even at older ages.
The latter two models cannot explain the data point as well.
So, it could indicate some contribution to the UV by a none-stellar (e.g. AGN) source.
For the SFG ULTRAVISTA99938, the Spitzer IRAC $3.6$ and $4.5\micron$ data are slightly below the $95\%$ credible region of the posterior model SED.
Since the nebular and dust emissions are not considered in the SED model, the bands harbor contributions from emission lines may artificially boost the observed brightness and push the model fit up. 
However, given the error bar, the data is basically consistent with the model without the contribution from dust emission.
So, this suggests that the contribution of dust emission to the two IRAC bands could be ignored.
This is consistent with the relatively strong UV emission and low dust extinction ($A_{\rm v}=0.28^{+0.42}_{-0.20}$) as shown in Table \ref{tab:twoG} .


\begin{table}
	\centering
	\begin{tabular}{@{}lll@{}}
		\toprule
		Parameter & PEG & SFG \\
		\midrule
		{\boldmath$z              $} & $0.82^{+0.13}_{-0.05}$ & $0.60^{+0.03}_{-0.07}$ \\
		{\boldmath$\sigma_{\rm sys}      $} & $0.06^{+0.02}_{-0.02}$ & $0.07^{+0.04}_{-0.03}$ \\
		{\boldmath${\rm log}(age/\rm{yr})$} & $9.63^{+0.12}_{-0.30}$ & $9.45^{+0.24}_{-0.47}$ \\
		{\boldmath${\rm log}(\tau/\rm{yr})$} & $7.33^{+0.90}_{-0.87}$ & $10.58^{+0.94}_{-1.21}$ \\
		{\boldmath${\rm log}(Z/{\rm Z_{\odot}})$} & $-1.49^{+0.54}_{-0.36}$ & $-1.60^{+0.49}_{-0.38}$ \\
		{\boldmath$A_{\rm v}/{\rm mag}$} & $1.05^{+0.23}_{-0.56}$ & $0.28^{+0.42}_{-0.20}$ \\
		$z_{form}                     $ & $2.65^{+3.86}_{-1.03}$ & $1.22^{+1.04}_{-0.49}$ \\
		${\rm log}(SFR/[\rm M_{\odot}/\rm{yr}])$ & $-67.72^{+61.42}_{-931.28}$ & $0.01^{+0.26}_{-0.15}$ \\
		$log(M_*/\rm M_{\odot})    $ & $10.78^{+0.10}_{-0.15}$ & $9.54^{+0.10}_{-0.21}$ \\
		${\rm log}(L_{\rm bol}/[\rm erg/s])$ & $44.60^{+0.06}_{-0.05}$ & $43.97^{+0.24}_{-0.12}$ \\
		\bottomrule
	\end{tabular}\caption{The estimation of free parameters (in bold font) and derived parameters for the PEG ULTRAVISTA114558 and the SFG ULTRAVISTA99938 with the Uniform prior for all free parameters.}
	\label{tab:twoG}
\end{table}


\subsection{Bayesian discrimination of SSP, SFH and DAL for the SED modeling of individual galaxies} \label{ss:1SED_model}
In \S \ref{ss:1SED_par}, we have demonstrated the results that can be obtained with the Bayesian parameter estimation methods by the application to two example galaxies.
We have assumed the standard model ($M_0^1$) with the most widely used BC03 SSP with a Chabrier03 IMF (bc03\_ch), Exp-dec SFH , and SB-like DAL. 
There are many other possible choices for SSP, SFH and DAL, and they will result in very different estimations for the physical parameters of a galaxy.
So, it is very important to find out the best choice for SSP, SFH and DAL when modeling the SED of a galaxy.
Here, we present a Bayesian discrimination of their different choices when modeling the SED of the PEG ULTRAVISTA114558, and the SFG ULTRAVISTA99938.

\subsubsection{The case for SSP, SFH and DAL being fixed}\label{sss:1SED_model_0}
We firstly consider the cases where the SSP, SFH and DAL are all fixed to a specific choice.
The standard model ($M_0^1$) mentioned above is just a special example of this kind of SED modelings of a galaxy.
With the Bayesian comparison of this kind of SED modelings, we can find out the best combination of SSP, SFH and DAL for an individual galaxy.

\begin{figure*}[]
  \begin{center}
	\includegraphics[scale=0.739]{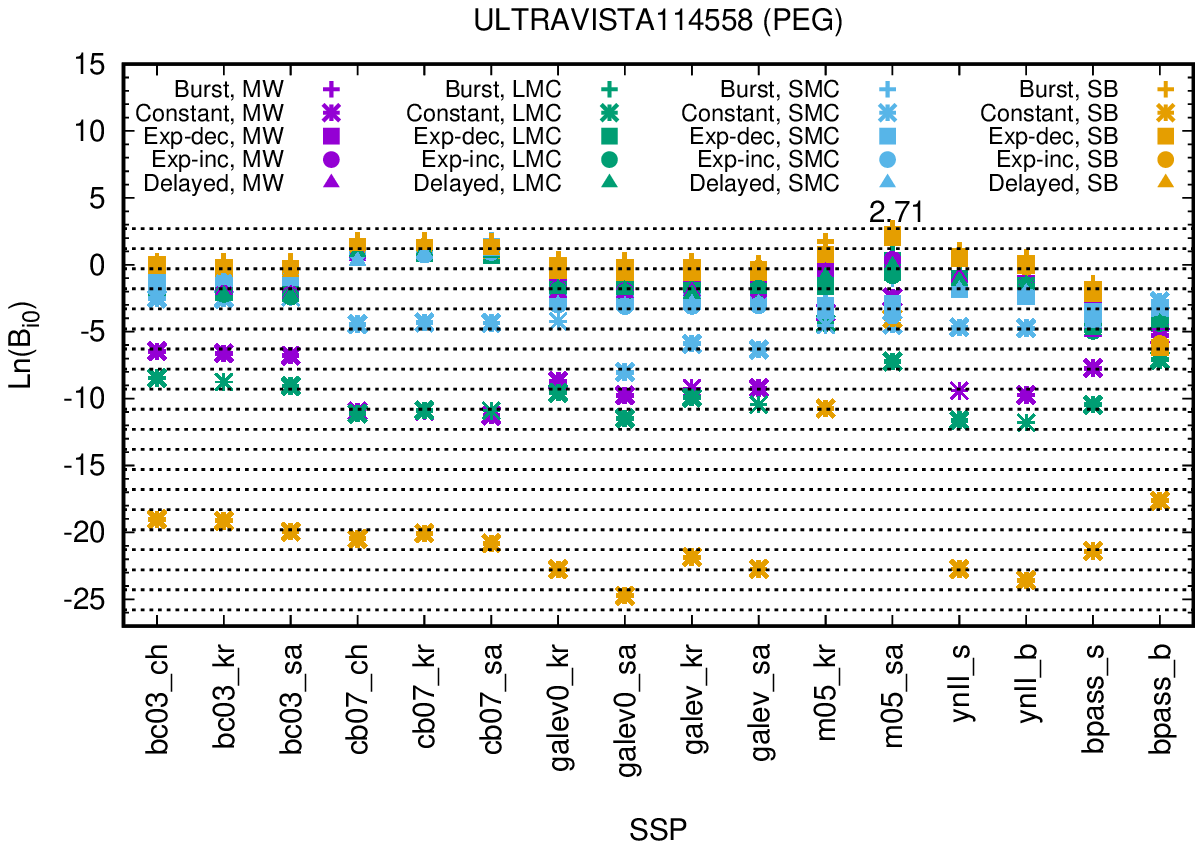}
	\includegraphics[scale=0.739]{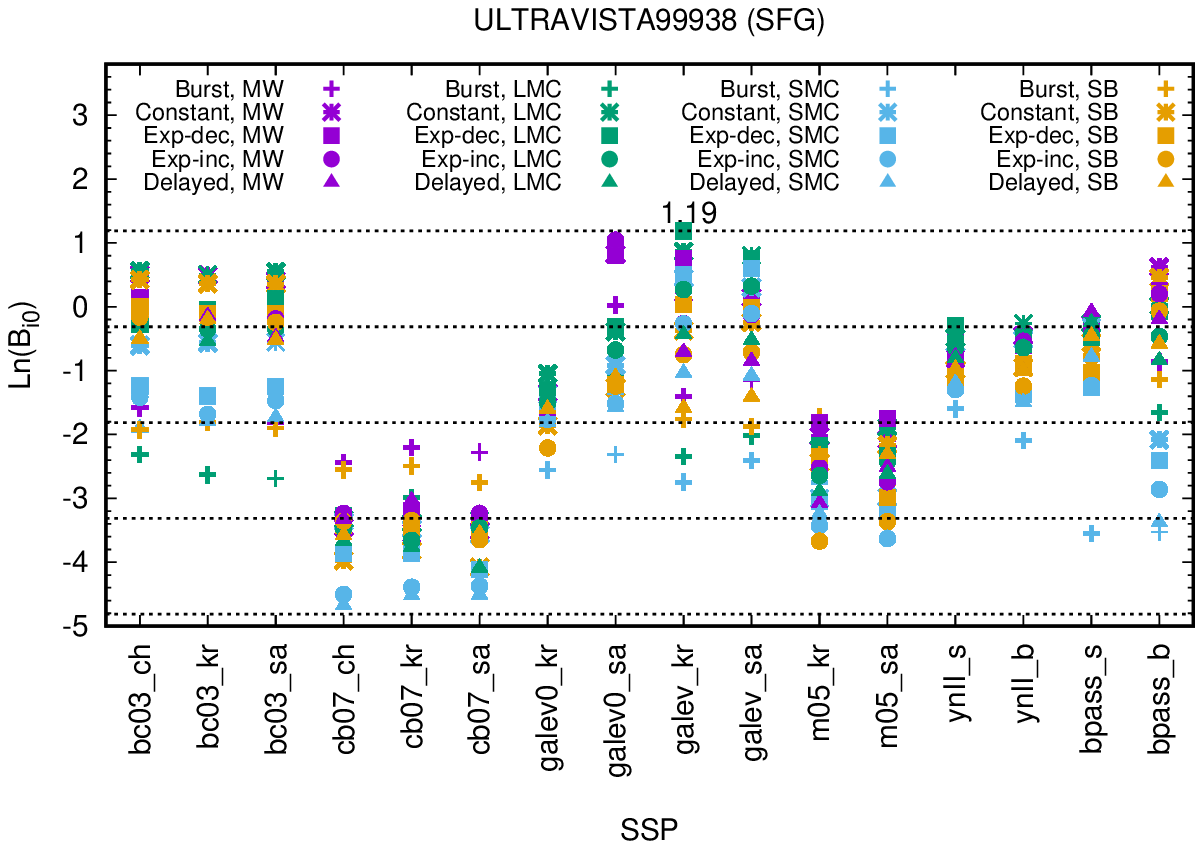}
  \end{center}
  \caption{The Bayes factors with respect to the standard model ($M_0^1$, which assumes the BC03 SSP with a Chabrier03 IMF, Exp-dec SFH and SB-like DAL) for the ${\bm M}({ssp}_{0},{sfh_0},{dal_0})$-like SED modelings of the PEG ULTRAVISTA114558 (left) and the SFG ULTRAVISTA99938 (right) where SSP (see Table \ref{tab:SSPs} for the meaning of each SSP model), SFH and DAL are all fixed to a particular choice. The dot lines show the values of the Bayes factor with a step of $1.5$. The value of Bayes factor for the model with the highest Bayes factor is also shown in the figure. For the PEG ULTRAVISTA114558, the combination of the \ma\ SSP with a Salpeter55 IMF (m05\_sa), the Burst SFH and the SB-like DAL has the highest value ($2.71$) of Bayes factor. For the SFG ULTRAVISTA99938, the combination of the version of \galev\ SSP with the consideration of emission lines and with a Kroupa01 IMF (galev\_kr), the Exp-dec SFH and the LMC-like DAL has the highest value ($1.19$) of Bayes factor. The positive value of Bayes factor indicates that the model has higher Bayesian evidence than the standard model $M_0^1$.}
  \label{fig:BF_1SED_0_sp}
\end{figure*}

In Figure \ref{fig:BF_1SED_0_sp}, we show the Bayes factors with respect to the standard model ($M_0^1$) for the SED modelings of the PEG ULTRAVISTA114558 and the SFG ULTRAVISTA99938 with all possible combinations of the  $16$ SSP, $5$ SFH and $4$ DAL.
It is clear from the figure that the Bayes factors for the SED modeling of galaxy with different SSP models could be very different, even if the same SFH and DAL are assumed.
For the PEG ULTRAVISTA114558, the combination of the M05 model with a Salpeter55 IMF (m05\_sa), the Burst SFH, and the SB-like DAL has the highest value (2.71) of Bayes factor.
For the SFG ULTRAVISTA99938, the combination of the version of the GALEV model with the consideration of emission lines and a Kroupa01 IMF (galev\_kr), the Exp-dec SFH, and the LMC-like DAL has the highest value (1.19) of Bayes factor.

\begin{figure*}[]
  \begin{center}
	\includegraphics[scale=0.73]{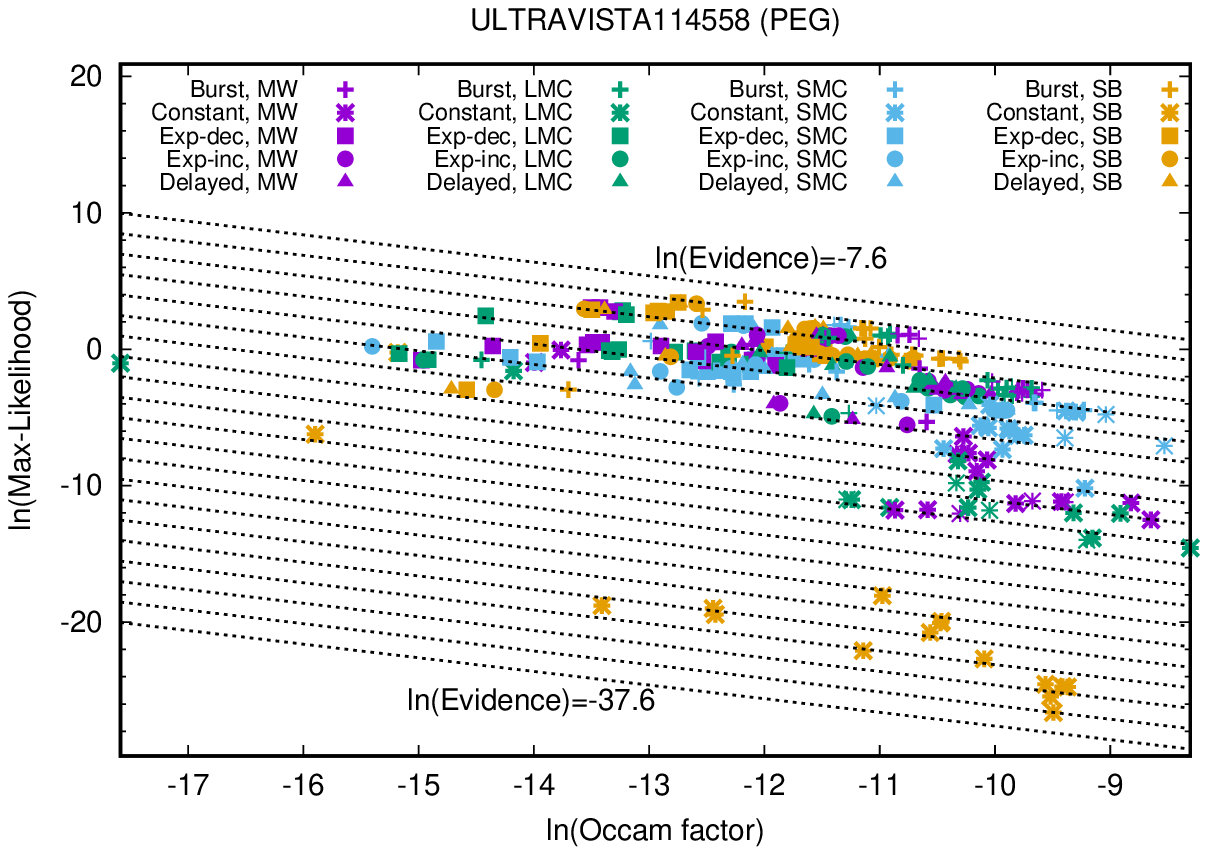}
	\includegraphics[scale=0.73]{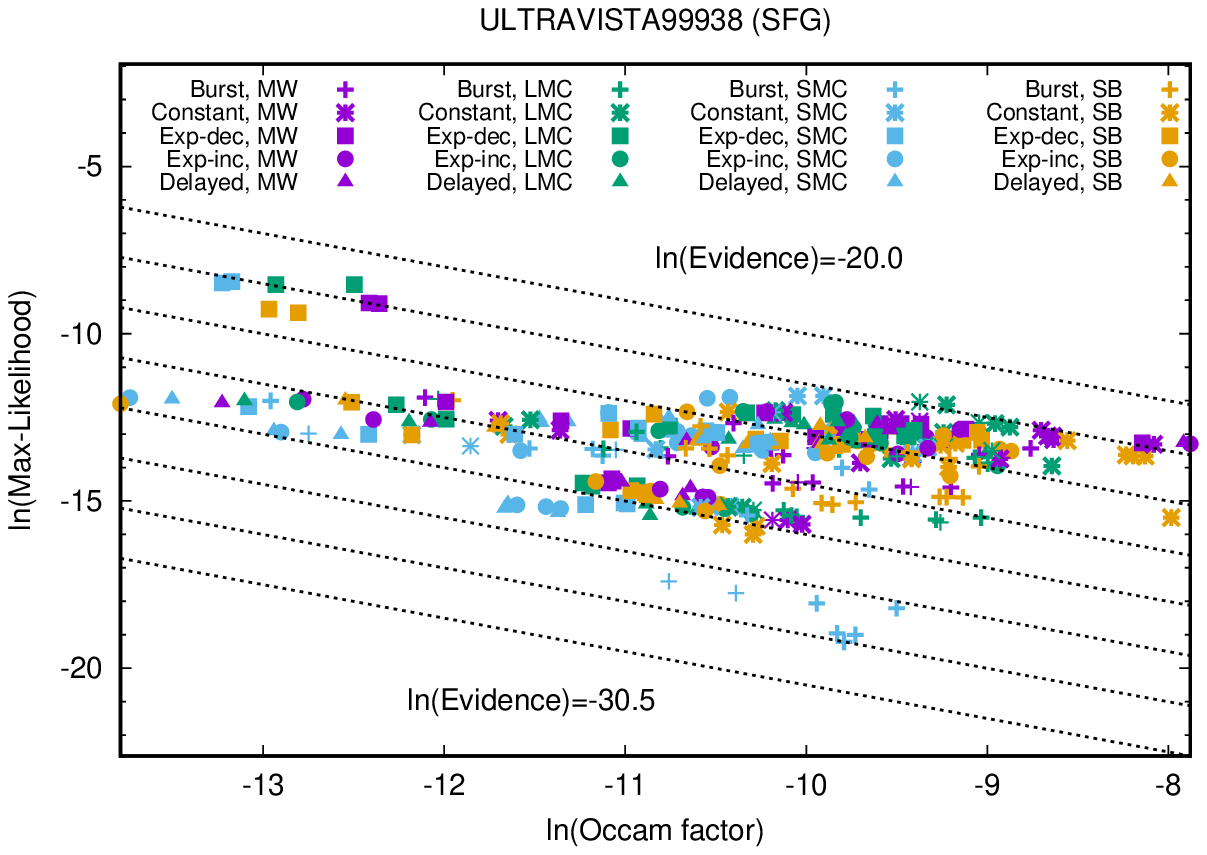}
  \end{center}
  \caption{The max-likelihood vs. Occam factor diagram vs. Bayesian evidence diagram (hereafter the ML-OF-BE diagram) for the ${\bm M}({ssp}_{0},{sfh_0},{dal_0})$-like SED modelings of the PEG ULTRAVISTA114558 (left) and the SFG ULTRAVISTA99938 (right) where SSP, SFH and DAL are all fixed to a particular choice. The max-likelihood, which is defined in Equation \ref{eq:MaxL_1SED_000} and directly related to the $\chi^2_{\rm min}$, represent the goodness-of-fit to the data of a model. The Occam factor, which is defined in Equation \ref{eq:Omega_1SED_000}, represents the complexity of a model. The Bayesian evidence, which is defined in Equation \ref{eq:bayes_ev_1SED_000} and indicated as dot lines with a step of $1.5$, is just the product of the max-likelihood and Occam factor. The $4$ different colors represent the models with different assumptions about the DAL, while the $5$ different shapes represent the models with different assumptions about the SFH. For a given color and shape, there are $16$ points, representing models with different assumptions about the SSP model.}
  \label{fig:ML-OF-BE_1SED_0_sp}
\end{figure*}

It is also worth noticing that, for the PEG ULTRAVISTA114558, the SED modeling of it assuming a constant SFH has the lowest Bayes factors for almost all combinations of SSP and SFH. 
The max-likelihoods and Occam factors for these models shown in Figure \ref{fig:ML-OF-BE_1SED_0_sp} reveal the reason for this trend.
The SED modelings of the PEG ULTRAVISTA114558 assuming a constant SFH are mainly located at the bottom right of the figure, which represent low goodness-of-fit to the data and low model complexity.
This result indicates that the constant SFH is too simple to be able to explain the observational SED of the PEG ULTRAVISTA114558.
Contrarily, most of the modelings assuming a Burst SFH are located at the top right of the figure, which represents high goodness-of-fit to the data and low model complexity.
For the SFG ULTRAVISTA99938, it is not easy to find out a clear trend in favor of a particular SFH or DAL.
However, it can be noticed in the right panel of Figure \ref{fig:BF_1SED_0_sp} that the \cb\ and \ma\ set of SSP models are less suitable for the SFG ULTRAVISTA99938 than other SSPs, which seems to be the opposite of what is the case for the PEG ULTRAVISTA114558 in the left panel of Figure \ref{fig:BF_1SED_0_sp}.

\subsubsection{The case for one of the SSP, SFH and DAL being fixed}\label{sss:1SED_model_1}
In \S \ref{sss:1SED_model_0}, we present a Bayesian comparison of the SED modelings of a galaxy for the cases where SSP, SFH and DAL are all being fixed.
This is useful for finding out the best combination of SSP, SFH and DAL for a galaxy.
However, it is not very helpful to find out the best SSP, SFH, or DAL, respectively.
Actually, we are more interested in questions such as which SSP is the best independently of the choices of SFH and DAL, which SFH is the best independently of the choices of SSP and DAL, and which DAL is the best independently of the choices of SSP and SFH.
These more interesting questions can be answered by considering the cases where only one of the SSP, SFH and DAL is fixed to a specific choice while the other two are allowed to change within a given sets.
For the computation of Bayes factors, we have used the same standard model ($M_0^1$) as in \S \ref{sss:1SED_model_0}.
It is worth to mention that the structure of the SED modelings considered here (See Figures \ref{fig:HBM011},\ref{fig:HBM101}, and \ref{fig:HBM110}) are diffrent from that of the standard model ($M_0^1$, with a structure shown in Figure \ref{fig:HBM000}).
So, the value of Bayes factor is determined not only by the selection of the physical components (SSP, SFH, DAL), but also by the modeling structure.
However, only the differences between Bayes factors are meaningful for the comparison of the different selections of the physical components (SSP/SFH/DAL).

\begin{figure*}[]
  \begin{center}
	\includegraphics[scale=0.73]{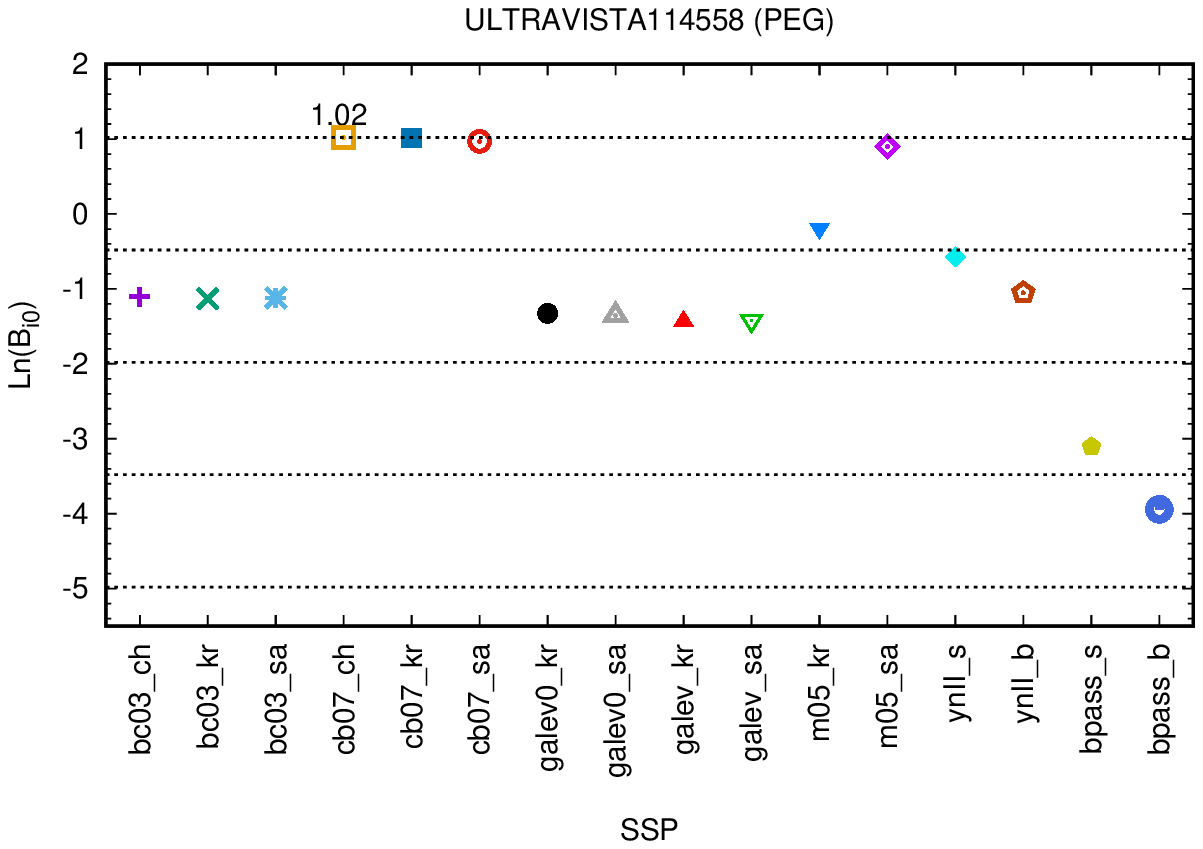}
	\includegraphics[scale=0.73]{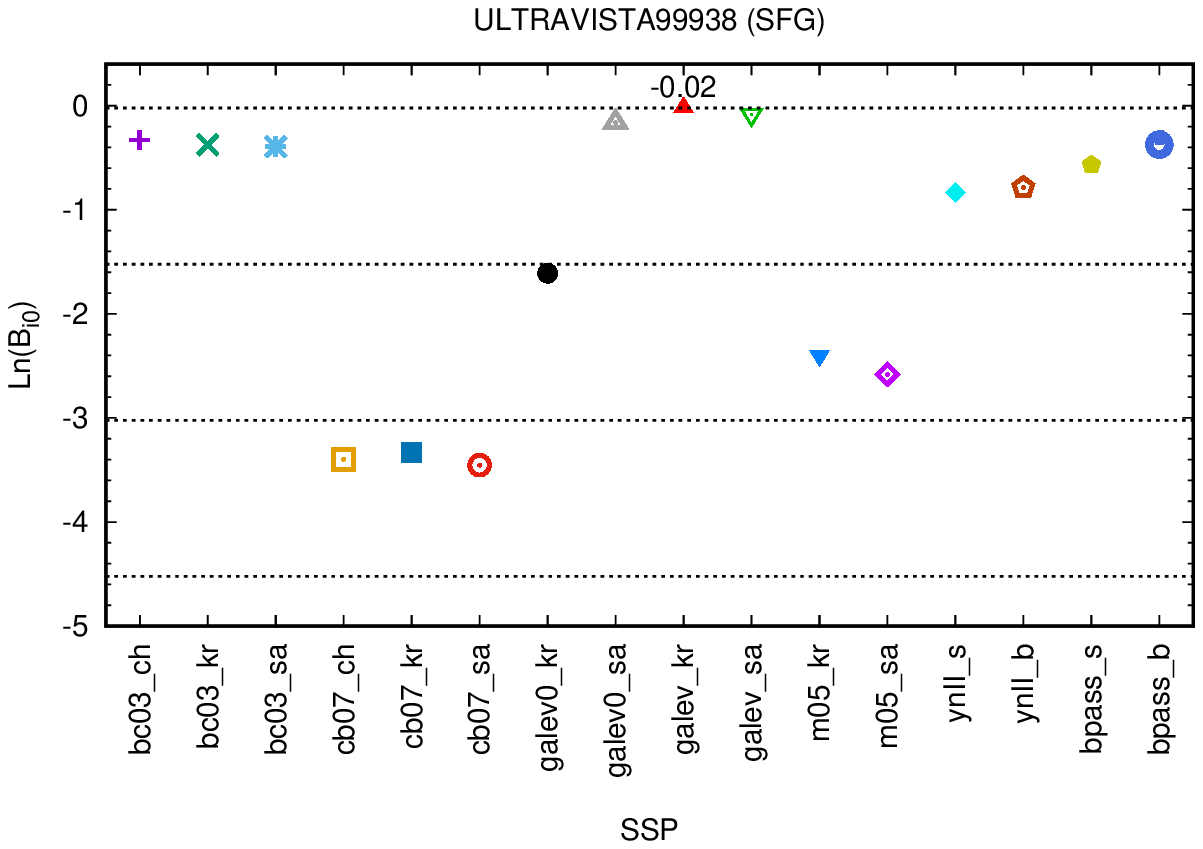}
  \end{center}
  \caption{Similar to Figure \ref{fig:BF_1SED_0_sp}, but for the ${\bm M}({ssp}_{0},{sfh},{dal})$-like SED modelings where a fixed SSP, free SFH and DAL are assumed. The \cb\ SSP with a Chabrier03 IMF (cb07\_ch) has the highest value of Bayes factor ($1.02$) for the PEG ULTRAVISTA114558, while the version of GALEV model with the consideration of emission lines and a Kroupa01 IMF (galev\_kr) has the highest value of Bayes factor ($-0.02$) for the SFG ULTRAVISTA99938. The negative Bayes factor indicates that  even the best one of the ${\bm M}({ssp}_{0},{sfh},{dal})$-like SED modelings is not better than the standard model $M_0^1$ which is a ${\bm M}({ssp}_{0},{sfh}_{0},{dal}_{0})$-like SED modeling. So, for the SFG ULTRAVISTA99938, the additional complexity as introduced by the two additional free parameters ($sfh$ and $dal$) is not justified by a much better fit to the observational data. However, the two additional parameters are still useful to make sure the comparison of SSP models is independent of the selection of SFH and DAL. For the comparison of any two SSP models, only the difference of their Bayes factors is meaningful to us.}
  \label{fig:BF_1SED_011_sp}
\end{figure*}

In Figure \ref{fig:BF_1SED_011_sp}, we show the Bayes factors with respect to the standard model ($M_0^1$) for the SED modelings of the PEG ULTRAVISTA114558 and the SFG ULTRAVISTA99938, where a fixed SSP, free SFH and DAL are assumed.
This can be used to answer the question: Which SSP is the best for the particular galaxy independently of the choices of SFH and DAL?
It is clear from the figure that the \cb\ SSP with a Chabrier03 IMF (cb07\_ch) has the highest value of Bayes factor ($1.02$) for the PEG ULTRAVISTA114558, while the version of the GALEV model with the consideration of emission lines and a Kroupa01 IMF (galev\_kr) has the highest value of Bayes factor ($-0.02$) for the SFG ULTRAVISTA99938.
It is interesting to notice that, the more ``TP-AGB heavy'' SSP models of \cb\ and \ma\ systematically have much larger Bayes factor than others for the PEG ULTRAVISTA114558, while they systematically have much smaller Bayes factor than others for the SFG ULTRAVISTA99938.
On the other hand, for the PEG, the performance of the GALEV models are not sensitive to the consideration of emission lines and the selection of IMF.
For the SFG, the performance of the version of the GALEV models with the consideration of emission lines (galev\_kr, galev\_sa) are not sensitive to the selection of IMF.
Contrarily, the performance of the version of the GALEV models without the consideration of emission lines (galev0\_kr, galev0\_sa) are very sensitive to the selection of IMF.

\begin{figure*}[]
  \begin{center}
	\includegraphics[scale=0.73]{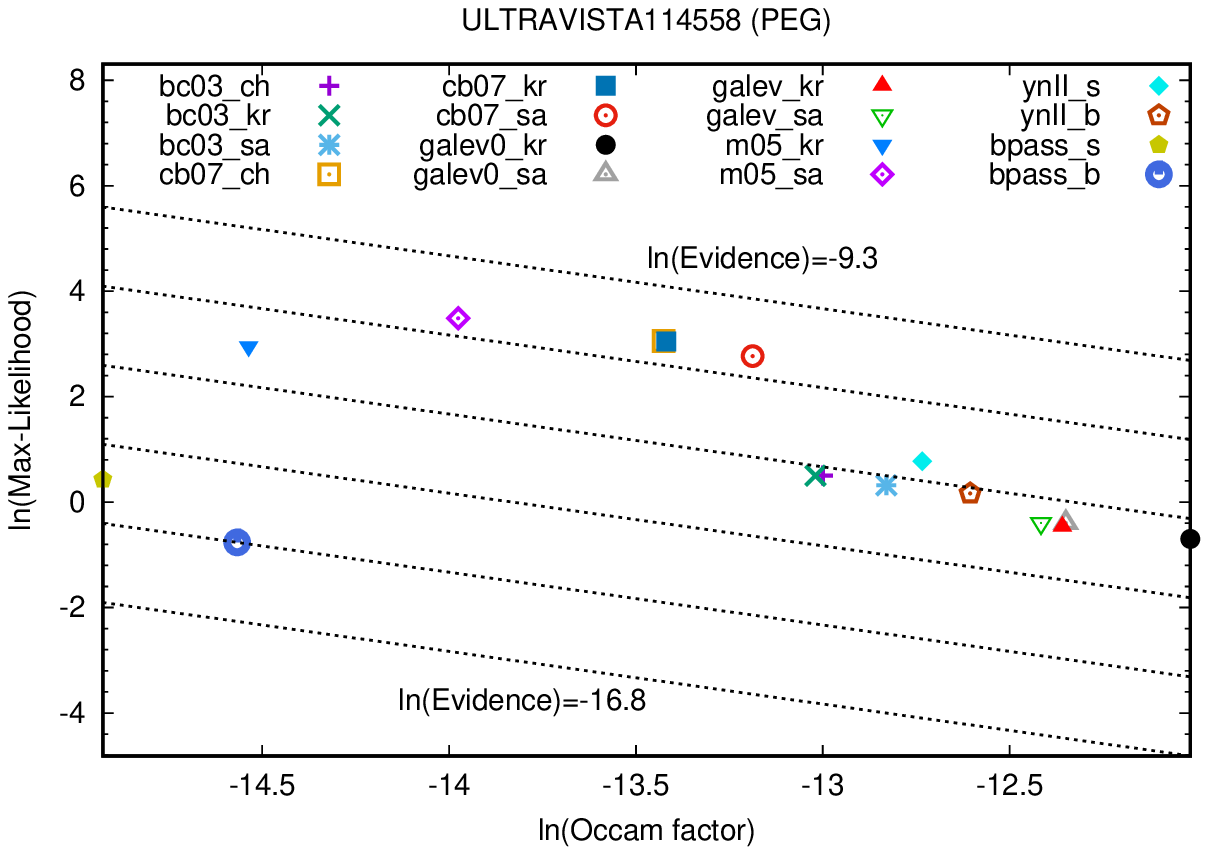}
	\includegraphics[scale=0.73]{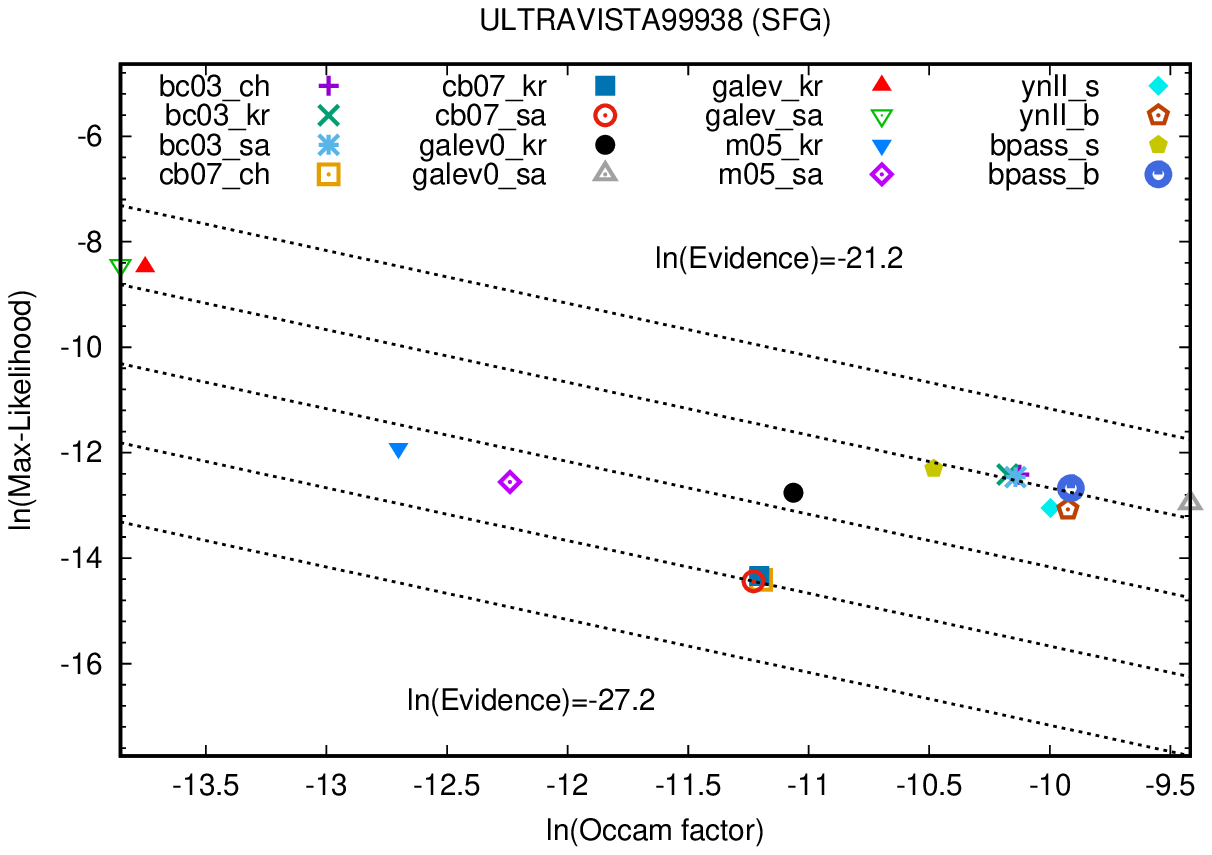}
  \end{center}
  \caption{Similar to Figure \ref{fig:ML-OF-BE_1SED_0_sp}, but for the ${\bm M}({ssp}_{0},{sfh},{dal})$-like SED modelings where a fixed SSP, free SFH and DAL are assumed. Here, the Occam factor and Max-likelihood are defined in Equations \ref{eq:Omega_total_1SED_011} and \ref{eq:MaxL_1SED_011}, respectively. The Bayesian evidence is defined in Equation \ref{eq:bayes_ev_1SED_011_a}, and calculated with Equation \ref{eq:bayes_ev_1SED_011}. The more ``TP-AGB heavy'' SSP models of \cb\ (cb07\_ch, cb07\_kr, cb07\_sa) and \ma\ (m05\_kr, m05\_sa) provide much better fits to the observational data of the PEG ULTRAVISTA114558, while the version of GALEV model with the consideration of emission lines (galev\_kr, galev\_sa) provide much better fits to the observational data of the SFG ULTRAVISTA99938.}
  \label{fig:ML-OF-BE_1SED_011_sp}
\end{figure*}

In Figure \ref{fig:ML-OF-BE_1SED_011_sp}, we show the max-likelihoods, Occam factors, and Bayesian evidences for the same set of SED modelings.
It is clear from the figure that the more ``TP-AGB heavy'' SSP models of \cb\ and \ma\ can provide a better fit (as indicated by the much larger value of max-likelihoods) to the observational data than other SSP models for the PEG ULTRAVISTA114558.
This is the main reason why they have much larger Bayesian evidence and Bayes factor than that of other SSP models as shown in Figure \ref{fig:BF_1SED_011_sp}.
Besides, both the version of \bpass\ model with and without the consideration of binaries are located at the bottom left of the ML-OF-BE diagram (indicating a low goodness-of-fit to the data and large model complexity), which suggest that the model is not very suitable for this PEG. 
For the SFG ULTRAVISTA99938, the results in Figure \ref{fig:ML-OF-BE_1SED_011_sp} show that the version of GALEV model with the consideration of emission lines can provide a significantly better explanation to the data than other SSP models which have not included the contribution of emission lines.
Given this result, it is clear that the consideration of nebular emission lines is very necessary for the SFG.
It is also interesting to notice that the Bayesian evidences and Bayes factors of the \bc\ models are only slightly smaller than that of the \galev\ models for the SFG, although the former cannot provide similar goodness-of-fit to the data. 
This is mainly because the \bc\ models have much larger Occam factors than the \galev\ models for the SFG, which indicate lower model complexity of the former.

\begin{figure*}[]
  \begin{center}
	\includegraphics[scale=0.73]{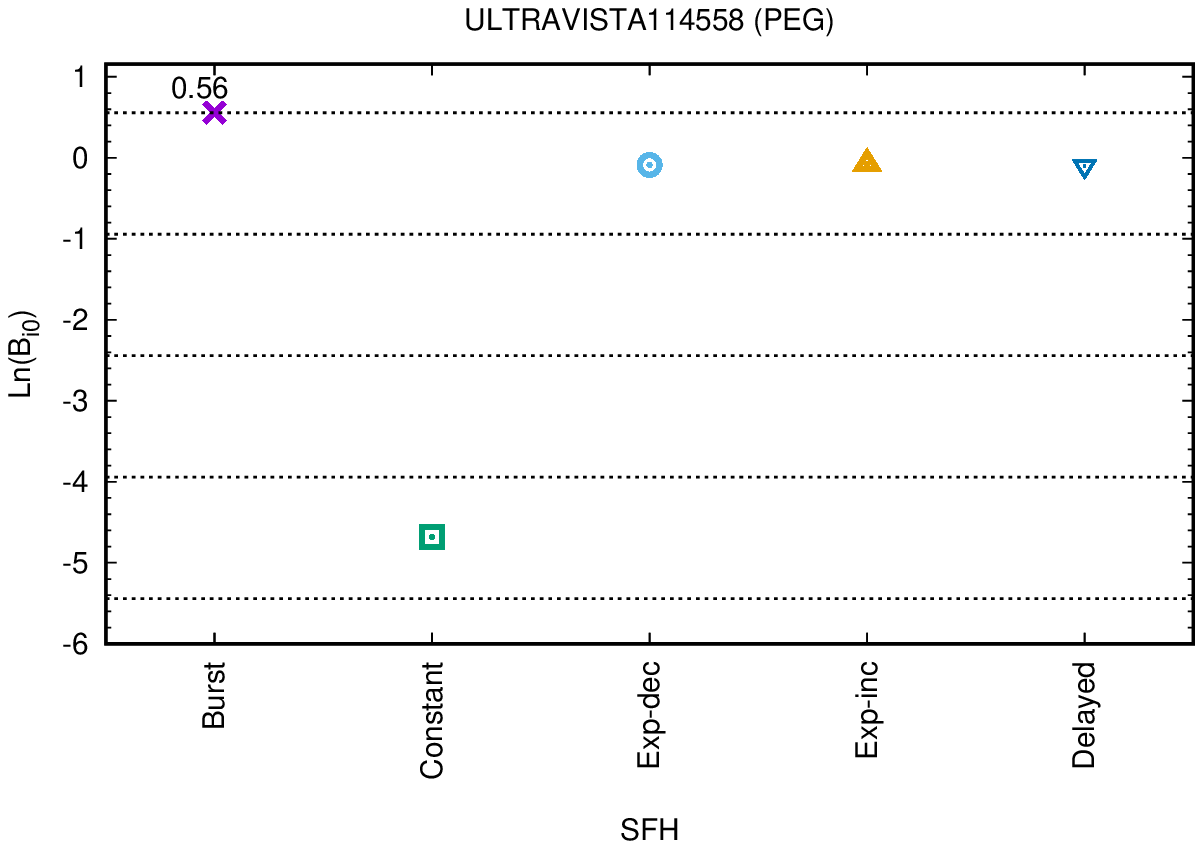}
	\includegraphics[scale=0.73]{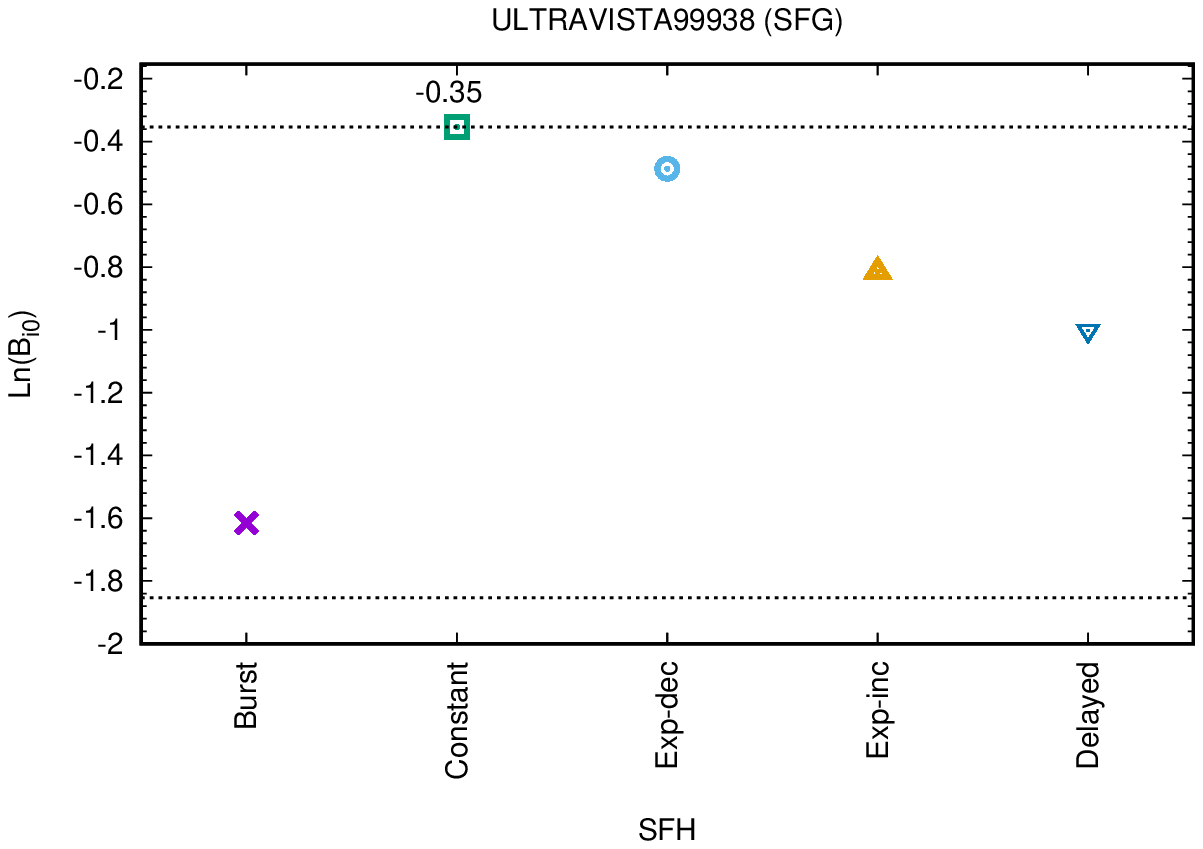}
  \end{center}
  \caption{Similar to Figure \ref{fig:BF_1SED_011_sp}, but for the ${\bm M}({ssp},{sfh}_{0},{dal})$-like SED modelings where a fixed SFH, free SSP and DAL are assumed. The Burst SFH has the largest value of Bayes factor ($0.56$) for the PEG ULTRAVISTA114558 (left), while the constant SFH has the largest value of Bayes factor ($-0.35$) for the SFG ULTRAVISTA99938 (right).}
  \label{fig:BF_1SED_101_sp}
\end{figure*}

\begin{figure*}[]
  \begin{center}
	\includegraphics[scale=0.73]{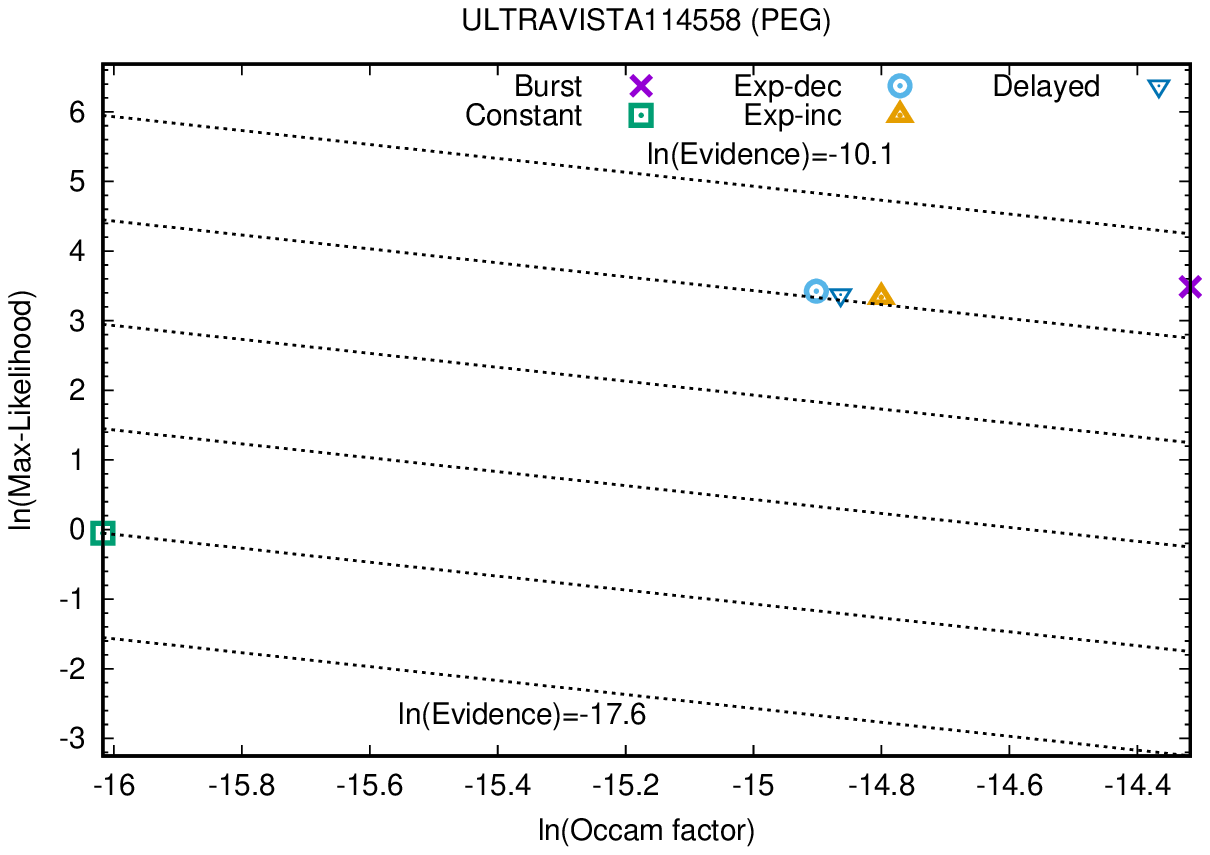}
	\includegraphics[scale=0.73]{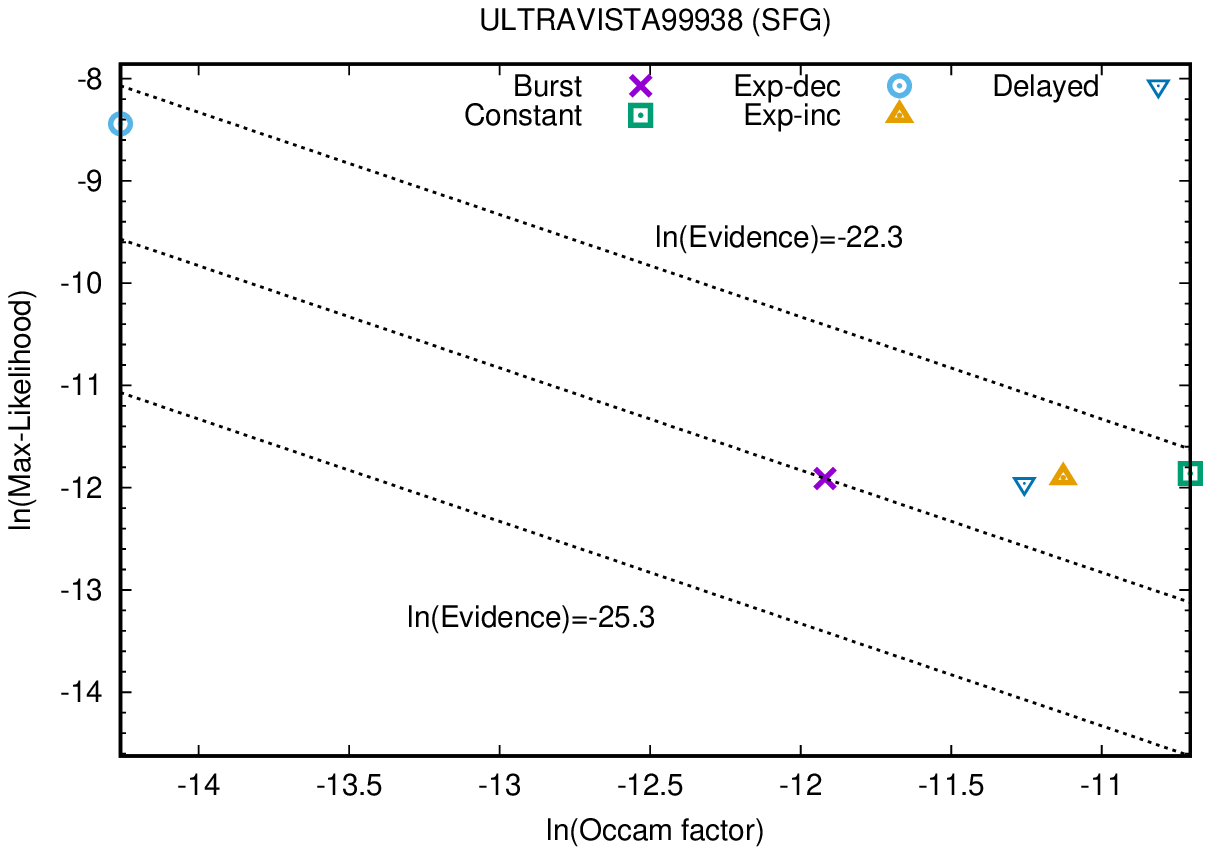}
  \end{center}
  \caption{Similar to Figure \ref{fig:ML-OF-BE_1SED_011_sp}, but for the ${\bm M}({ssp},{sfh}_{0},{dal})$-like SED modelings where a fixed SFH, free SSP and DAL are assumed. For the PEG ULTRAVISTA114558, the Burst SFH provides the best fit to the data and has the highest value of Occam factor (i.e. the lowest model complexity). For the SFG ULTRAVISTA99938, the Exp-dec SFH provides the best fit to the data but has the lowest value of Occam factor (i.e. the highest model complexity).}
  \label{fig:ML-OF-BE_1SED_101_sp}
\end{figure*}

Similarly, in Figure \ref{fig:BF_1SED_101_sp}, we show the Bayes factors with respect to the standard model ($M_0^1$) for the SED modelings of the PEG ULTRAVISTA114558 and the SFG ULTRAVISTA99938, where a fixed SFH, free SSP and DAL are assumed.
The comparison of this kind of SED modeling can be used to answer the question: Which SFH is the best for the particular galaxy independently of the choices of SSP and DAL?
It is very clear from the figure that the  Burst SFH has the highest value of Bayes factor ($0.56$) for the PEG ULTRAVISTA114558, while the  constant SFH has the highest value of Bayes factor ($-0.35$) for the SFG ULTRAVISTA99938.
For the PEG ULTRAVISTA114558, the Bayes factor of the Burst SFH is significantly larger than that of the constant SFH.
This is just the opposite of what is the case for the SFG ULTRAVISTA99938, which indicates very different nature of the two galaxies.
Meanwhile, the ML-OF-BE diagram in Figure \ref{fig:ML-OF-BE_1SED_101_sp} show that the  Burst SFH also provides the best explanation to the observational data of the PEG ULTRAVISTA114558, while the Exp-dec SFH, instead of the constant SFH, provides the best explanation to the observational data of the SFG ULTRAVISTA99938.
This is mainly because Burst SFH has the largest value of Occam factor (i.e. the lowest model complexity) for the PEG.
On the other hand, although the Exp-dec SFH can provide the best explanation to the data of the SFG, it also has the lowest value of Occam factor (i.e. the highest model complexity).
Since the more model complexity cannot be balanced by the better goodness-of-fit to the data, it actually has lower Bayesian evidence and Bayes factor than the simpler constant SFH.


\begin{figure*}[]
  \begin{center}
	\includegraphics[scale=0.73]{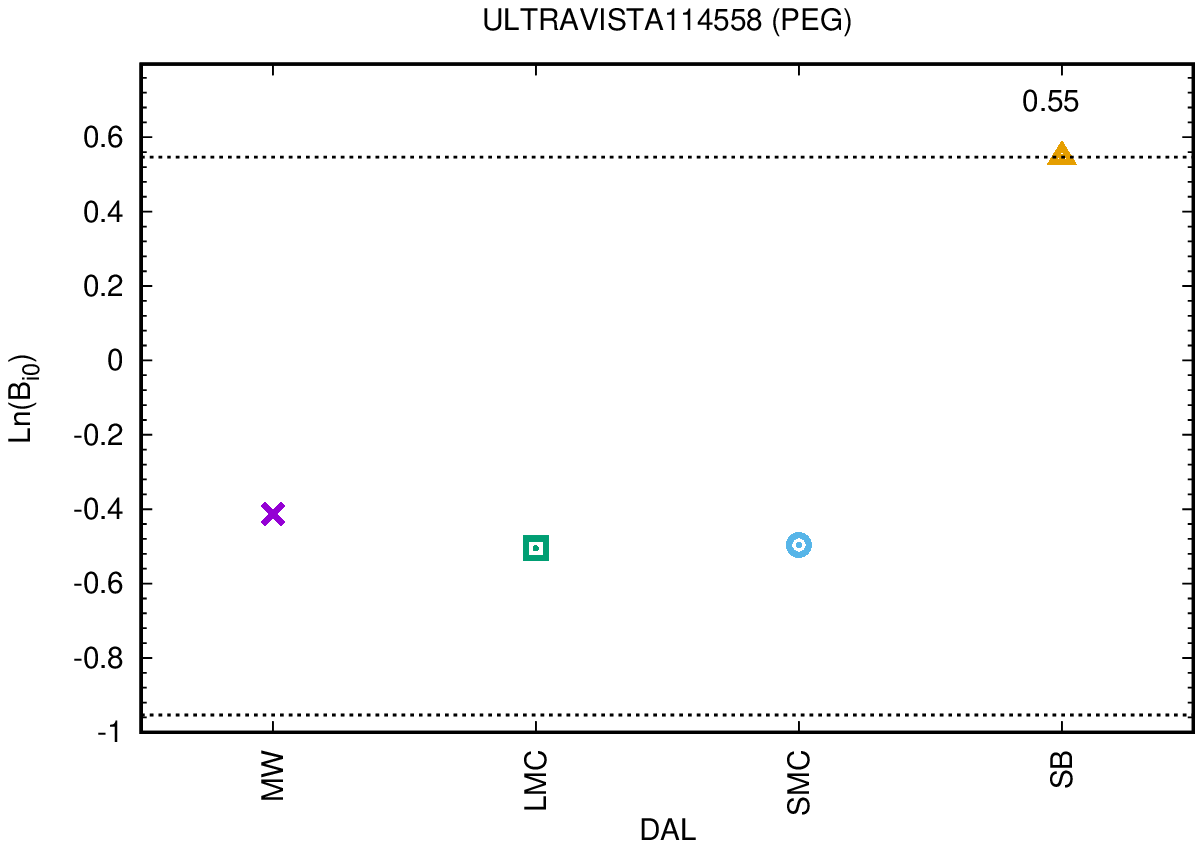}
	\includegraphics[scale=0.73]{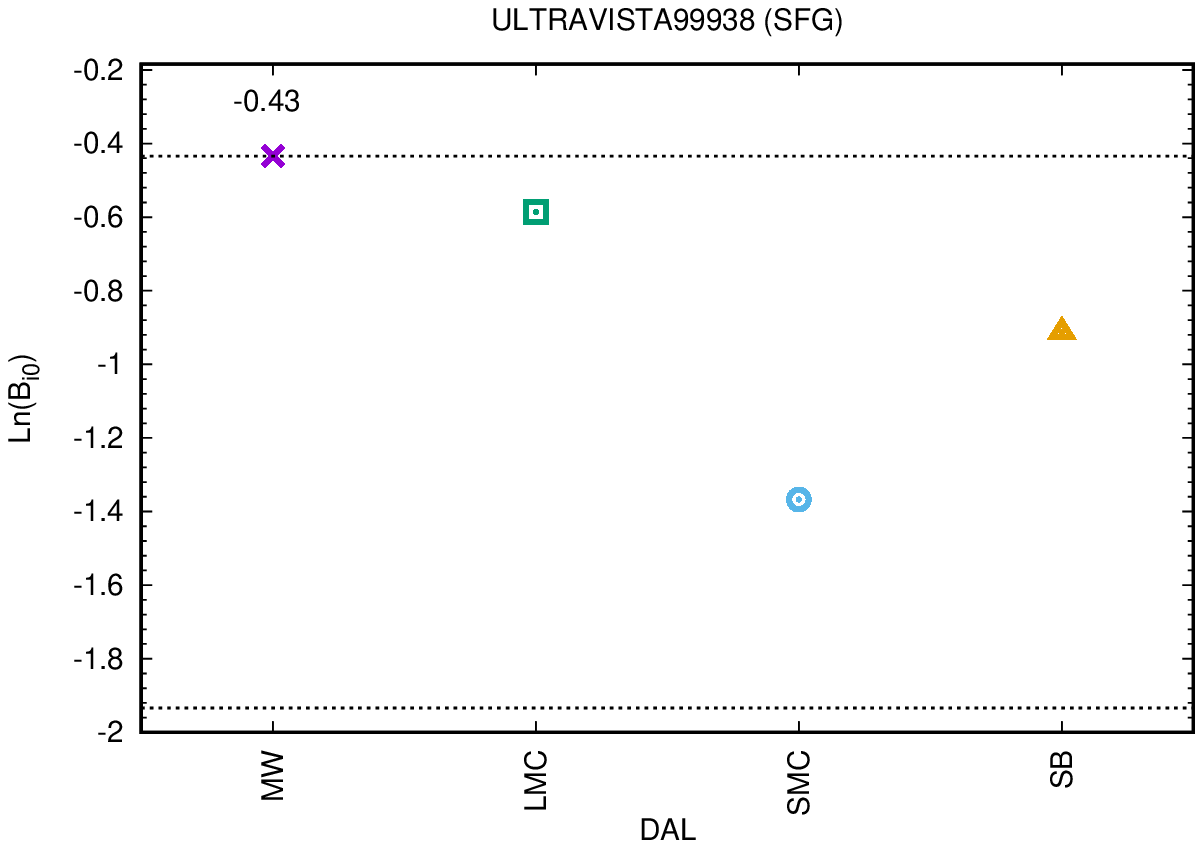}
  \end{center}
  \caption{Similar to Figure \ref{fig:BF_1SED_011_sp}, but for the ${\bm M}({ssp},{sfh},{dal}_{0})$-like SED modeling where a fixed DAL, free SSP and SFH are assumed. The SB-like DAL has the highest value of Bayes factor ($0.55$) for the PEG ULTRAVISTA114558 (left) , while the MW-like DAL has the highest value of Bayes factor ($-0.43$) the SFG ULTRAVISTA99938 (right).}
  \label{fig:BF_1SED_110_sp}
\end{figure*}

\begin{figure*}[]
  \begin{center}
	\includegraphics[scale=0.73]{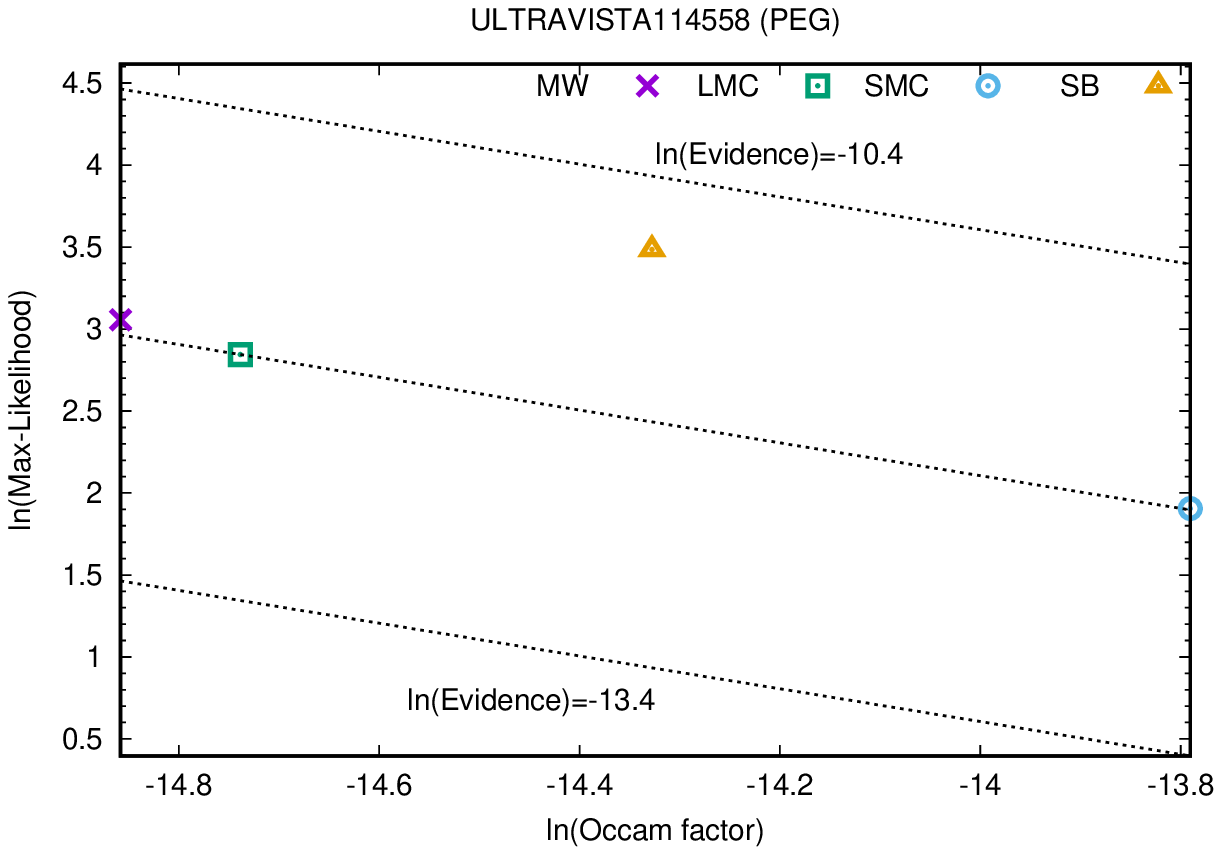}
	\includegraphics[scale=0.73]{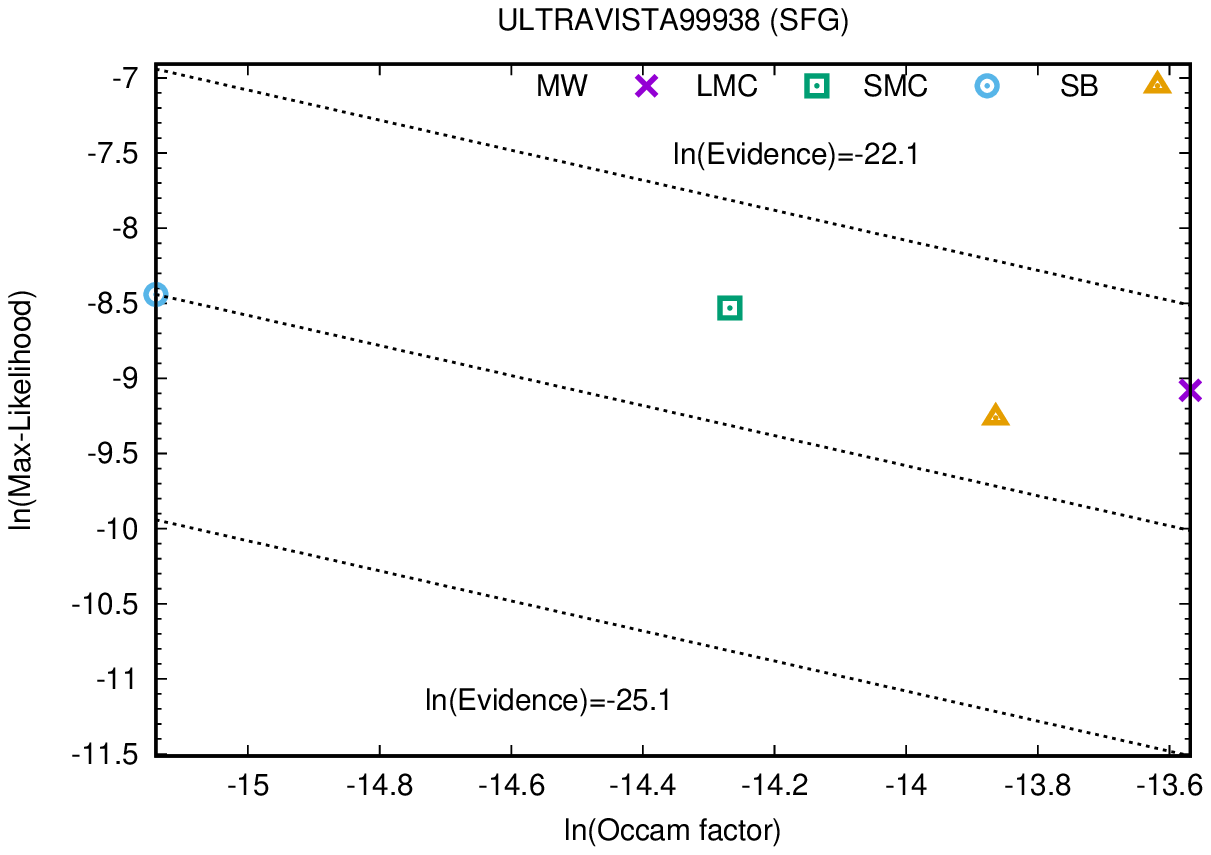}
  \end{center}
  \caption{Similar to Figure \ref{fig:ML-OF-BE_1SED_011_sp}, but for the ${\bm M}({ssp},{sfh},{dal}_{0})$-like SED modelings where a fixed DAL, free SSP and SFH are assumed. For the PEG ULTRAVISTA114558, the SB-like DAL provides better a fit to the data and has the largest Bayesian evidence. For the SFG ULTRAVISTA99938, the SMC-like and LMC-like DAL provide better fits to the data, but the MW-like DAL has the largest Bayesian evidence.}
  \label{fig:ML-OF-BE_1SED_110_sp}
\end{figure*}

Finally, in Figure \ref{fig:BF_1SED_110_sp}, we show the Bayes factors with respect to the standard model ($M_0^1$) for the SED modelings of the PEG ULTRAVISTA114558 and the SFG ULTRAVISTA99938, where a particular DAL is assumed but the SSP and SFH are set to be free to vary.
The comparison of this kind of SED modeling can be used to answer the question: Which DAL is the best for the particular galaxy independently of the choices of SSP and SFH?
It is clear from the figure that the SB-like DAL  has the highest value of Bayes factor ($0.55$) for the PEG ULTRAVISTA114558, while the MW-like DAL has the highest value of Bayes factor ($-0.43$) for the SFG ULTRAVISTA99938.
Besides, the ML-OF-BE diagram in Figure \ref{fig:ML-OF-BE_1SED_110_sp} show that the SB-like DAL also provides the best explanation to the observational data of the PEG, while the SMC-like and LMC-like DAL, instead of the MW-like DAL, provide the best explanation to the observational data of the SFG.
This is simply due to the much larger Bayes factor of the MW-like DAL than that of the SMC-like and LMC-like DAL for the SFG.




\subsection{Bayesian discrimination of SSP, SFH and DAL for the SED modeling of a sample of galaxies}\label{ss:NSED_model}
In \S \ref{ss:1SED_model}, we presented a detailed Bayesian discrimination of SSPs, SFHs, and DALs for the SED modeling of a PEG and an SFG respectively.
This kind of study is useful for investigating the particular characteristic of a galaxy.
However, since only one object is involved in each case, the conclusions obtained for it are not necessarily suitable for other objects of the same type.
So, in many cases, we are more interested in comparing the performance of different SSPs, SFHs and DALs for a sample of galaxies.
In this subsection, based on the method of calculating the Bayesian evidence for the SED modeling of a sample of galaxies in \S \ref{s:ev_NSED}, we present a detailed Bayesian discrimination of different assumptions about SSP, SFH and DAL for the SED modeling of a sample of galaxies for the first time.

\subsubsection{The case for all  the SSP, SFH and DAL being universal and fixed}\label{sss:NSED_model_U_0}
A fundamental difference between the SED modeling of an individual galaxy and the SED modeling of a sample of galaxies is that for the latter, we can assume either the same SSP, SFH and/or DAL for all objects in the sample (the universal case), or different SSPs, SFHs, and/or DALs for different objects (the object-dependent case).
So, with the Bayesian discrimination of different assumptions in the SED modelings of a sample of galaxies, it is possible to test the universality of different SSPs, SFHs and DALs.
Here, we firstly consider the cases where SSPs, SFHs and DALs are all assumed to be universal.

With SSP, SFH and DAL being all assumed to be universal, we still have the freedom of selecting them from the many possible choices.
So, we firstly consider the cases where SSP, SFH and DAL are all fixed to a specific choice.
This is the most widely used assumption when modeling and interpreting the SEDs of a sample of galaxies, the structure of which is shown in Figure \ref{fig:HBM_N0}.
For example, in many works, people often assume the standard model ($M_0^{\rm N}$) with the bc03\_ch SSP, the Exp-dec SFH, and the SB-like DAL for all galaxies in their samples. 
By the Bayesian comparison of this kind of SED modelings, we can find out the specific combination of SSP, SFH and DAL with the best universality for a sample of galaxies.

\begin{figure*}[]
  \begin{center}
	\includegraphics[scale=0.73]{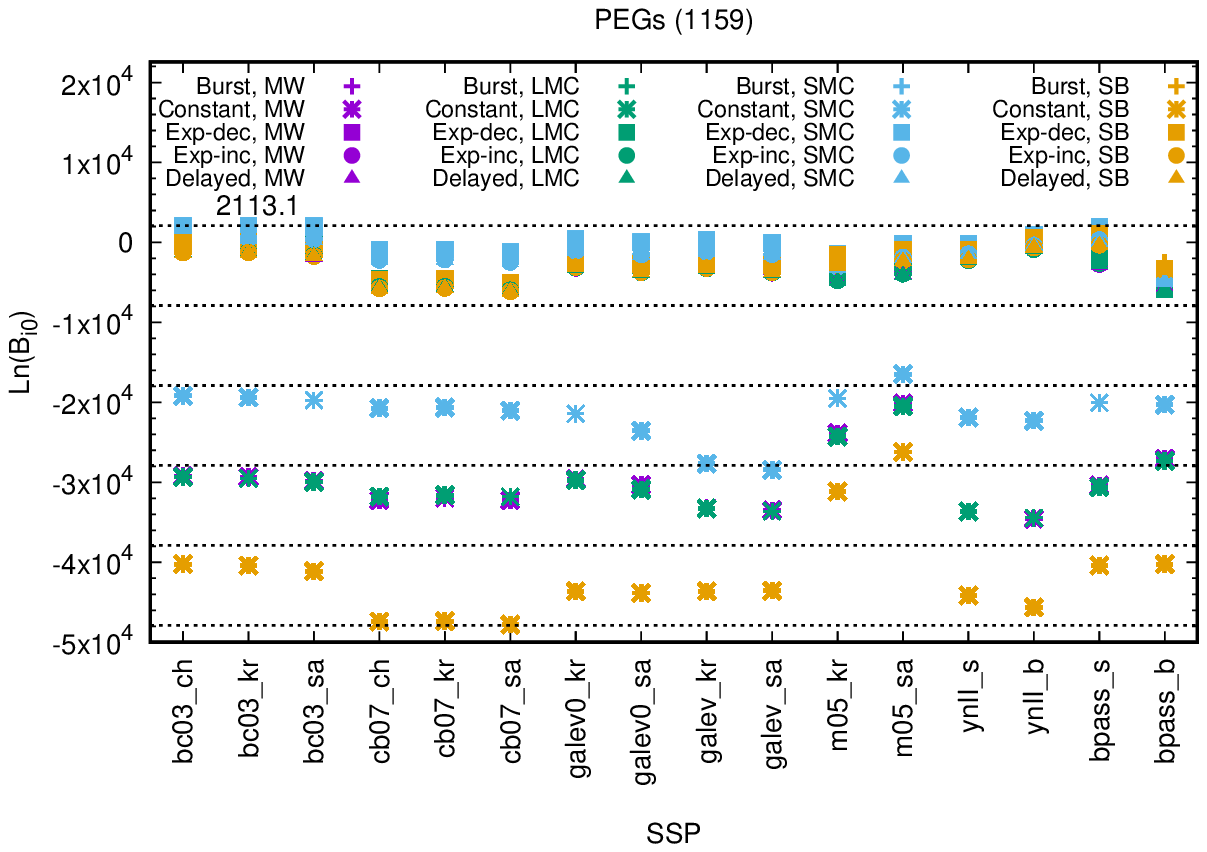}
	\includegraphics[scale=0.73]{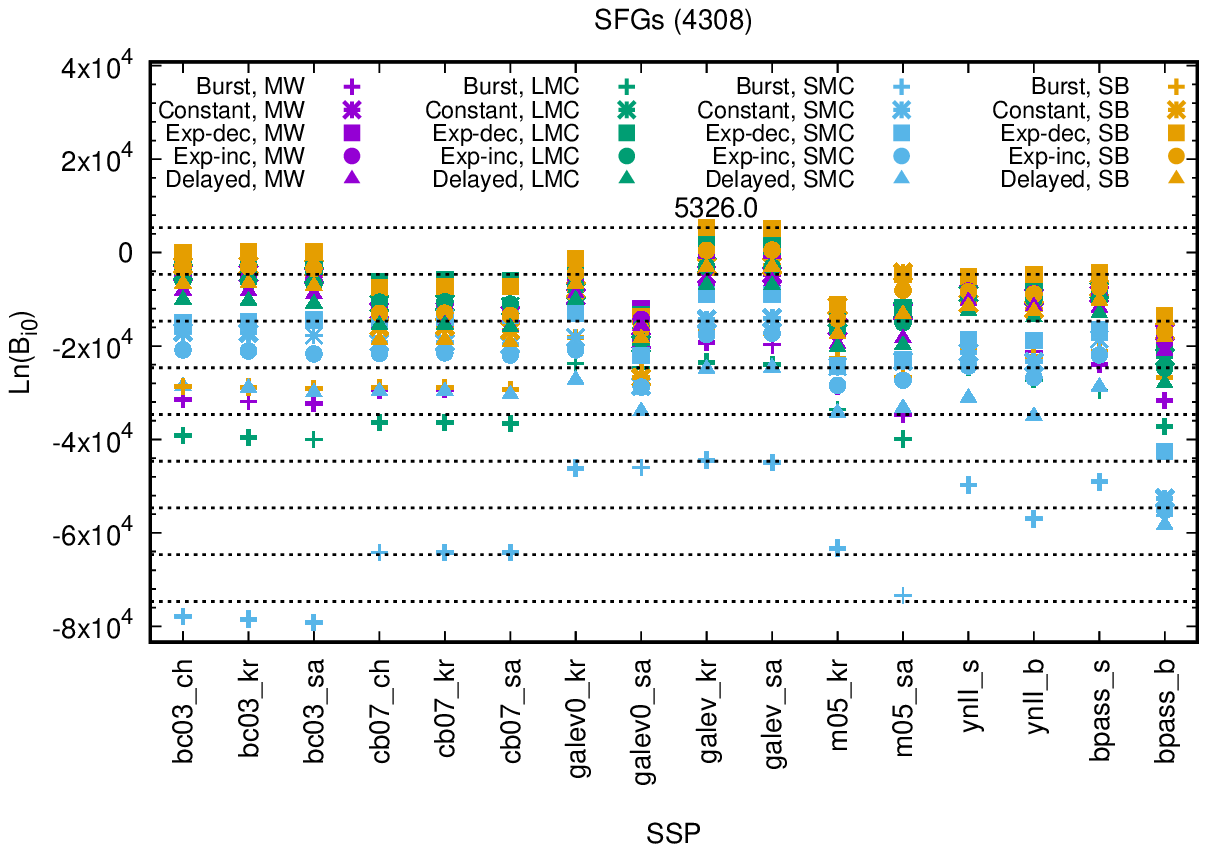}
  \end{center}
  \caption{The Bayes factors with respect to the standard model ($M_0^{\rm N}$, which assumes the BC03 SSP with a Chabrier03 IMF, Exp-dec SFH and SB-like DAL for all galaxies in the sample) for the ${\bm M}({ssp}_{0},{sfh_0},{dal_0})$-like SED modelings of PEGs (left) and SFGs  (right), where the SSP, SFH and DAL are all assumed to be universal and fixed to a particular choice. The dot lines show the values of the Bayes factor with a step of $10000$. For the PEGs, the combination of the \bc\ SSP with a Kroupa01 IMF (bc03\_kr), the Exp-dec SFH and the SMC-like DAL has the highest value ($2113.10$) of Bayes factor. For the SFGs, the combination of the version of \galev\ SSP with the consideration of emission lines and a Kroupa01 IMF (galev\_kr), the Exp-dec SFH and the SB-like DAL has the highest value ($5326.00$) of Bayes factor. Since a sample of galaxies, instead of just one object (as in Figure \ref{fig:BF_1SED_0_sp}), is involved, the conclusions obtained here are with respect to the sample as a whole.}
  \label{fig:BF_NSED_U_0_sp}
\end{figure*}

\begin{figure*}[]
  \begin{center}
	\includegraphics[scale=0.73]{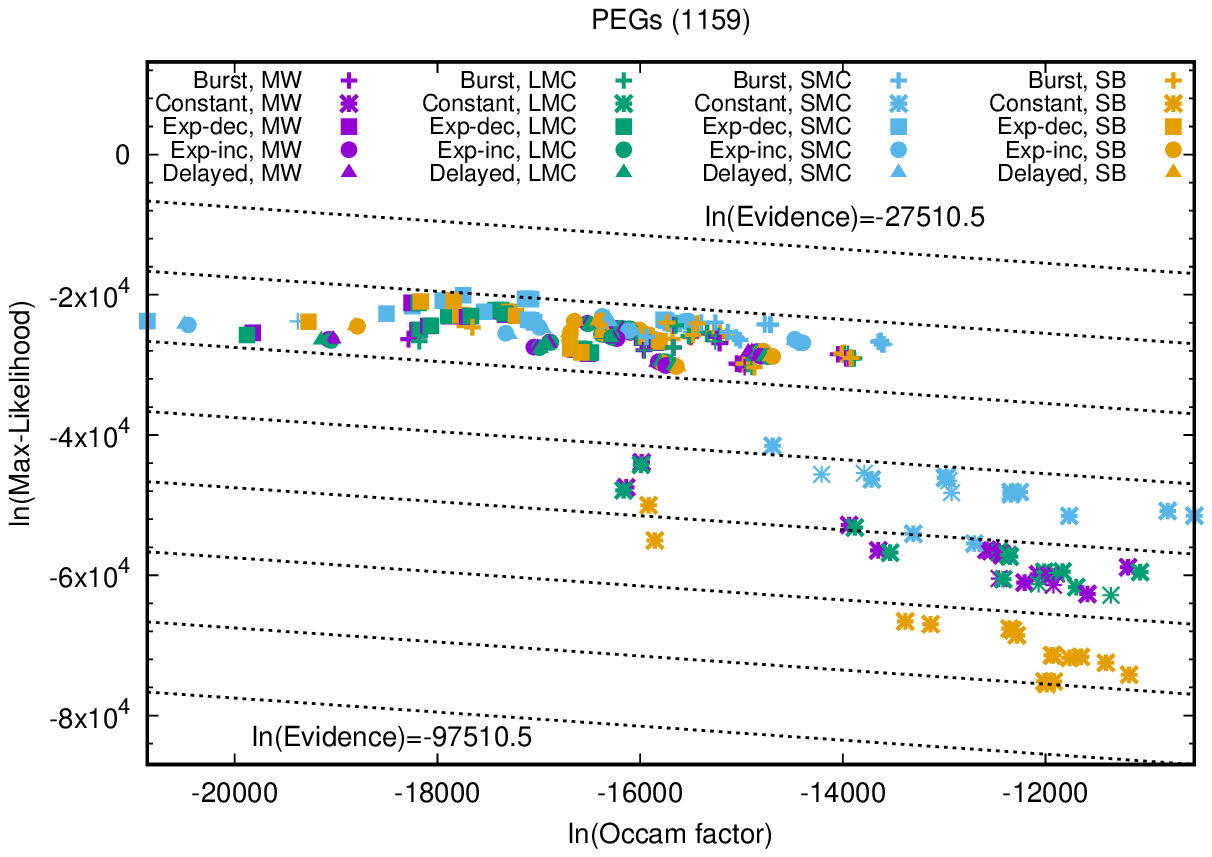}
	\includegraphics[scale=0.73]{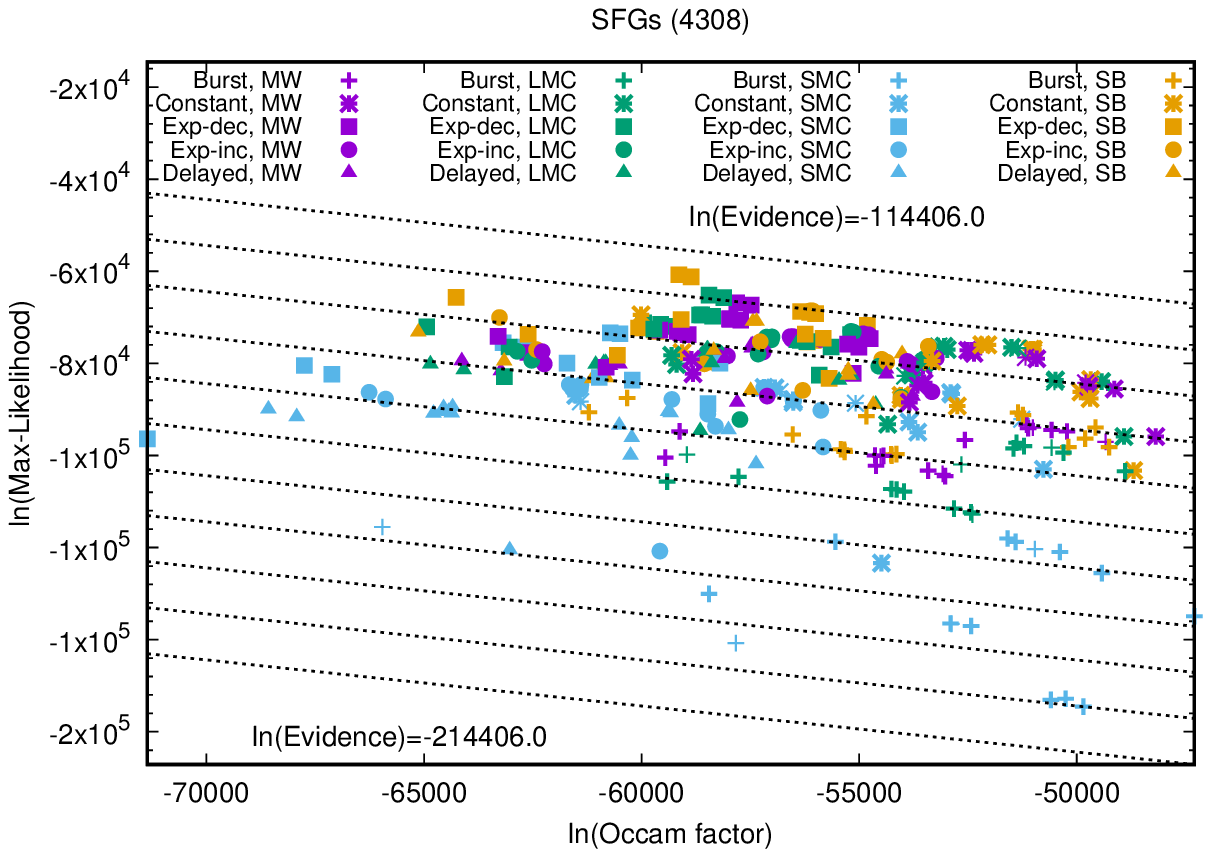}
  \end{center}
  \caption{The ML-OF-BE diagram for the ${\bm M}({ssp}_{0},{sfh_0},{dal_0})$-like SED modelings of PEGs (left) and SFGs  (right), where SSP, SFH and DAL are all assumed to be universal and fixed to a particular choice. Here, The Occam factor and max-likelihood are defined in Equations \ref{eq:Omega_total_NSED_UUU_000} and \ref{eq:MaxL_NSED_UUU_000}, respectively. The Bayesian evidence is defined in Equation \ref{eq:bayes_ev_NSED_UUU_000_a}, and calculated with Equation \ref{eq:bayes_ev_NSED_UUU_000}. The dot lines show the values of the Bayesian evidence with a step of $10000$.}
  \label{fig:ML-OF-BE_NSED_U_0_sp}
\end{figure*}

In Figure \ref{fig:BF_NSED_U_0_sp}, we show the Bayes factors with respect to the standard model ($M_0^{\rm N}$) for all possible combinations of the  $16$ SSP, $5$ SFH, and $4$ DAL when modeling the SEDs of a sample of PEGs and SFGs respectively.
The combination of the \bc\ SSP with a Kroupa01 IMF (bc03\_kr), the Exp-dec SFH and the SMC-like DAL has the highest value ($2113.1$) of Bayes factor for the PEGs, while the combination of the version of \galev\ SSP with the consideration of emission lines and a Kroupa01 IMF (galev\_kr), the Exp-dec SFH and the SB-like DAL has the highest value ($5326.0$) of Bayes factor for the SFGs. 
This is very different from the results for individual galaxies in Figure \ref{fig:BF_1SED_0_sp}.
Since a sample of galaxies, instead of just one object, is involved, the conclusions obtained here are  with respect to the sample as a whole.
Similar to the results for individual galaxies in Figure \ref{fig:BF_1SED_0_sp}, the Bayes factors for the SED modeling of a sample of galaxies with different SSP models could be very different, even if the same SFH and DAL are assumed.
Besides, for the PEGs, the form of SFH has the lowest value of Bayes factors for almost all combinations of SSP and DAL.
For the SFGs, the Burst SFH  has the lowest value of Bayes factors for almost all combinations of SSP and DAL. 
These general trends can be understood from the max-likelihoods, Occam factors and Bayesian evidences of these models in Figure \ref{fig:ML-OF-BE_NSED_U_0_sp}.
It can be noticed that for the PEGs, most of the models assuming a Burst SFH are located at the top right of the figure, which indicates a low model complexity and high goodness-of-fit to the data, while most of the models assuming a constant SFH are located at the bottom right of the figure, which indicates low model complexity but low goodness-of-fit to the data.
On the other hand, for the SFGs, most of the models assuming a Burst SFH are located at the bottom right of the figure, which indicates a low model complexity but low goodness-of-fit to the data, while most of the models assuming a constant SFH are located at the top right of the figure, which indicates low model complexity and low goodness-of-fit to the data.

\subsubsection{The case for SSP, SFH and DAL being universal but only one of them being fixed}\label{sss:NSED_model_U_1}
In \S \ref{sss:NSED_model_U_0}, we presented a Bayesian comparison of the SED modelings of a sample of galaxies where SSP, SFH and DAL are all assumed to be universal and fixed to a specific choice.
This is useful for finding out the combination of SSP, SFH and DAL with the best universality for a sample of galaxies.
However, we are more interested in questions such as which SSP is the best independently of the choices of SFH and DAL, which SFH is the best independently of the choices of SSP and DAL, and which DAL is the best independently of the choices of SSP and SFH.
This is somewhat similar to the case for individual galaxies in \S \ref{sss:1SED_model_1}.
However, here, we want to obtain the conclusions for a sample of galaxies instead of that for an individual galaxy.

\begin{figure*}[]
  \begin{center}
	\includegraphics[scale=0.73]{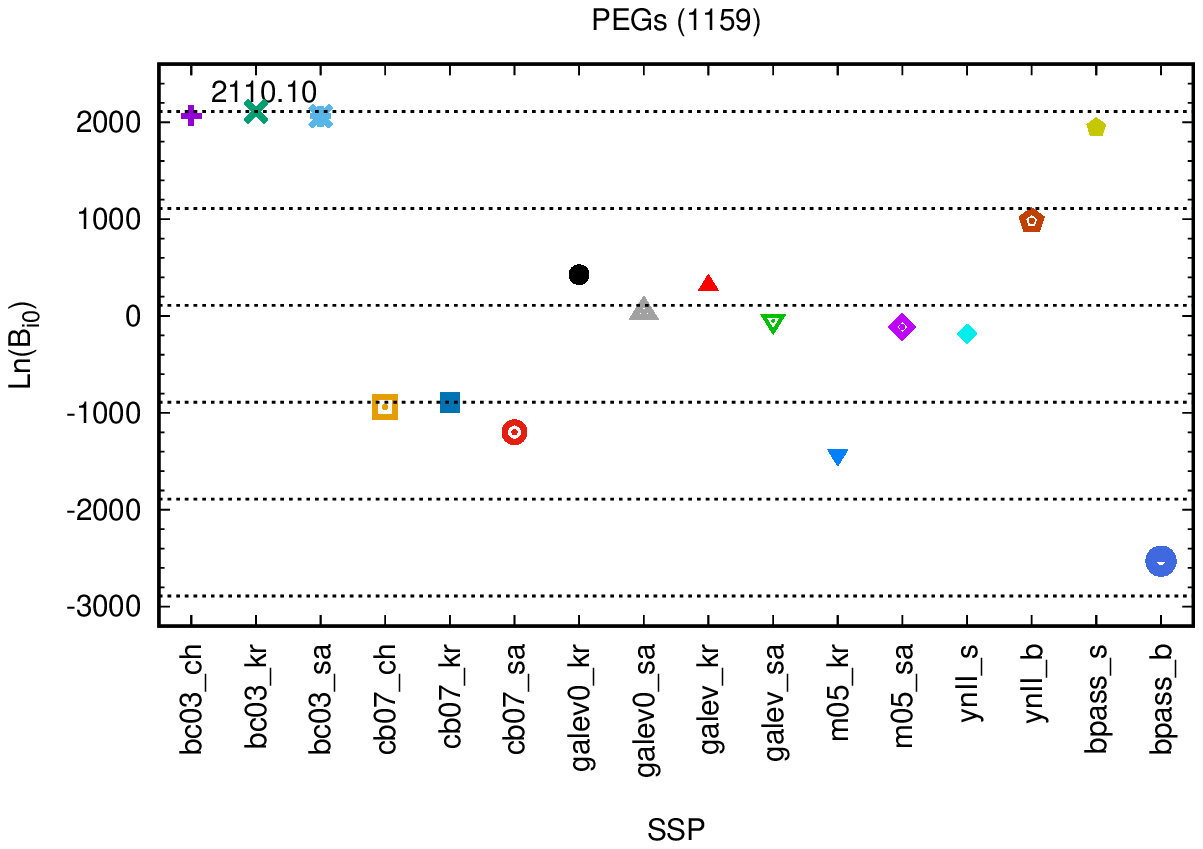}
	\includegraphics[scale=0.73]{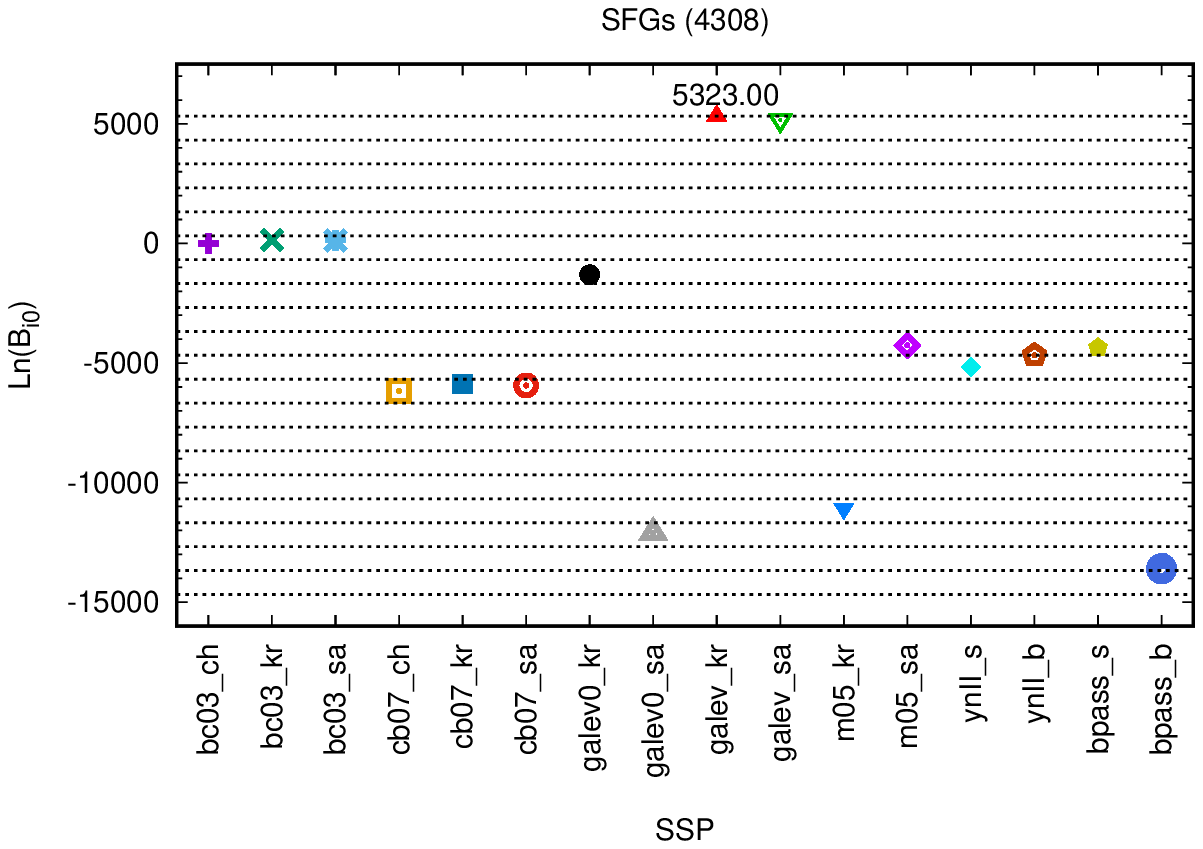}
  \end{center}
  \caption{Similar to Figure \ref{fig:BF_NSED_U_0_sp}, but for the ${\bm M}({ssp}_{0},{sfh},{dal})$-like SED modelings, where the SSP, SFH and DAL are all assumed to be universal, but a fixed SSP, free SFH and DAL are assumed. The dot lines show the values of the Bayes factor with a step of $1000$. The \bc\ SSP with a Kroupa01 IMF (bc03\_kr) has the highest value of Bayes factor ($2110.10$) for the PEGs (left), while the version of GALEV SSP with the consideration of emission lines and a Kroupa01 IMF (galev\_kr) has the highest value of Bayes factor ($5323.00$) for the SFGs (right).}
  \label{fig:BF_NSED_U_011_sp}
\end{figure*}

\begin{figure*}[]
  \begin{center}
	\includegraphics[scale=0.73]{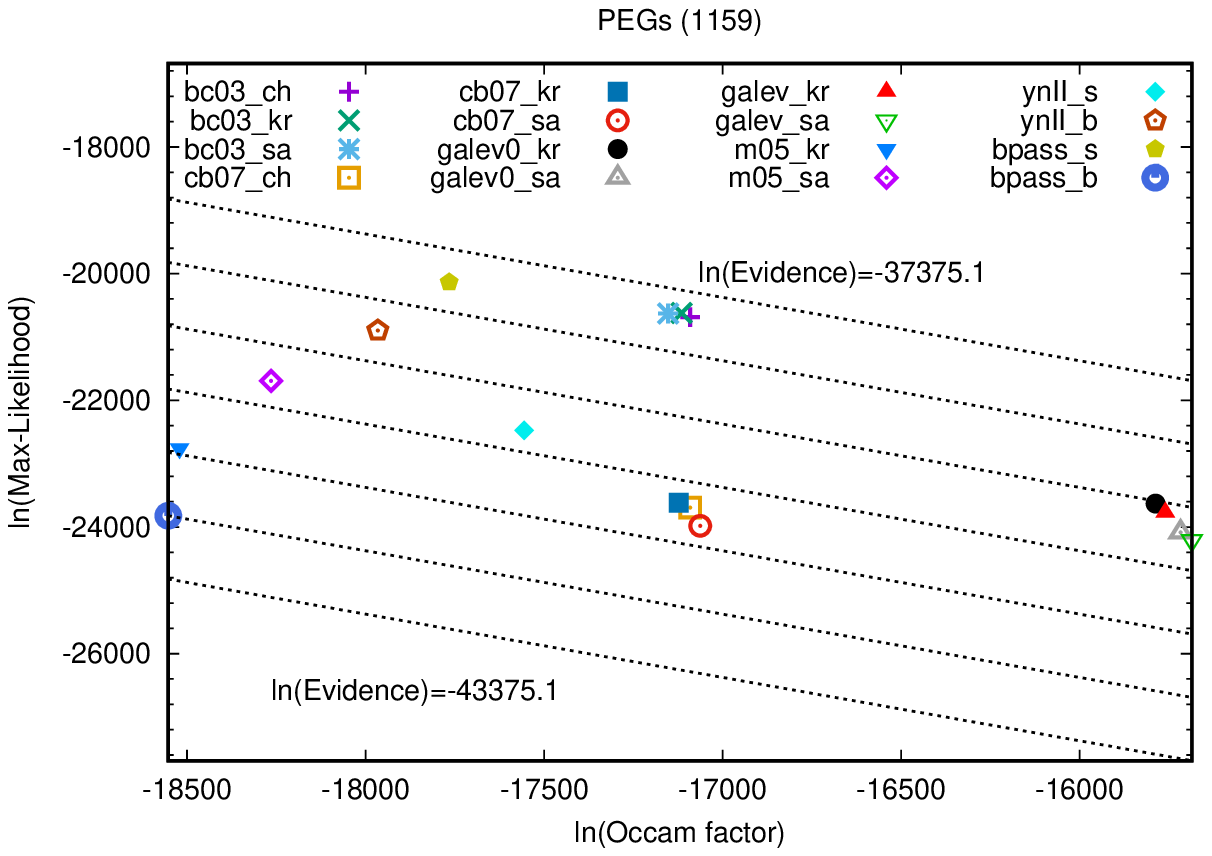}
	\includegraphics[scale=0.73]{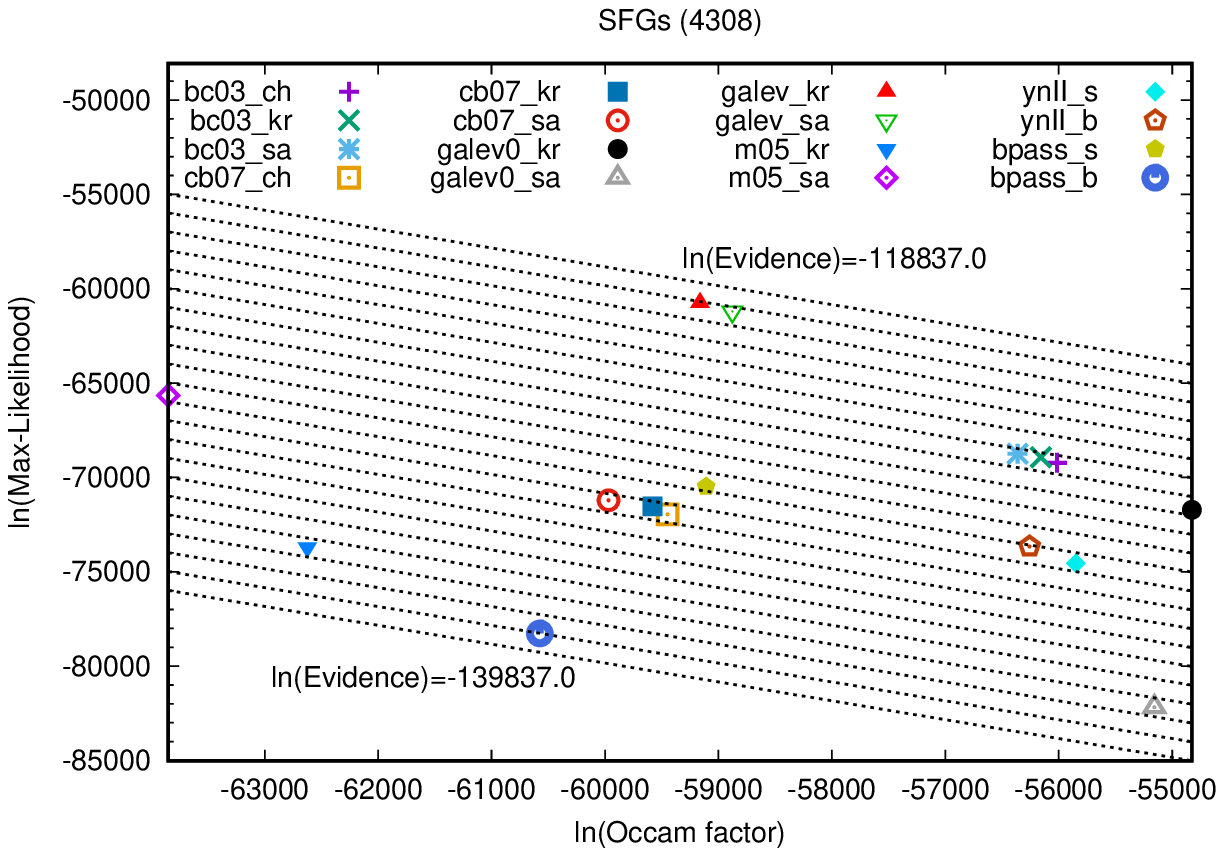}
  \end{center}
  \caption{Similar to Figure \ref{fig:ML-OF-BE_NSED_U_0_sp}, but for the ${\bm M}({ssp}_{0},{sfh},{dal})$-like SED modelings, where the SSP, SFH and DAL are all assumed to be universal, but a fixed SSP, free SFH and DAL are assumed. Here, the Occam factor and max-likelihood are defined in Equations \ref{eq:Omega_total_NSED_UUU_011} and \ref{eq:MaxL_NSED_UUU_011}, respectively. The Bayesian evidence is defined in Equation \ref{eq:bayes_ev_NSED_UUU_011_a}, and calculated with Equation \ref{eq:bayes_ev_NSED_UUU_011}. The dot lines show the values of the Bayesian evidence with a step of $1000$. The version of \bpass\ SSP model without the consideration of binaries (bpass\_s) provides the best fit to the observational data of the PEGs (left), while the version of GALEV SSP models with the consideration of emission lines (galev\_kr, galev\_sa) provides the best fits to the observational data of the SFGs (right).}
  \label{fig:ML-OF-BE_NSED_U_011_sp}
\end{figure*}

In Figure \ref{fig:BF_NSED_U_011_sp}, we show the Bayes factors with respect to the standard model ($M_0^{\rm N}$) for the SED modelings of the PEGs and the SFGs, where a particular SSP is assumed but the SFH and DAL are set to be free to vary.
The comparison of this kind of SED modeling can be used to answer the question: Which SSP is the best for all galaxies in the sample and independently of the choices of SFH and DAL?
It is very clear from the figure that the \bc\ SSP with a Kroupa01 IMF has the highest value of Bayes factor ($2110.10$) for the PEGs, while the version of GALEV model with the consideration of emission lines and a Kroupa01 IMF has the highest value of Bayes factor ($5323.00$) for the SFGs.
Besides, the result for all PEGs in the sample is very different from that for the particular PEG ULTRAVISTA114558, for which the more ``TP-AGB heavy'' SSP models of \cb\ and \ma\ have much larger Bayes factor than other SSPs as shown in Figure \ref{fig:BF_1SED_011_sp}.
Both the results for PEGs and SFGs suggest that the more ``TP-AGB heavy'' SSP models of \cb\ and \ma\ are not universally better than other ``TP-AGB light'' models.
For the PEGs, assuming the version of \bpass\ SSP without the consideration of binaries leads to a Bayes factor that is very close to that of assuming the \bc\ SSPs.
It can be noticed in Figure \ref{fig:ML-OF-BE_NSED_U_011_sp} that the former actually leads to a better fit to the observational data as shown by the larger max-likelihood. 
However, the \bc\ SSPs can lead to larger Occam factors which implies lower model complexity.
It is also worth  noticing that the version of \bpass\ SSP with the consideration of binaries has the lowest Bayes factor.
As shown in Figure \ref{fig:ML-OF-BE_NSED_U_011_sp}, this SSP is located at the bottom left of the ML-OF-BE diagram, which implies low goodness-of-fit to the observational data of PEGs and relatively high model complexity.
On the other hand, the results for the SFGs are more consistent with that for the particular SFG ULTRAVISTA99938 shown in Figure \ref{fig:BF_1SED_011_sp} and \ref{fig:ML-OF-BE_1SED_011_sp}.
However, it becomes even clearer that the version of \galev\ SSP with the consideration of nebular emission lines not only has the highest value of Bayes factor but also provides the best explanation to the observational data of the SFGs.
These results suggest that the consideration of nebular emission lines is indispensable for explaining the photometric observations of the SFGs.

\begin{figure*}[]
  \begin{center}
	\includegraphics[scale=0.73]{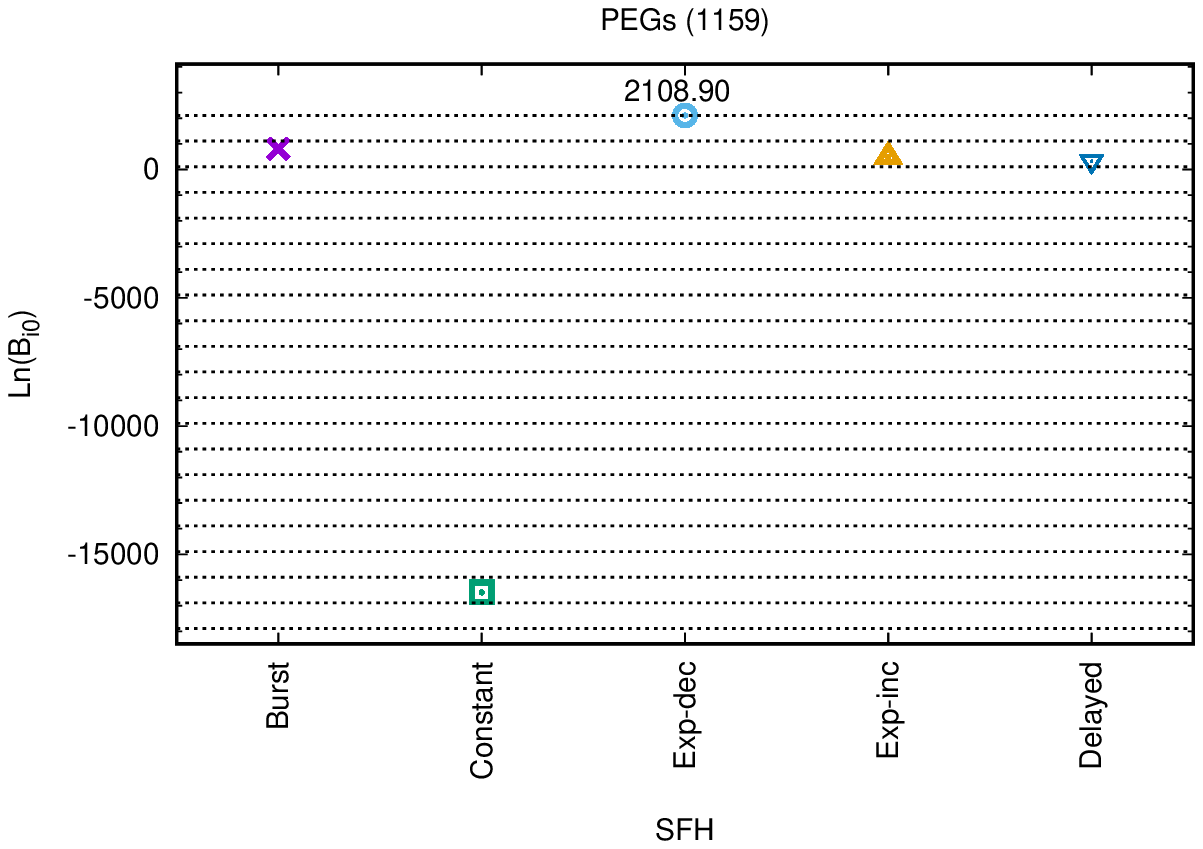}
	\includegraphics[scale=0.73]{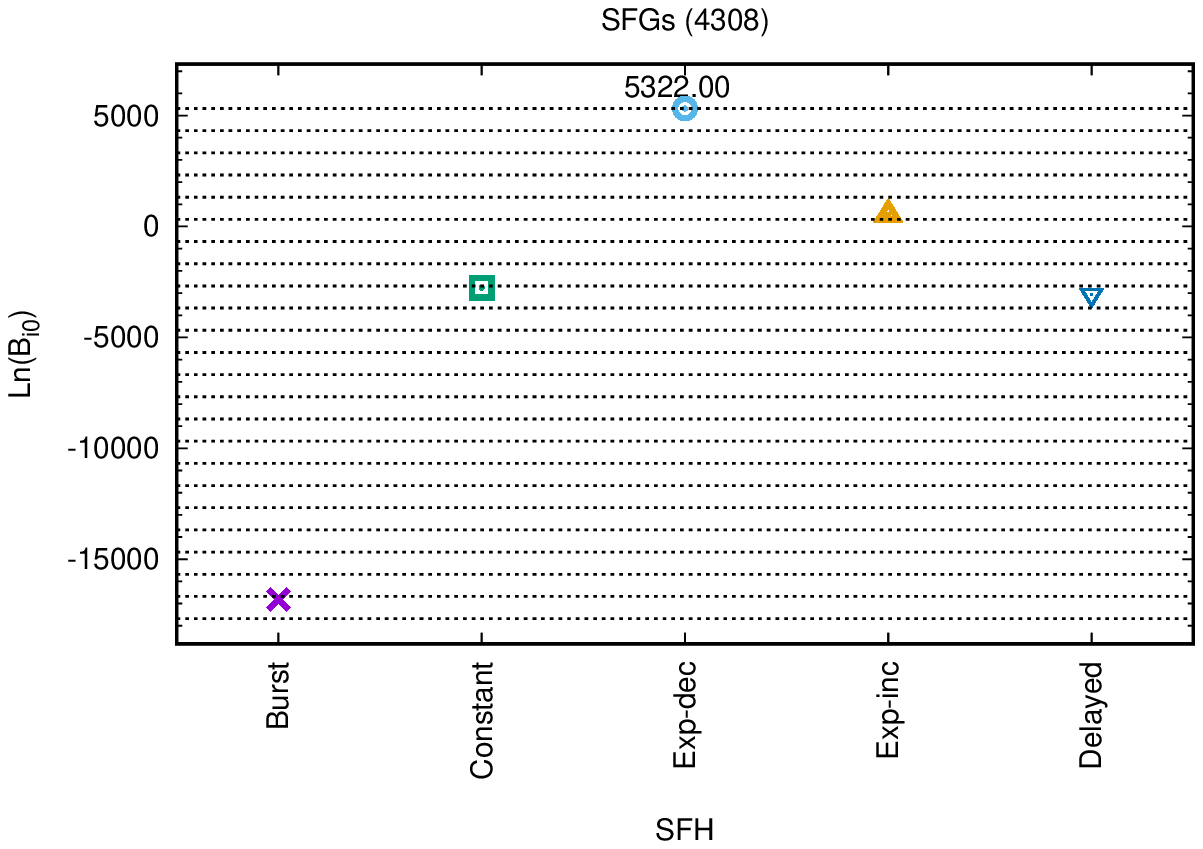}
  \end{center}
  \caption{Similar to Figure \ref{fig:BF_NSED_U_011_sp}, but for the ${\bm M}({ssp},{sfh}_{0},{dal})$-like case, where a fixed SFH, free SSP and DAL are assumed. The commonly assumed Exp-dec SFH has the highest values of Bayes factor for both PEGs (left, $2108.90$) and SFGs (right, $5322.00$).}
  \label{fig:BF_NSED_U_101_sp}
\end{figure*}

\begin{figure*}[]
  \begin{center}
	\includegraphics[scale=0.73]{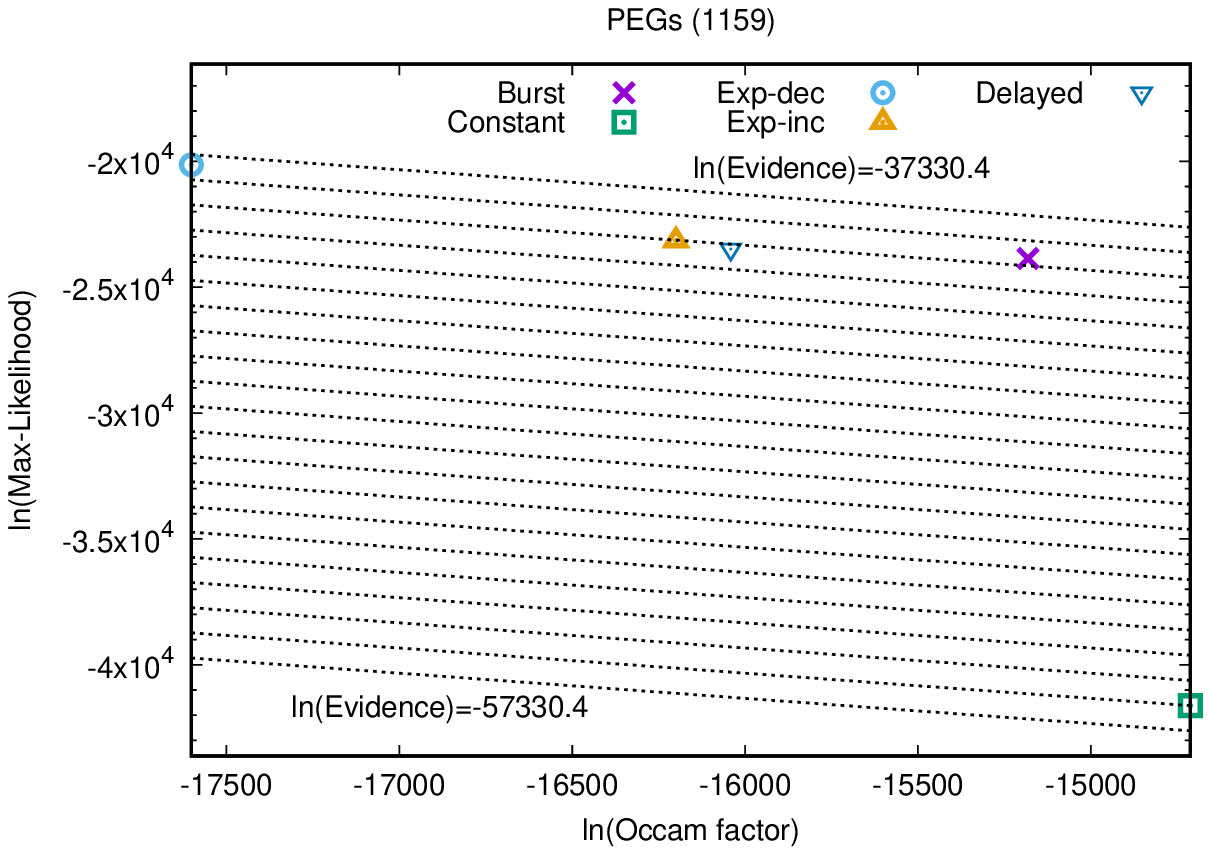}
	\includegraphics[scale=0.73]{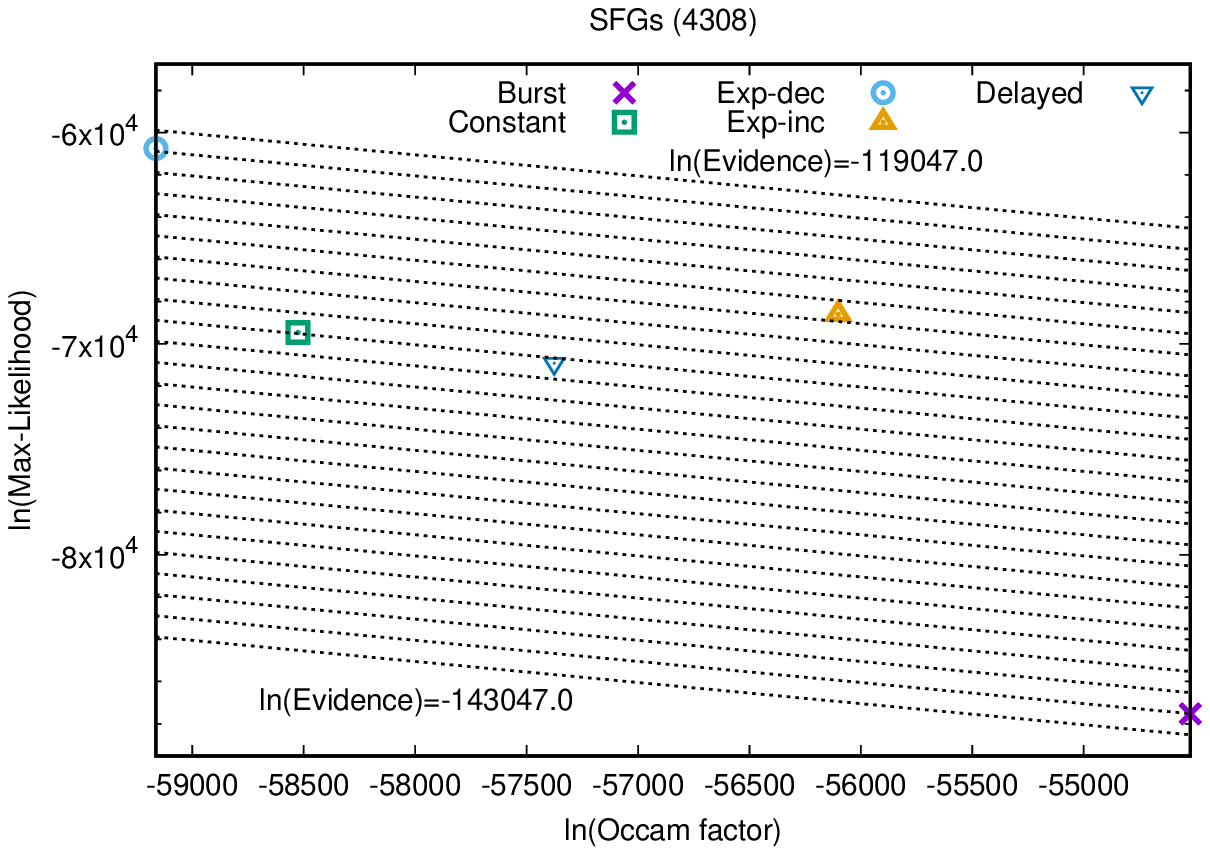}
  \end{center}
  \caption{Similar to Figure \ref{fig:ML-OF-BE_NSED_U_011_sp}, but for the ${\bm M}({ssp},{sfh}_{0},{dal})$-like case, where a fixed SFH, free SSP and DAL are assumed. For both PEGs and SFGs, the widely used Exp-dec SFH provides the best explanation to the observational data, although it has the lowest value of Occam factor (i.e. the highest model complexity).}
  \label{fig:ML-OF-BE_NSED_U_101_sp}
\end{figure*}

\begin{figure*}[]
  \begin{center}
	\includegraphics[scale=0.73]{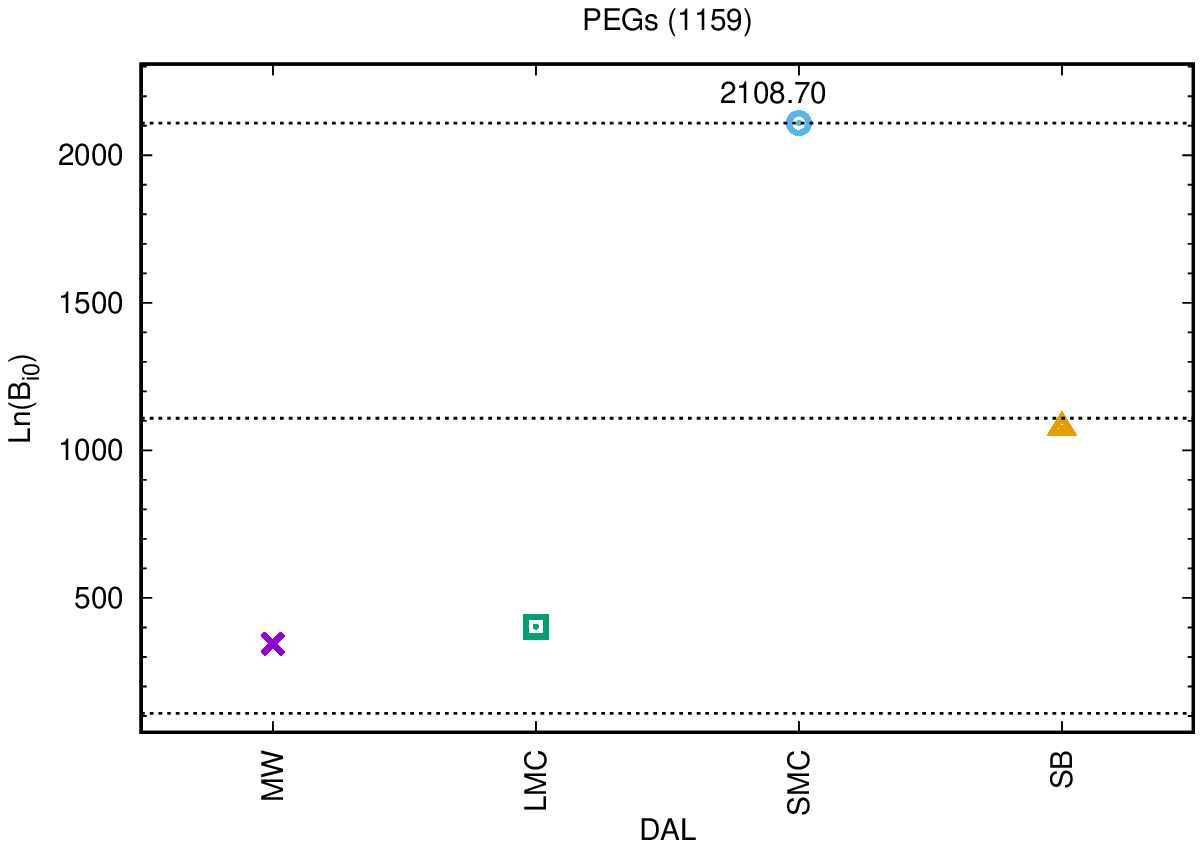}
	\includegraphics[scale=0.73]{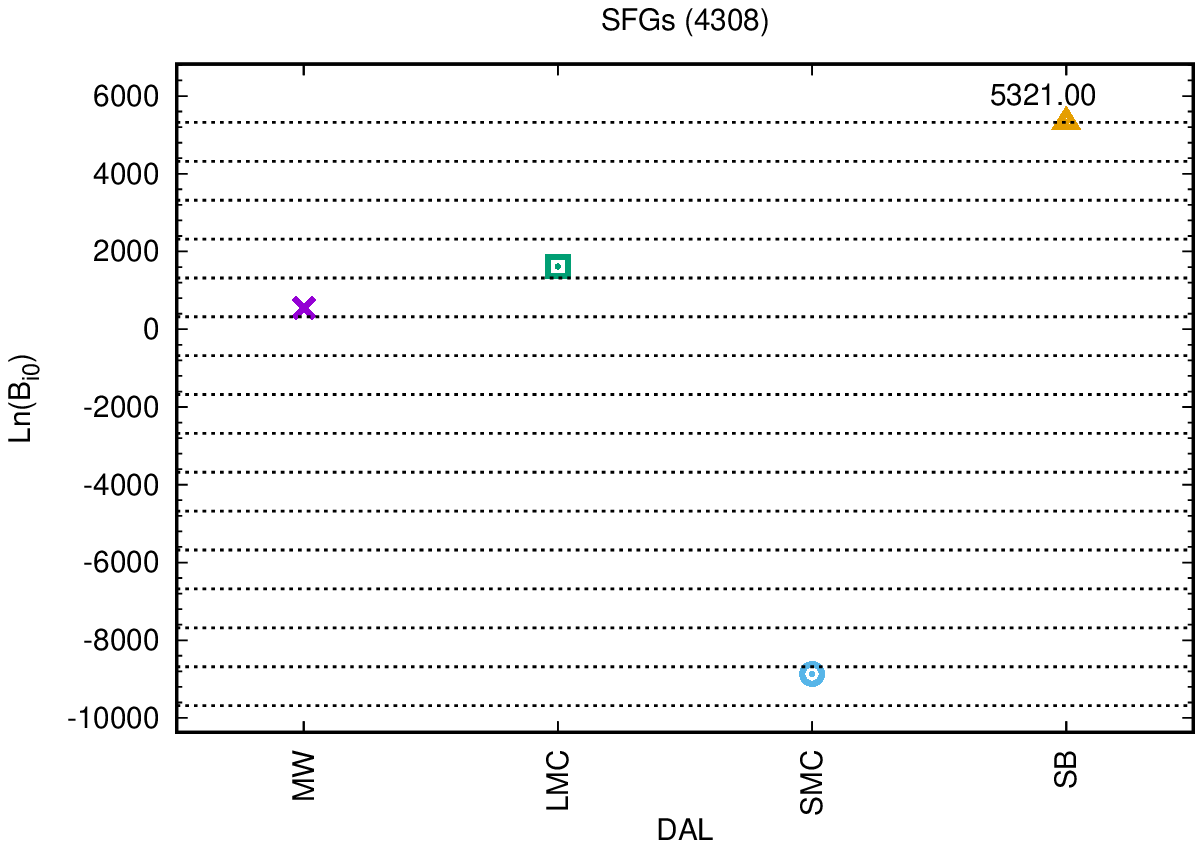}
  \end{center}
  \caption{Similar to Figure \ref{fig:BF_NSED_U_011_sp}, but for the ${\bm M}({ssp},{sfh},{dal}_{0})$-like case, where a fixed DAL, free SSP and SFH are assumed. The SMC-like DAL has the highest value of Bayes factor ($2108.70$) for the PEGs, while the SB-like DAL has the highest value of Bayes factor ($5321.00$) for the SFGs.}
  \label{fig:BF_NSED_U_110_sp}
\end{figure*}

\begin{figure*}[]
  \begin{center}
	\includegraphics[scale=0.73]{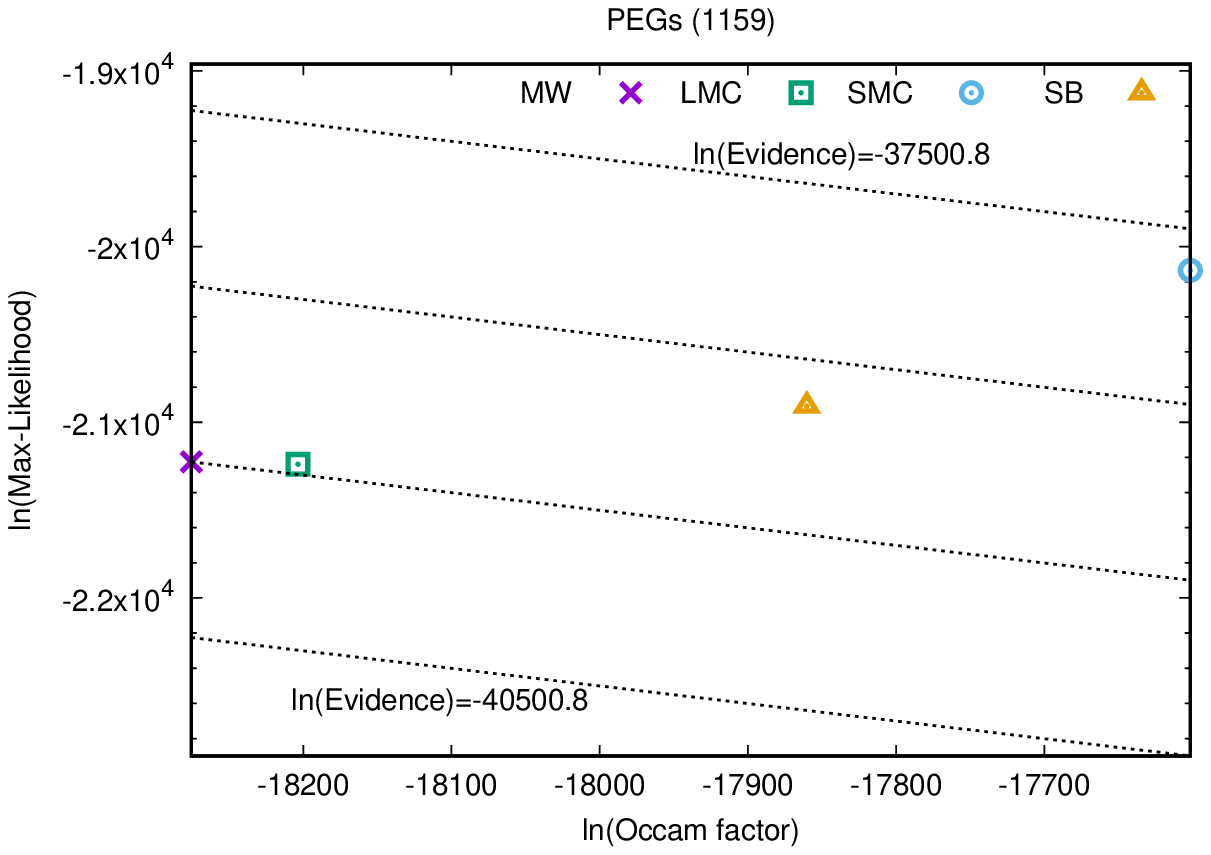}
	\includegraphics[scale=0.73]{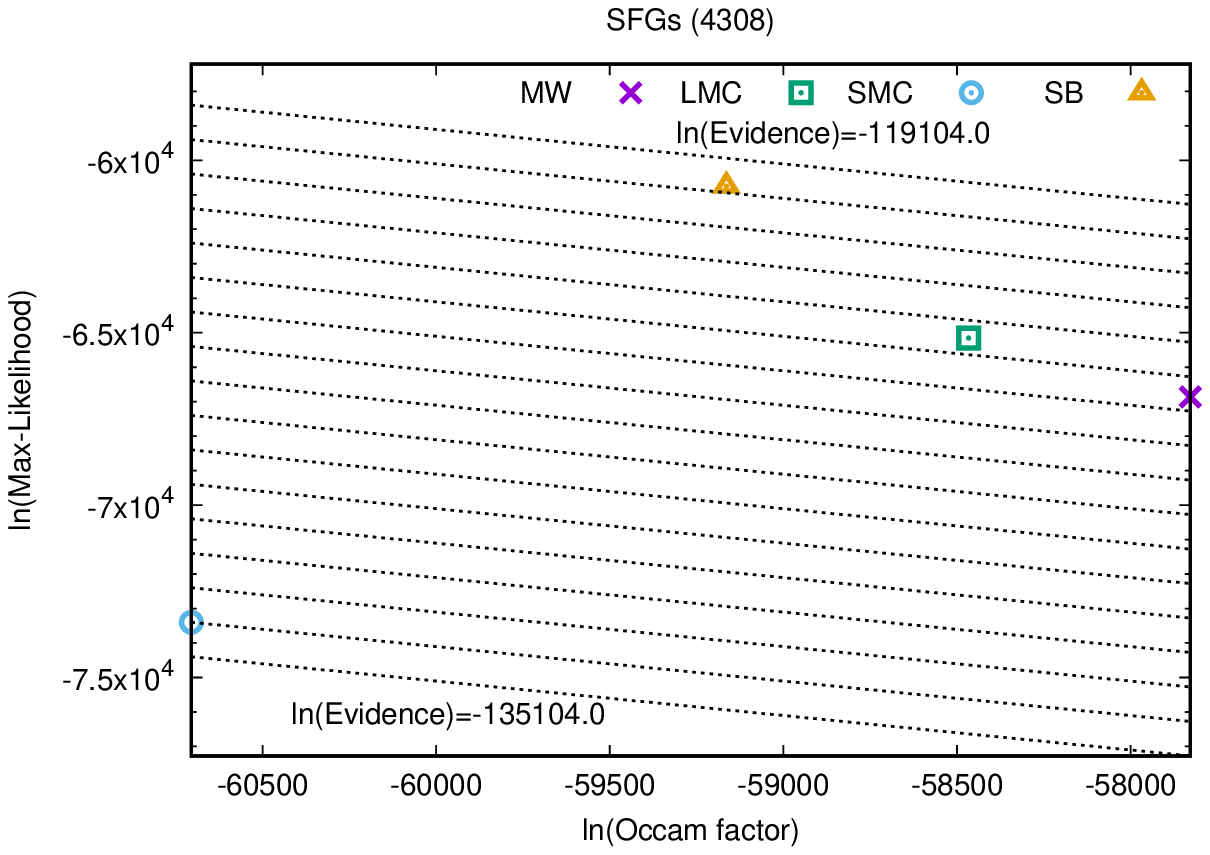}
  \end{center}
  \caption{Similar to Figure \ref{fig:ML-OF-BE_NSED_U_011_sp}, but for the ${\bm M}({ssp},{sfh},{dal}_{0})$-like case, where a fixed DAL, free SSP and SFH are assumed. The SMC-like DAL provides the best explanation to the observational data of PEGs (left), while the SB-like DAL provides the best explanation to the observational data of SFGs (right).}
  \label{fig:ML-OF-BE_NSED_U_110_sp}
\end{figure*}

In Figures \ref{fig:BF_NSED_U_101_sp} and \ref{fig:ML-OF-BE_NSED_U_101_sp}, we present a Bayesian comparison of the different forms of SFHs for the PEGs and SFGs.
The results show that the commonly assumed Exp-dec SFH provides the best explanation to the observational data of both PEGs and SFGs, and has the highest value of Bayes factor, although it has the lowest value of Occam factor and consequently the highest model complexity.
So, the Exp-dec SFH has the best universality for both PEGs and SFGs in our sample, although it is not necessarily the best for all galaxies. 
Besides, the performance of the Burst SFH is much better than the constant SFH for the PEGs, while the opposite is true for the SFGs.
Similarly, in Figures \ref{fig:BF_NSED_U_110_sp} and \ref{fig:ML-OF-BE_NSED_U_110_sp}, we present a Bayesian comparison of the different forms of DALs for the PEGs and SFGs.
The results show that the SMC-like DAL provides the best explanation to the observational data of PEGs and has the highest value of Bayes factor ($2108.90$).
For the SFGs, the SB-like DAL provides the best explanation to the observational data and has the highest value of Bayes factor ($5322.00$). 
The very different SFH and DAL suggest that formation mechanism for the PEGs and the SFGs are generally very different.



\subsubsection{The case for only one of the SSP, SFH and DAL being universal and fixed}\label{sss:NSED_model_U_2}
As demonstrated in \S \ref{sss:NSED_model_U_1}, by the Bayesian comparison of the SED modelings of a sample of galaxies where the SSP, SFH and DAL are all assumed to be universal but only one of them is fixed to a specific choice, we can investigate the universality of different SSPs, SFHs, and DALs, respectively.
However, it is not necessary to assume that SSP, SFH and DAL are all universal when investigating the universality of only one of them.
Actually, it could be even more interesting to find out which SSP model has the best universality for all galaxies in a sample without assuming a universal SFH and DAL, which form of SFH has the best universality for all galaxies in a sample without assuming a universal SSP and DAL, and which form of DAL has the best universality for all galaxies in a sample without assuming a universal SSP and SFH.
So, by the Bayesian comparison of the SED modelings of a sample of galaxies where only one of the SSP, SFH and DAL is assumed to be universal and fixed to a specific choice, we can better understand the universality of different SSPs, SFHs, and DALs, respectively.

\begin{figure*}[]
  \begin{center}
	\includegraphics[scale=0.73]{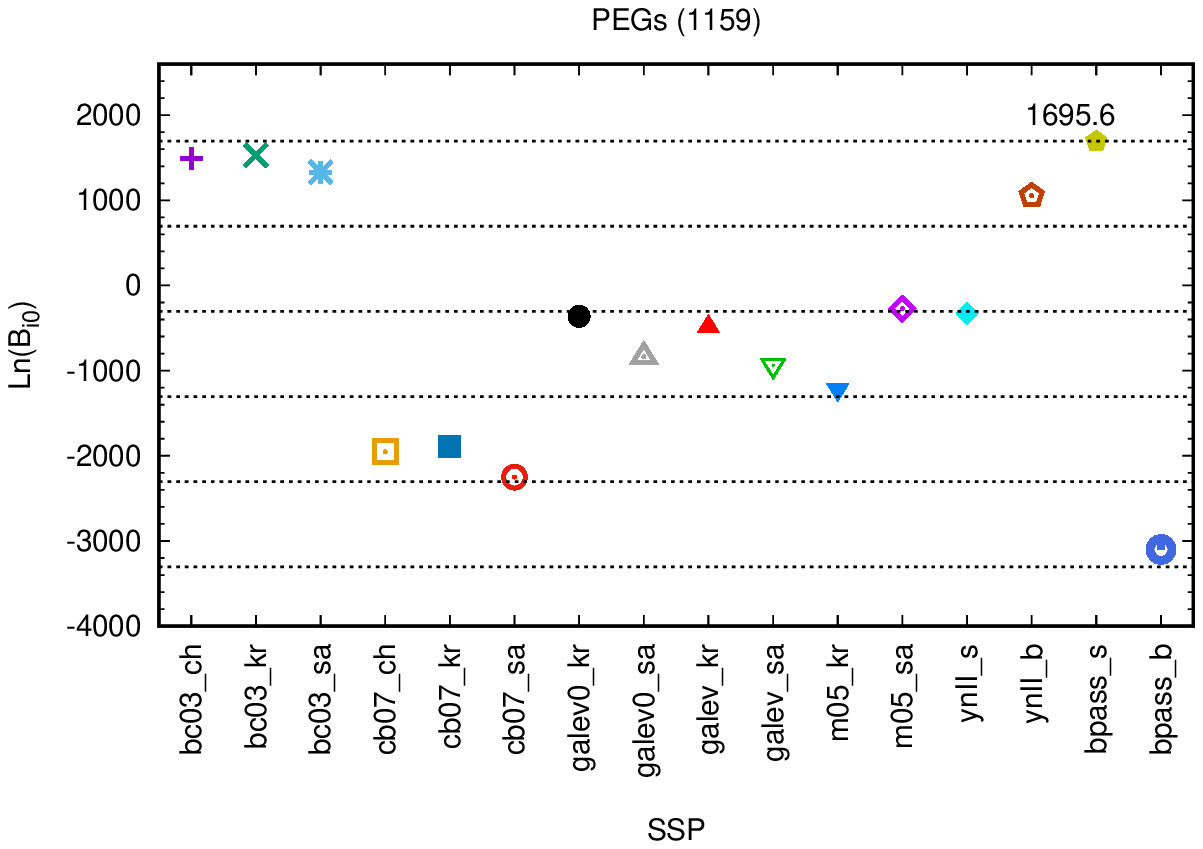}
	\includegraphics[scale=0.73]{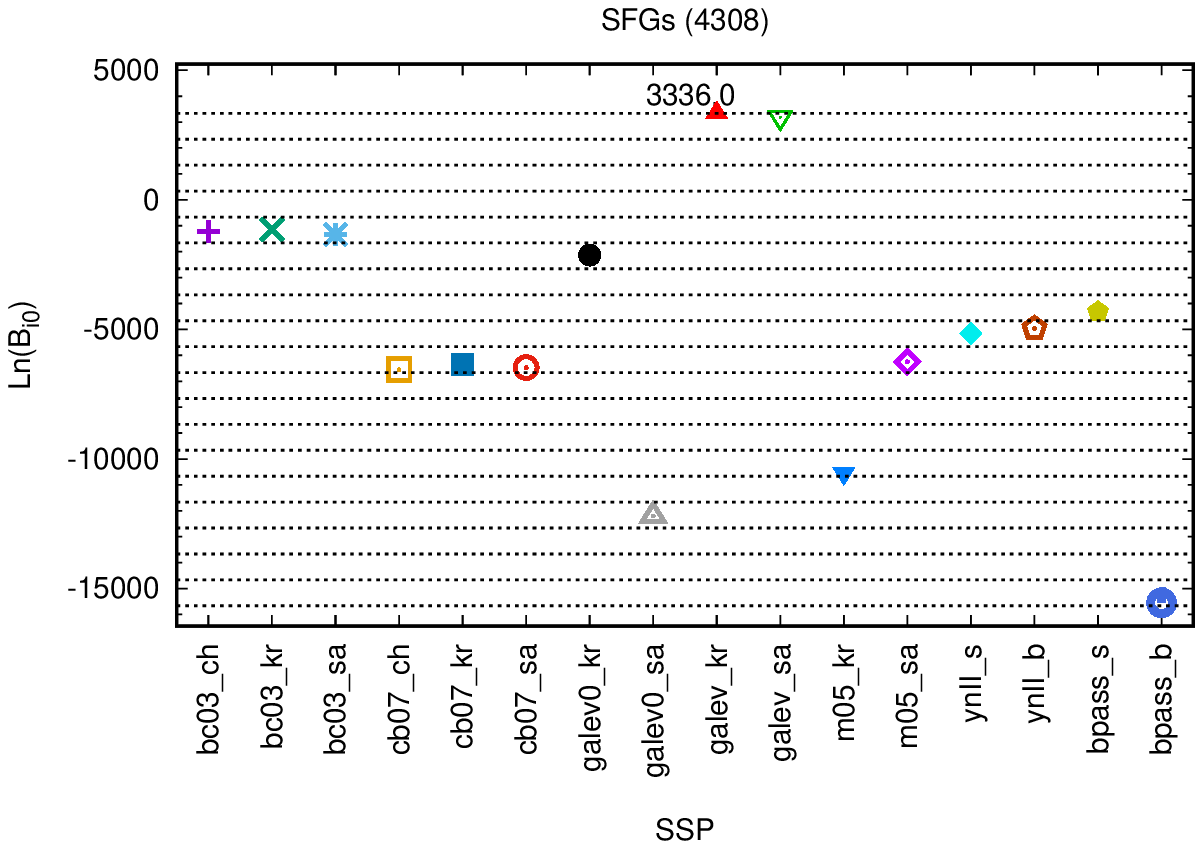}
  \end{center}
  \caption{Similar to Figure \ref{fig:BF_NSED_U_011_sp}, but for the ${\bm M}({ssp}_{0},{sfh_1},{sfh_2},\dots,{sfh_N},{dal_1},{dal_2},\dots,{dal_N})$-like SED modelings where a universal and fixed SSP, object-dependent and free SFH and DAL are assumed. The version of \bpass\ SSP without the consideration of binaries (bpass\_s) has the highest value ($1695.60$) of Bayes factor for the PEGs (left), while the version of GALEV SSP with the consideration of emission lines and a Kroupa01 IMF (galev\_kr) has the highest value ($3336.00$) of Bayes factor for the SFGs (right).}
  \label{fig:BF_NSED_UDD_011_sp}
\end{figure*}

\begin{figure*}[]
  \begin{center}
	\includegraphics[scale=0.72]{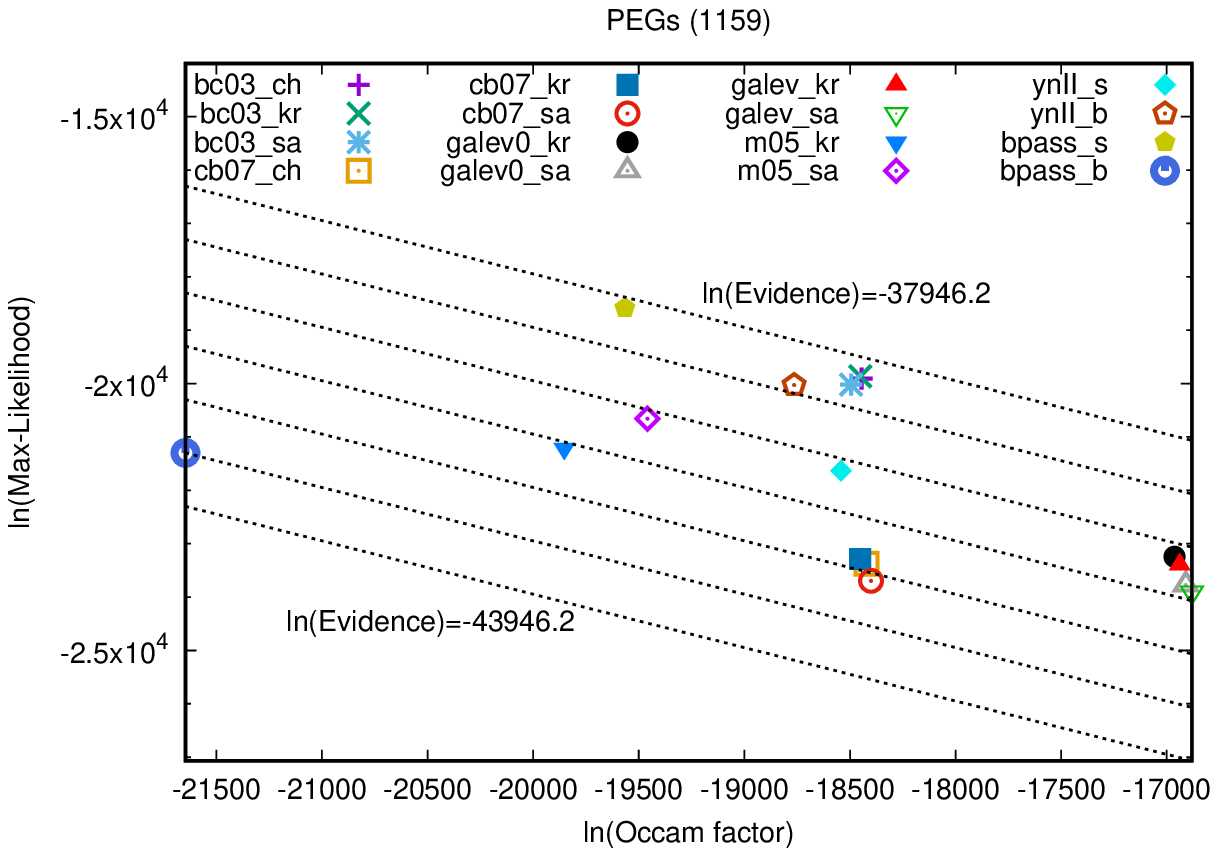}
	\includegraphics[scale=0.72]{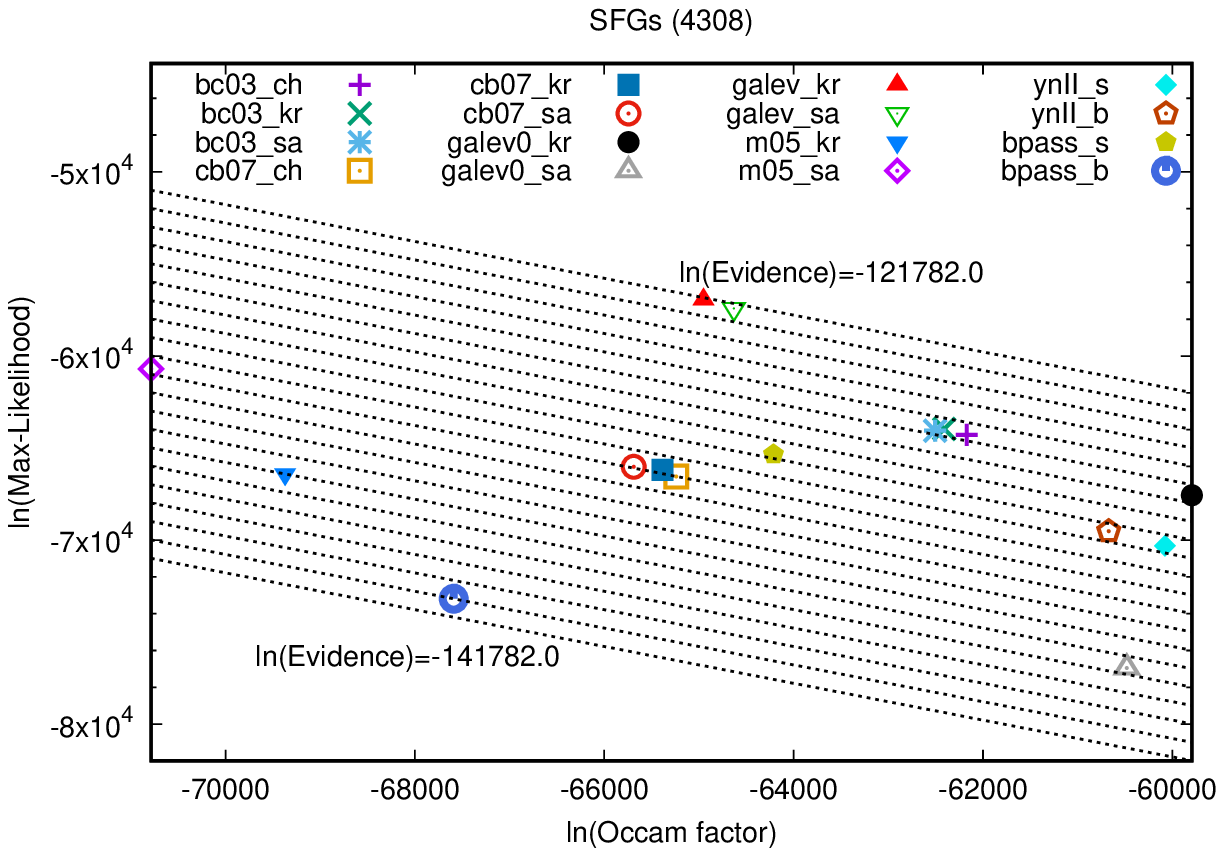}
  \end{center}
  \caption{Similar to Figure \ref{fig:ML-OF-BE_NSED_U_011_sp}, but for the ${\bm M}({ssp}_{0},{sfh_1},{sfh_2},\dots,{sfh_N},{dal_1},{dal_2},\dots,{dal_N})$-like SED modelings where a universal and fixed SSP, object-dependent and free SFH and DAL are assumed. Here, the Occam factor and max-likelihood are defined in Equations \ref{eq:Omega_total_NSED_UDD_011} and \ref{eq:MaxL_NSED_UDD_011}, respectively. The Bayesian evidence is defined in Equation \ref{eq:bayes_ev_NSED_UDD_011_a}, and calculated with Equation \ref{eq:bayes_ev_NSED_UDD_011}. The version of \bpass\ SSP without the consideration of binaries (bpass\_s) provides the best fit to the observational data of  PEGs (left), while the version of GALEV SSP models with the consideration of emission lines (galev\_kr, galev\_sa) provides the best fits to the observational data of SFGs (right).}
  \label{fig:ML-OF-BE_NSED_UDD_011_sp}
\end{figure*}

The Bayesian comparison of the ${\bm M}({ssp}_{0},{sfh_1},{sfh_2},\dots,{sfh_N},{dal_1},{dal_2},\dots,{dal_N})$-like SED modelings of a sample of galaxies can be used to answer the question: Which SSP model has the best universality for all galaxies in the sample and independently of the SFH and DAL assumed for different galaxies?
In Figure \ref{fig:BF_NSED_UDD_011_sp}, we show the Bayes factors with respect to the standard model ($M_0^{\rm N}$) for the ${\bm M}({ssp}_{0},{sfh_1},{sfh_2},\dots,{sfh_N},{dal_1},{dal_2},\dots,{dal_N})$-like SED modelings of the PEGs and the SFGs where only the SSP is assumed to be universal and fixed to a particular choice while the SFH and DAL are assumed to be object-dependent and free.
For the PEGs, it is clear that the version of \bpass\ SSP without the consideration of binaries has the highest value of Bayes factor ($1695.60$), which is only slightly larger than that for the \bc\ SSPs.
The max-likelihoods and Occam factors in the left panel of Figure \ref{fig:ML-OF-BE_NSED_UDD_011_sp} show that the version of \bpass\ SSP without the consideration of binaries provides a much better explanation to the observational data of PEGs than the \bc\ SSPs, while the latter have  much larger Occam factors and consequently much lower model complexity.
For the SFGs, the version of GALEV SSP with the consideration of emission lines and a Kroupa01 IMF has the highest value of Bayes factor ($3336.00$), which is much larger than that of all the other SSPs.
The max-likelihoods and Occam factors in the right panel of Figure \ref{fig:ML-OF-BE_NSED_UDD_011_sp} show that the version of GALEV SSP with the consideration of emission lines and a Kroupa01 IMF provides a much better explanation to the observational data of SFGs than the \bc\ SSPs, while the latter have  much larger Occam factors and consequently much lower model complexity.
A more detailed discussion about the performance of different SSP models will be presented in \S \ref{sec:disc}.

\begin{figure*}[]
  \begin{center}
	\includegraphics[scale=0.73]{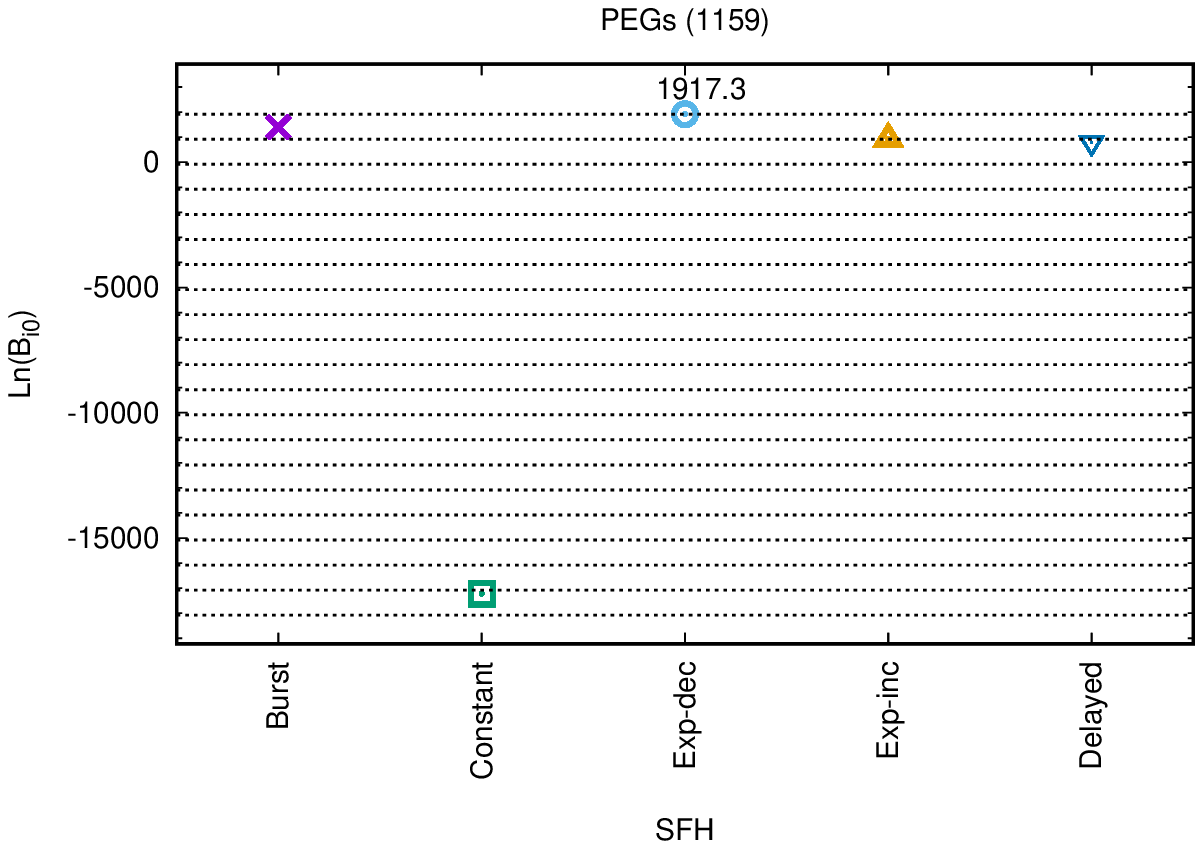}
	\includegraphics[scale=0.73]{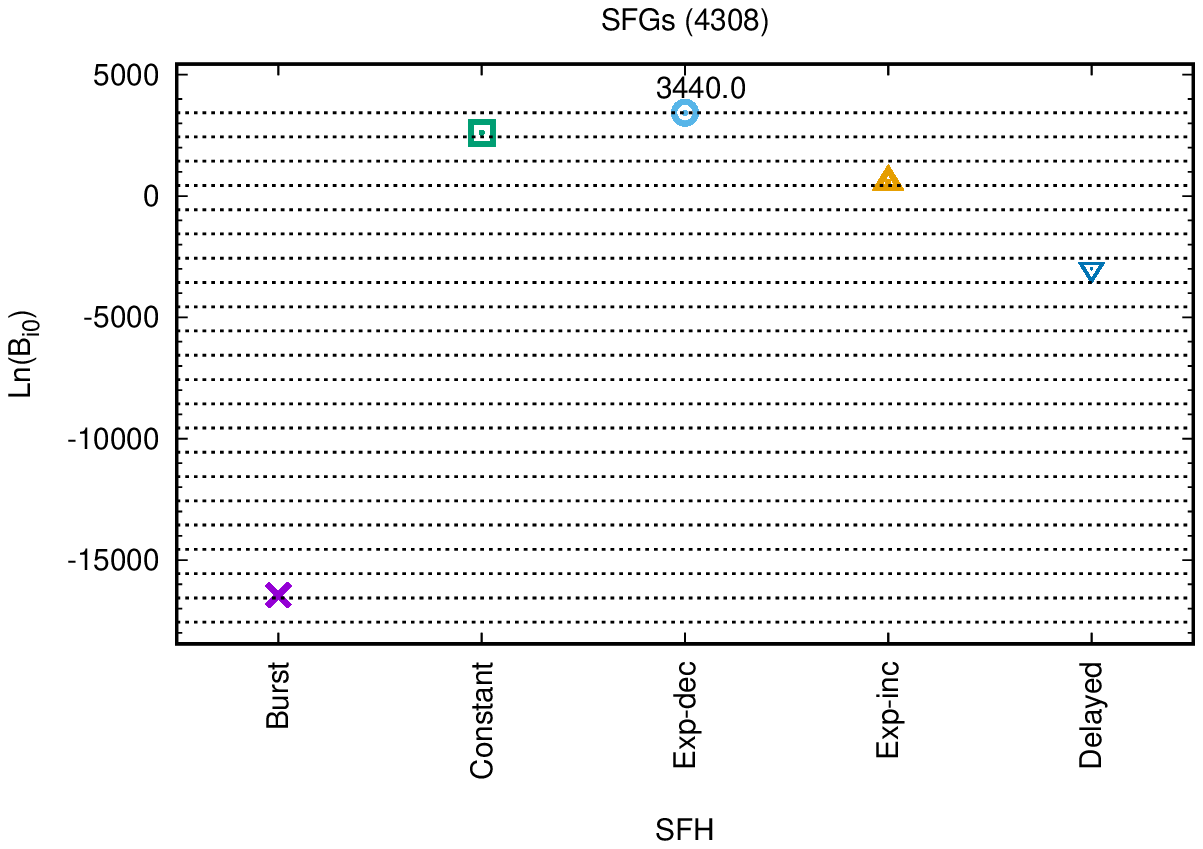}
  \end{center}
  \caption{Similar to \ref{fig:BF_NSED_UDD_011_sp}, but for the ${\bm M}({ssp_1},{ssp_2},\dots,{ssp_N},{sfh}_{0},{dal_1},{dal_2},\dots,{dal_N})$-like case where a universal and fixed SFH, object-dependent and free SSP and DAL are assumed. The commonly assumed Exp-dec SFH has the highest values of Bayes factor for both PEGs (left, $1917.30$) and SFGs (right, $3440.00$).}
  \label{fig:BF_NSED_DUD_101_sp}
\end{figure*}

\begin{figure*}[]
  \begin{center}
	\includegraphics[scale=0.72]{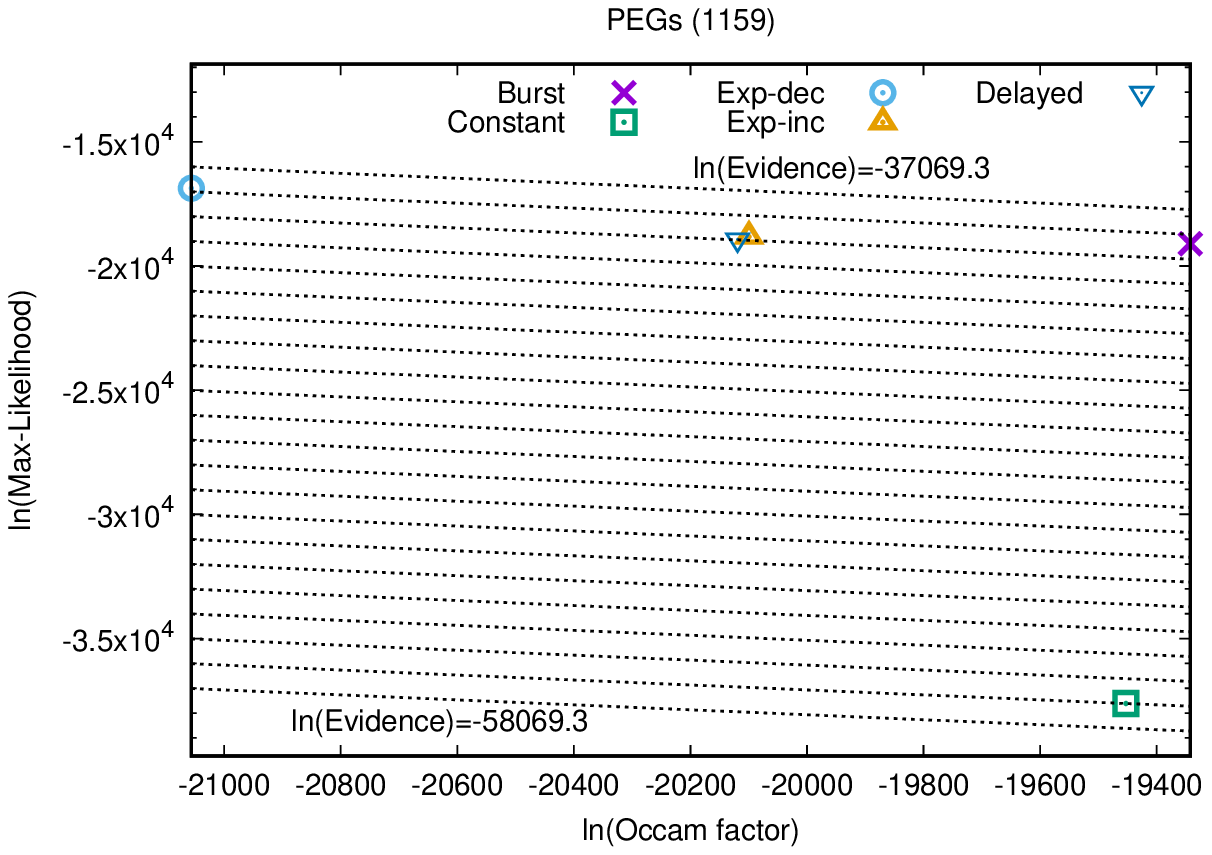}
	\includegraphics[scale=0.72]{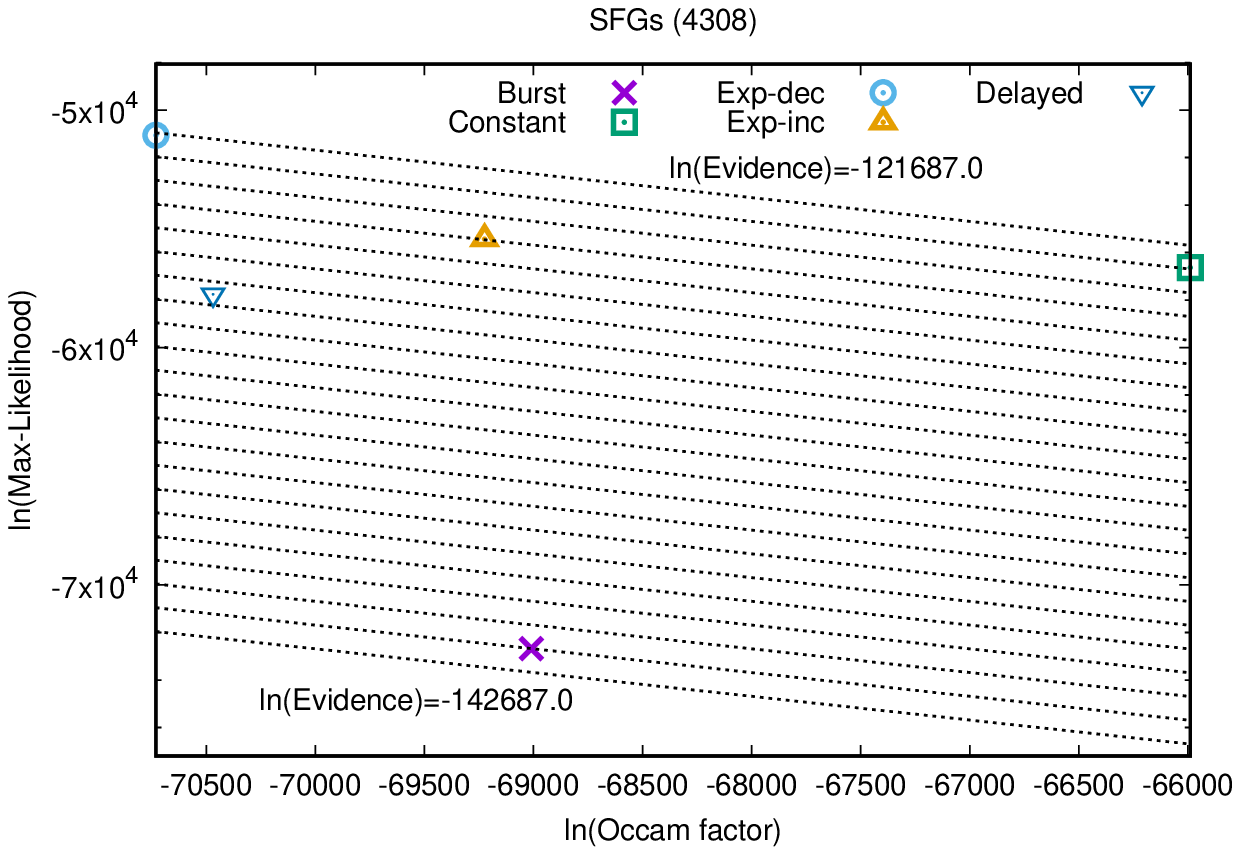}
  \end{center}
  \caption{Similar to Figure \ref{fig:ML-OF-BE_NSED_UDD_011_sp}, but for the ${\bm M}({ssp_1},{ssp_2},\dots,{ssp_N},{sfh}_{0},{dal_1},{dal_2},\dots,{dal_N})$-like case where a universal and fixed SFH, object-dependent and free SSP and DAL are assumed. For both PEGs (left) and SFGs (right), the widely used Exp-dec SFH provides the best explanation to the observational data, although it has the lowest value of Occam factor (i.e. the highest model complexity).}
  \label{fig:ML-OF-BE_NSED_DUD_101_sp}
\end{figure*}

Similarly, the Bayesian comparison of the ${\bm M}({ssp_1},{ssp_2},\dots,{ssp_N},{sfh}_{0},{dal_1},{dal_2},\dots,{dal_N})$-like SED modelings of a sample of galaxies can be used to answer the question: Which form of SFH has the best universality for all galaxies in the sample and independently of the SSP and DAL assumed for different galaxies?
In Figure \ref{fig:BF_NSED_DUD_101_sp}, we show the Bayes factors with respect to the standard model ($M_0^{\rm N}$) for the ${\bm M}({ssp_1},{ssp_2},\dots,{ssp_N},{sfh}_{0},{dal_1},{dal_2},\dots,{dal_N})$-like SED modelings of the PEGs and the SFGs where only the SFH is assumed to be universal and fixed to a particular choice while the SSP and DAL are assumed to be object-dependent and free.
For the PEGs, the Exp-dec SFH has the largest Bayes factor, while the Burst SFH has the second largest Bayes factor. 
For the SFGs, the Exp-dec SFH still has the largest Bayes factor, while the constant SFH has the second largest Bayes factor. 
The ML-OF-BE diagram in the left panel of Figure \ref{fig:ML-OF-BE_NSED_DUD_101_sp} show that the Exp-dec SFH provides a much better explanation to the observational data of PEGs than other form of SFHs.
The Burst SFH has a much larger Occam factor and consequently a much lower model complexity, although it is not as good as the Exp-dec SFH for fitting the observational data of PEGs.
Meanwhile, the ML-OF-BE diagram in the right panel of Figure \ref{fig:ML-OF-BE_NSED_DUD_101_sp} show that the Exp-dec SFH also provides a much better explanation to the observational data of SFGs than other form of SFHs.
The constant SFH has a much larger Occam factor and consequently a much lower model complexity, although it is not as good as the Exp-dec SFH for fitting the observational data of SFGs.

\begin{figure*}[]
  \begin{center}
	\includegraphics[scale=0.73]{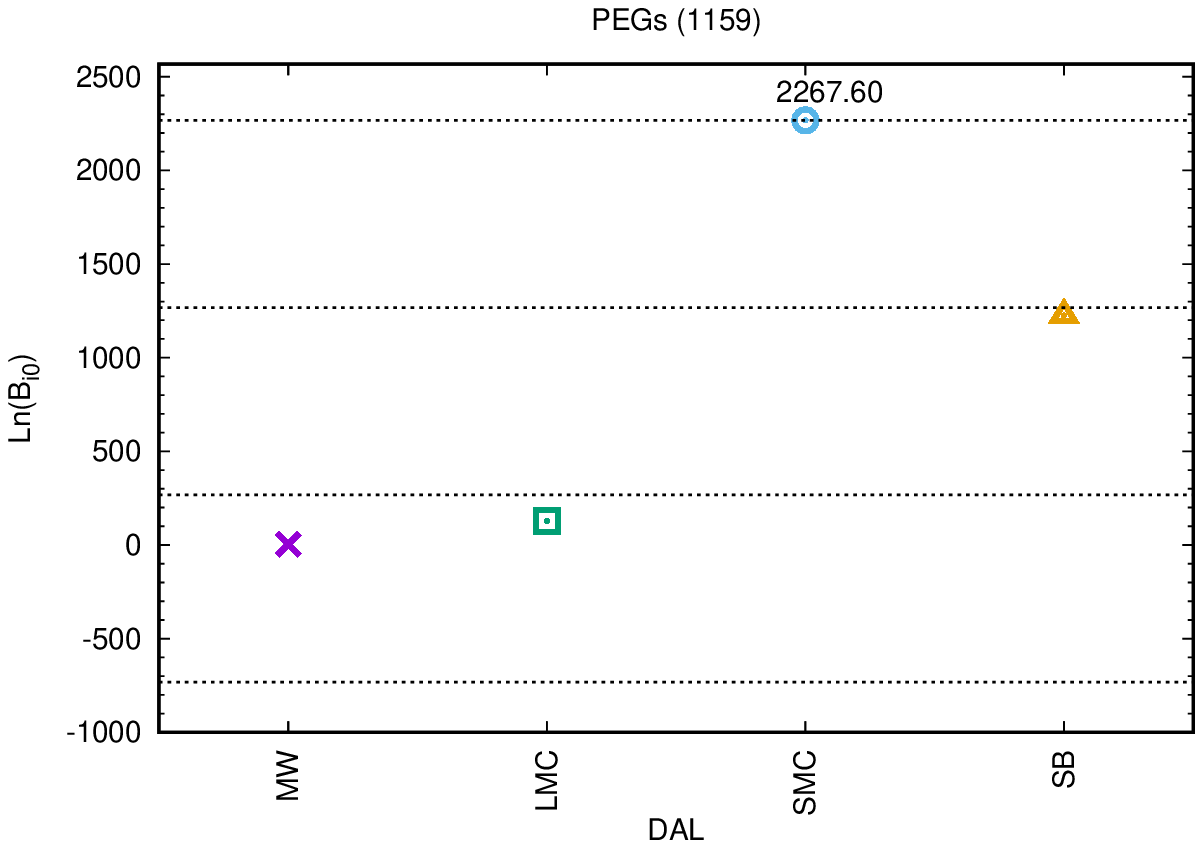}
	\includegraphics[scale=0.73]{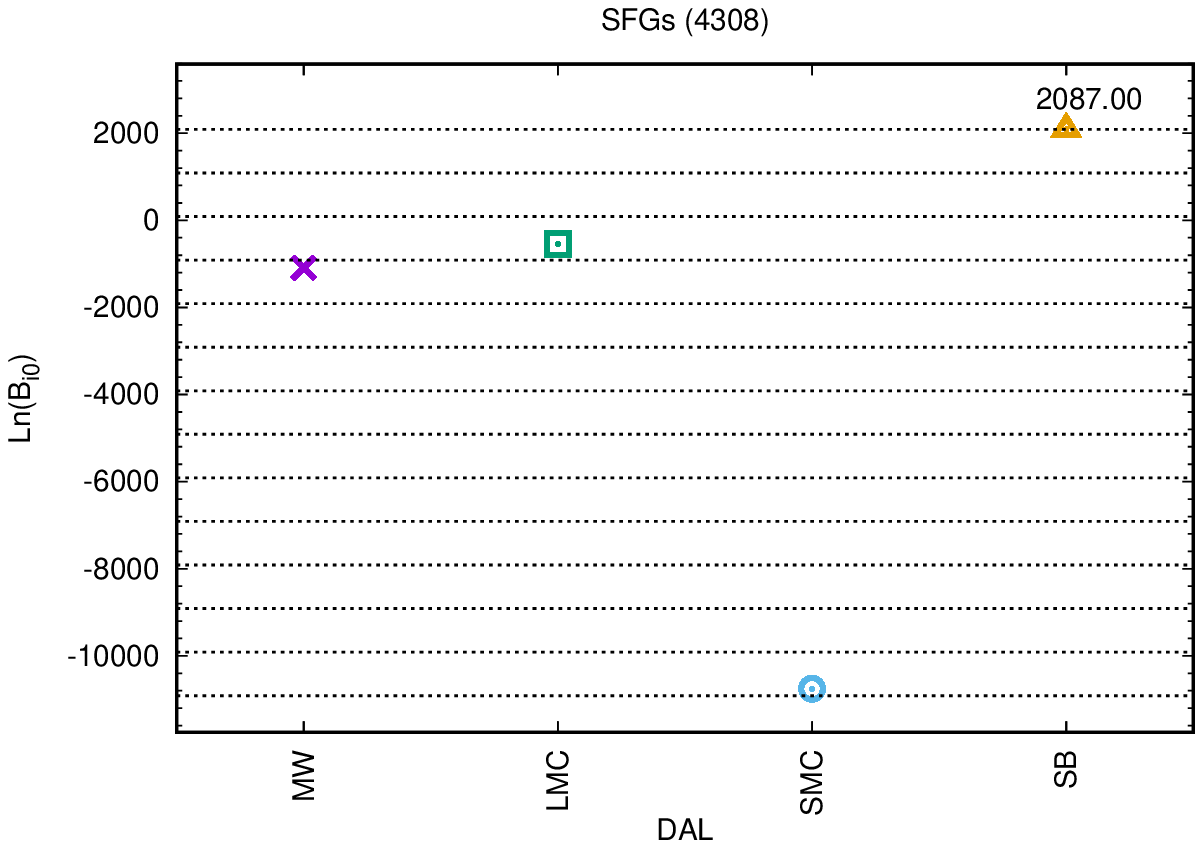}
  \end{center}
  \caption{Similar to \ref{fig:BF_NSED_UDD_011_sp}, but for the ${\bm M}({ssp_1},{ssp_2},\dots,{ssp_N},{sfh_1},{sfh_2},\dots,{sfh_N},{dal}_{0})$-like case where a universal and fixed DAL, object-dependent and free SSP and SFH are assumed. The SMC-like DAL has the highest value ($2267.60$) of Bayes factor for the PEGs (left), while the SB-like DAL has the highest value ($2087.00$) of Bayes factor for the SFGs (right).}
  \label{fig:BF_NSED_DDU_110_sp}
\end{figure*}

\begin{figure*}[]
  \begin{center}
	\includegraphics[scale=0.72]{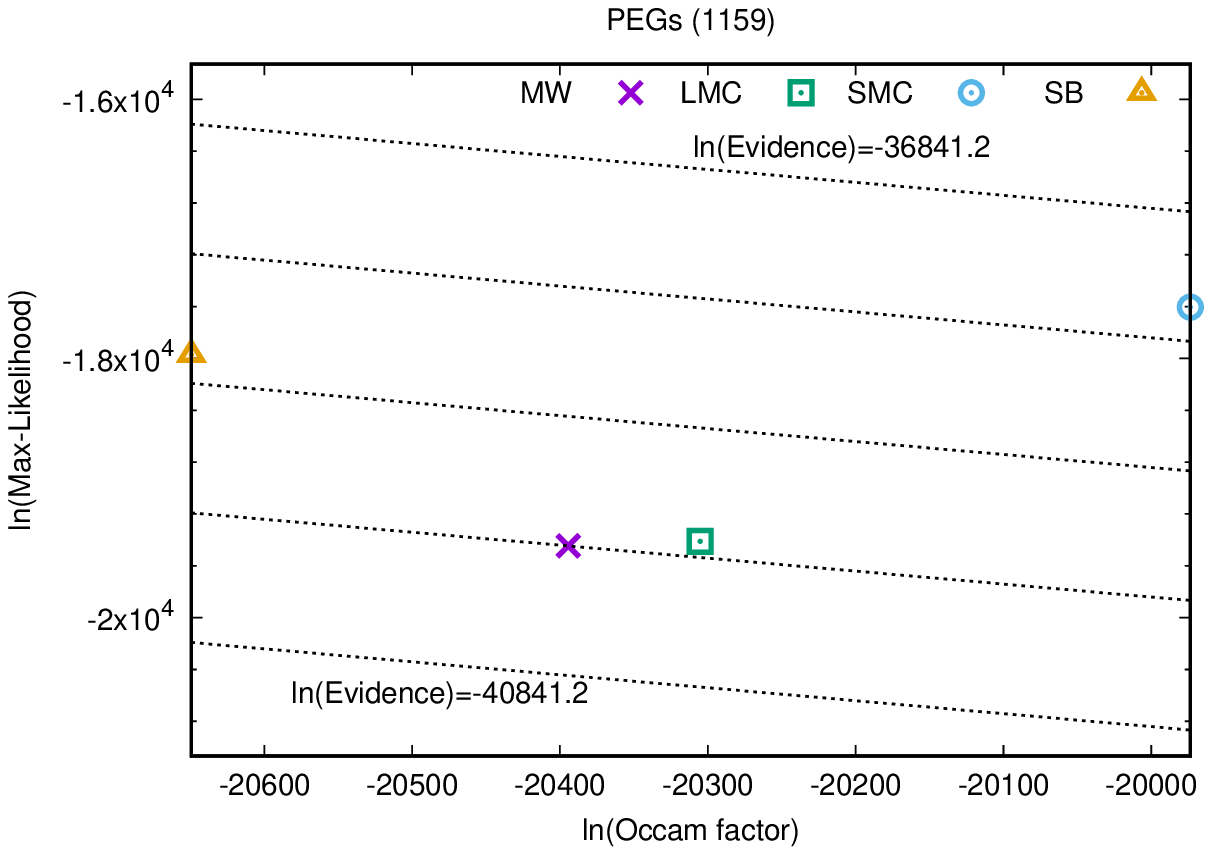}
	\includegraphics[scale=0.72]{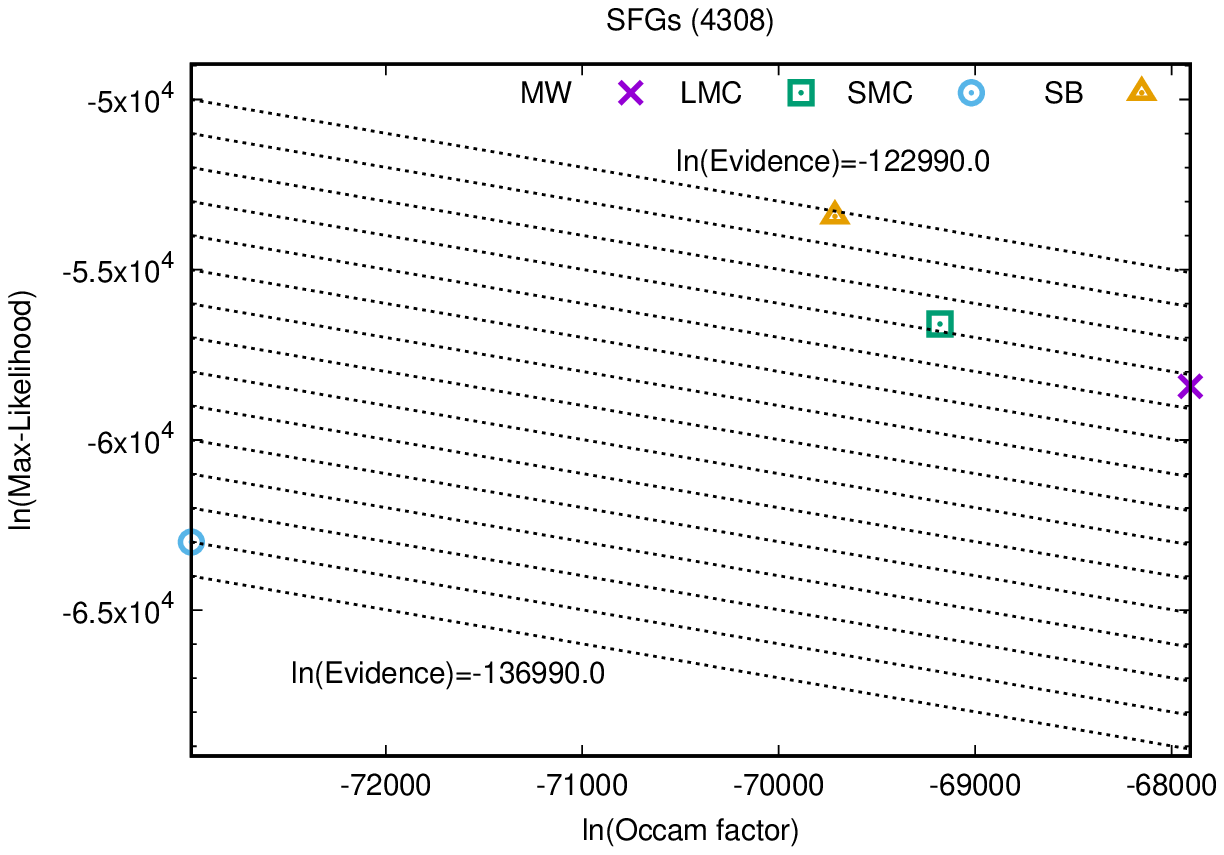}
  \end{center}
  \caption{Similar to Figure \ref{fig:ML-OF-BE_NSED_UDD_011_sp}, but for the ${\bm M}({ssp_1},{ssp_2},\dots,{ssp_N},{sfh_1},{sfh_2},\dots,{sfh_N},{dal}_{0})$-like case where a universal and fixed DAL, object-dependent and free SSP and SFH are assumed. The SMC-like DAL provides the best explanation to the observational data of PEGs (left) and has the lowest model complexity, while the SB-like DAL provides the best explanation to the observational data of SFGs (right) but has a relatively large model complexity.}
  \label{fig:ML-OF-BE_NSED_DDU_110_sp}
\end{figure*}

Finally, the Bayesian comparison of the ${\bm M}({ssp_1},{ssp_2},\dots,{ssp_N},{sfh_1},{sfh_2},\dots,{sfh_N},{dal}_{0})$-like SED modelings of a sample of galaxies can be used to answer the question: Which form of DAL has the best universality for all galaxies in the sample and independently of the SSP and SFH assumed for different galaxies?
In Figure \ref{fig:BF_NSED_DDU_110_sp}, we show the Bayes factors with respect to the standard model ($M_0^{\rm N}$) for the ${\bm M}({ssp_1},{ssp_2},\dots,{ssp_N},{sfh_1},{sfh_2},\dots,{sfh_N},{dal}_{0})$-like SED modelings of the PEGs and the SFGs where only the DAL is assumed to be universal and fixed to a particular choice while the SSP and SFH are assumed to be object-dependent and free.
It is clear from the figure that the SMC-like DAL has the largest Bayes factor for the PEGs, while SB-like DAL has the largest Bayes factor for the SFGs. 
The ML-OF-BE diagram in the left panel of Figure \ref{fig:ML-OF-BE_NSED_DDU_110_sp} show that the SMC-like DAL provides a better explanation to the observational data of PEGs than other form of DALs and has the lowest model complexity.
Meanwhile, the ML-OF-BE diagram in the right panel of Figure \ref{fig:ML-OF-BE_NSED_DDU_110_sp} show that the SB-like DAL provides a much better explanation to the observational data of SFGs than other forms of DAL, although it has a relatively large model complexity.


\section{Discussion}\label{sec:disc}
As mentioned in \S \ref{s:intro} and \S \ref{s:model}, there are many uncertain components in the SED modeling of galaxies.
Different considerations of these uncertain components will result in very different SED predictions and very different estimations about the physical parameters of galaxies \citep{Conroy2009a,Lee2009a,Longhetti2009a,AbrahamseA2011a,MagrisC.2011a,Pforr2012a,Dolphin2012b,Kobayashi2013a,MichalowskiM2014a,PacificiC2015a}.
So, it is very important to find a valid method to discriminate among the different considerations about those uncertain components in the SED modeling of galaxies.
In this paper, we have proposed a new Bayesian framework to compare the SED modelings of a sample of galaxies with different assumptions about three of the major uncertain components: SSP, SFH and DAL.
We suggest that the Bayesian evidence, which is determined by the trade-off between the complexity of a model and its goodness-of-fit to the data, is a more reasonable and useful quantification for the performance of a model.
Besides, by calculating the Bayesian evidence for the SED modeling of a sample of galaxies instead of just one galaxy, this new Bayesian framework allow us to investigate the universality of different SSPs, SFHs and DALs. 
In this section, we discuss some results obtained with the first application of this new method.

\subsection{The universality of different SSP models} \label{ss:SSP}
One of the most important uncertainty in the SED modeling of galaxies is the modeling of a simple stellar population.
As mentioned in \S \ref{ss:ssp}, there are many uncertainties about the star formation (e.g. IMF)  and evolution, stellar spectral libraries and synthesis method in this procedure.
With the different treatments of these uncertainties, different SSP models may have different limitations.
However, for the study of the galaxy formation and evolution, the best SSP model should have the best universality to avoid the bias introduced by the employed SSP model.
Here, we discuss the results for the SSP models with a focus on their universality.

\subsubsection{The contribution of TP-AGB stars} \label{ss:tpagb}
While the importance of TP-AGB stars in the SED modeling of an SSP is well-established \citep{Maraston2005a,BruzualA2007a,Conroy2009a}, the appropriate treatment of them is still an open issue.
\cite{MarastonC2006a} presented a comparison between the performance of the \bc\ and \ma\ models, which are very different in the treatment of TP-AGB phase, for a sample of seven passively evolving high-z galaxies.
They found that TP-AGB phase is very important for the interpretation of rest-frame near-IR data, and the \ma\ models gives better fits to these galaxies than the \bc\ models.
In \cite{Kriek2010a} and \cite{Zibetti2013a}, two samples of (62 for the former and 16 for the latter) Post-starburst galaxies, where the contribution of TP-AGB stars are thought to be most prominent, have been used to discriminate the SSP models  with different considerations for the contribution of TP-AGB stars.
They found that the ``TP-AGB light'' \bc\ model are more favored  than the ``TP-AGB heavy'' \ma\ model, since the former can more consistently fit the rest-frame optical to near-infrared parts of the SEDs of these galaxies.
\cite{CapozziD2016a} presented a comparison of the performance of three SSP models with heavy, mild and light TP-AGB contribution for a sample of $51$ spectroscopically confirmed high-z passive galaxies.
They found that the observed SEDs of these galaxies can be best fitted by assuming a significant contribution from TP-AGB stars.
Different methods have been used in these works and sometimes lead to discrepant conclusions.
However, they are similar in that the performance of different models are mainly compared with their goodness-of-fit (as quantified by the $\chi^2$) to the observational data of a relative small sample of galaxies.

The Bayesian evidence, which is determined by the trade-off between the complexity of a model and its goodness-of-fit to the data, could be a more reasonable and useful quantification for the performance of a model.
In \S \ref{s:apply}, we have employed the Bayesian evidence to compare the performance of different SSP models for individual galaxies (\S \ref{ss:1SED_model}) and a sample of galaxies (\S \ref{ss:NSED_model}), respectively.
The results in Figures \ref{fig:BF_1SED_011_sp} and \ref{fig:ML-OF-BE_1SED_011_sp} show that the more ``TP-AGB heavy'' model of \cb\ and \ma\ have larger Bayesian evidence and goodness-of-fit to the data for the specific PEG ULTRAVISTA114558, which is just the opposite of what is the case for the specific SFG ULTRAVISTA99938.
Although this result is robust against the different choices of SFH and DAL for the two galaxies, it may not be representative of a population of galaxies.

In Figures \ref{fig:BF_NSED_U_011_sp} and \ref{fig:ML-OF-BE_NSED_U_011_sp}, we have compared the performance of different SSPs for a sample of $1159$ PEGs and a sample of $4308$ SFGs, where the SSP, SFH and DAL are all assumed to be universal but only the SSP is fixed to a particular choice.
For both the sample of PEGs and SFGs, the results suggest that the more ``TP-AGB heavy'' model of \cb\ and \ma\ are not universally better than other ``TP-AGB light'' models either in the sense of the Bayesian evidence or the goodness-of-fit to the data alone.
Furthermore, in Figures \ref{fig:BF_NSED_UDD_011_sp} and \ref{fig:ML-OF-BE_NSED_UDD_011_sp}, we have compared the performance of different SSPs for the sample of PEGs and SFGs without assuming a universal SFH and DAL for all galaxies to obtain more robust results.
Interestingly, the obtained results are basically the same.
So, the results of our Bayesian model comparison with a sample of galaxies do not support the more ``TP-AGB heavy'' model of either \cb\ or \ma.
It is worth noticing that the performance of the \cb\ and \ma\ models are somewhat similar, although the different stellar tracks and synthesis methods have been employed by them.
Besides, the \bc\  and \cb\  models are only different in the treatment of TP-AGB stars, while the \bc\  models obviously have better performance than the \cb\  models.
These results suggest that a universally appropriate treatment of the TP-AGB phase is still not well-established in the current SSP models.
This may not be so surprising given the large number of uncertainties involved in the modeling of TP-AGB phase \citep{Conroy2009a,MarigoP2013a,RosenfieldP2014a,RosenfieldP2016a}.

It is important to mention that the results obtained with Bayesian model comparison are always data-dependent as clearly shown in Equation \ref{eq:bayes_model}.
So, the above conclusion could depend on the sample of galaxies used in this paper.
Indeed, most galaxies in our sample are at low redshift (mostly with $z\lesssim 1$), where the contribution of TP-AGB stars are thought to be less important.
However, it is still not easy to understand why the more sophisticated treatments of TP-AGB stars result in SSP models that have a poorer performance for low redshift galaxies.
Since the more ``TP-AGB heavy'' model of \cb\ and \ma\ are primarily tested for galaxies at higher redshifts where the contribution of TP-AGB stars are thought to be very important, the models could be overly tuned for those galaxies.
We will test this with the comparison of the results obtained for low-redshift and high-redshift galaxies in a future work.

\subsubsection{The consideration of binary star interaction} \label{ss:binary}
The presence of a nearby companion may alter the evolution of a star significantly by their interactions.
It is observationally well-established that a large fraction of stars, especially the massive ones, are in binary or higher-order multiple systems \citep{Sana2012a,Duchene2013a}.
So, physically, it is very important to consider the effects of binary star interaction in the SED modeling of a stellar population.

We have employed two publicly available SSP models (\ynII\ and \bpass) which has included the effects of binary star interactions to test the importance of binaries.
Both the version with and without binaries of the two models have been considered.
In Figures \ref{fig:BF_NSED_U_011_sp} and \ref{fig:ML-OF-BE_NSED_U_011_sp}, we show the results obtained for the case that the SSP, SFH and DAL are all assumed to be universal but only the SSP is fixed to a particular choice for a sample of $1159$ PEGs and a sample of $4308$ SFGs.
It is clear from the figures that for both the sample of PEGs and SFGs, the version of \ynII\ model with binaries is much better than the version without binaries.
Surprisingly, the version of \bpass\ model with binaries is even worse than the version without binaries.
In Figures \ref{fig:BF_NSED_UDD_011_sp} and \ref{fig:ML-OF-BE_NSED_UDD_011_sp}, we further  considered the case without assuming a universal SFH and DAL for all galaxies in the sample of PEGs or SFGs.
It is even clearer that the version of \ynII\ model with binaries is much better than the version without binaries, especially for the sample of PEGs.
However, the version of \bpass\ model with binaries is still much worse than the version without binaries.
As shown in Figures \ref{fig:ML-OF-BE_NSED_U_011_sp} and \ref{fig:ML-OF-BE_NSED_UDD_011_sp}, the version of \bpass\ model with binaries always locate at the bottom left in the ML-OF-BE diagram, which indicate a low goodness-of-fit to the data and a high degree of model complexity.

Given the limitation of the BPASS V2.0 model as mentioned in \S \ref{ss:bpass}, the above results are not so surprising.
In \citep{EldridgeJ2017a}, the authors stated that the BPASS code was initially established for young stellar populations, and they do not recommend the current version of the code for fitting the stellar populations much older than 1 Gyr.
Since most galaxies in our sample are located at $z\lesssim1$, the contribution of the stellar populations much older than 1 Gyr cannot be ignored (see Tables \ref{tab:twoG} for the two examples).
Actually, we obtained the above results long before the publication of the \citep{EldridgeJ2017a} paper where the limitations of the model was firstly pointed out in detail.
So, the states in \cite{EldridgeJ2017a} is really an independent support for the effectiveness of our Bayesian model comparison method.
In \cite{StanwayE2018a}, the authors stated that some issues about binary evolution have been partly addressed in their recently released V2.2 models.
We would like to check this in a following work.

Meanwhile, it is important to notice in Figures \ref{fig:BF_NSED_UDD_011_sp} and \ref{fig:ML-OF-BE_NSED_UDD_011_sp} that the version of \bpass\ model without binaries is actually better than both the versions of \ynII\ model with and without binaries.
A possible reason for this result is that the BPASS model is based on a detailed stellar evolution calculation with the Cambridge STARS stellar evolution code instead of the approximate and rapid stellar evolution code of \cite{HurleyJ2000a,HurleyJ2002a} as employed by the \ynII\ model.
Besides, a Monte Carlo binary population synthesis technique has been employed in the \ynII\ model, which could drive the differences with the \bpass\ model.

\subsubsection{The universality of IMF} \label{ss:imf}
\begin{table}
	\centering
	\begin{tabular}{rrrr}
		\toprule
		SSP & PEGs(1159) & SFGs(4308) & \\
		\midrule
		bc03\_ch & 1494.00 & -1221.00 & \\
		bc03\_kr & 1529.40 & -1143.00 & \\
		bc03\_sa & 1329.60 & -1332.00 & \\
		cb07\_ch & -1953.70 & -6551.00 & \\
		cb07\_kr & -1893.60 & -6336.00 & \\
		cb07\_sa & -2250.50 & -6475.00 & \\
		galev0\_kr & -362.60 & -2131.00 & \\
		galev0\_sa & -837.60 & -12202.00 & \\
		galev\_kr & -489.60 & 3336.00 & \\
		galev\_sa & -938.50 & 3183.00 & \\
		m05\_kr & -1219.40 & -10541.00 & \\
		m05\_sa & -272.40 & -6250.00 & \\
		ynII\_s & -329.80 & -5155.00 & \\
		ynII\_b & 1053.70 & -4961.00 & \\
		bpass\_s & 1695.60 & -4322.00 & \\
		bpass\_b & -3100.80 & -15547.00 & \\
		\bottomrule
	\end{tabular}
	\caption{The detailed value of Bayes factor for the 16 SSPs as in Figure \ref{fig:BF_NSED_UDD_011_sp}.}
	\label{tab:BF_NSED_UDD_011_sp}
\end{table}
Some recent works \citep{vanDokkumP2008a,DaveR2008a,vanDokkumP2012a,Conroy2012b} suggest that the IMF might not be universal, but could be evolving with the mass and redshift of the galaxies.
By using the Bayesian model comparison method for a sample of galaxies, it is possible to compare the SED modeling assuming a universal IMF and that assuming an evolving IMF.
However, all the SSPs employed in this work  assume a universal IMF.
So, here, we just want to compare the degree of universality of different IMFs.

The results in Figure \ref{fig:BF_NSED_UDD_011_sp} show that it is possible to compare the degree of universality of SSPs with different assumptions about the IMF.
To make it clearer, in Table \ref{tab:BF_NSED_UDD_011_sp}, we show the detailed value of Bayes factors of different SSPs for PEGs and SFGs, which are just the same as in Figure \ref{fig:BF_NSED_UDD_011_sp}.
For all the \bc, \cb\ and \galev\ models, the version of them assuming a Kroupa01 IMF has a much larger Bayes factor than the version assuming a Salpeter55 IMF for both PEGs and SFGs. 
The only exception is the \ma\ model, which obviously favors the Salpeter55 IMF for both PEGs and SFGs.
A possible reason for this is that the population synthesis method employed by the \ma\ model is very different from that employed by other models.
It is also worth  noticing that the \ma\ model is more sensitive to the selection of IMF than other models, and has the lowest value of Occam factor as shown in Figure \ref{fig:ML-OF-BE_NSED_UDD_011_sp}.
Generally, our results suggest that the IMF of stellar population in PEGs and SFGs are not likely to be systematically different and the Kroupa01 IMF is more universal than the Salpeter55 IMF.

\subsubsection{The importance of nebular emissions} \label{ss:nebular}
The importance of including the contribution of nebular emission lines to the broadband fluxes of galaxies with active star formation has been well documented in the literature \citep{Charlot2001a,ZackrissonE2008a,Ilbert2009a,Schaerer2009a,SchenkerM2013a,Stark2013a,deBarros2014a}.
For example, \cite{Ilbert2009a} show that the flux of nebular emission lines can change the color by about 0.4 mag, and a reasonable treatment of emission lines can decrease the dispersion of photo-z estimation by a factor of 2.5.
Here, we discuss the results about nebular emission obtained with the Bayesian model comparison method for a sample of galaxies developed in this paper.

The results in Figure \ref{fig:BF_NSED_UDD_011_sp} and Table \ref{tab:BF_NSED_UDD_011_sp} show that the version of \galev\ SSP with the consideration of emission lines has a significantly larger Bayes factor than all the other SSPs without the consideration of emission lines for the SFGs.
The max-likelihoods in the right panel of Figure \ref{fig:ML-OF-BE_NSED_UDD_011_sp} show that this model can provide significantly better fit to the observational data of SFGs than others, although it has a relative smaller Occam factor and consequently a higher model complexity than most of the others.
So, it is clear that the nebular emission lines are indeed very important for the SFGs.
However, for the PEGs, the Bayes factors of the version of \galev\ SSP with and without the consideration of emission lines are not larger than most of the other models.
The results in the left panel of Figure \ref{fig:ML-OF-BE_NSED_UDD_011_sp} show that the \galev\ models provide a poorer fit to the observational data of PEGs than most of other models, although they have the largest value of Bayes factor.
These results suggest that, for the modeling of stellar emission, the \galev\ models are not more sophisticated than other models.
So, it is very likely that other SSP models would perform much better for SFGs when a reasonable consideration of nebular emission being included in them.
Unfortunately, without the version of them with nebular emissions self-consistently included, we cannot test this with the Bayesian model comparison method developed in this work.

\subsection{The universality of different forms of SFH} \label{ss:SFH}
In theory, due to the different environmental influences and formation conditions, the detailed SFHs of different galaxies are expected to be very different.
However, when the details in the SFHs being smoothed out, the general shape of them could be more similar.
In practice, the Exp-dec SFH has been widely employed in many works as if it is universal for all galaxies.
This assumption has been doubted in many works \citep{Maraston2010a,LeeS2010a,Pforr2012a,Reddy2012b,Lee2014a}, and many authors have suggested some more complicated \citep{GladdersM2013a,AbramsonL2016a,DiemerB2017a,CieslaL2017a,CarnallA2018a} or more physically motivated \citep{FinlatorK2007a,Pacifici2012a,IyerK2017a} form of SFHs.
In most of previous works, the different forms of SFH are mainly compared by their goodness-of-fit to the observational data. 
Apparently, a more complicated form of SFH tends to provide a better fit to the data.
However, this additional complexity is not necessarily well-supported by the data.

Here, we discuss the comparison of different forms of SFH with the Bayesian evidence, which is determined by the trade-off between the complexity of a model and its goodness-of-fit to the data.
In Figure \ref{fig:BF_1SED_101_sp}, we have compared the different forms of SFH for the PEG ULTRAVISTA114558 and the SFG ULTRAVISTA99938.
The Burst SFH has the largest Bayesian evidence for the PEG ULTRAVISTA114558 as shown in the left panel of Figure \ref{fig:BF_1SED_101_sp}.
However, the max-likelihoods in the left panel of ML-OF-BE diagram \ref{fig:ML-OF-BE_1SED_101_sp} show that its goodness-of-fit to the data is similar to that of the Exp-dec, Exp-inc and Delayed SFHs.
Actually, it has a much larger Occam factor and consequently much smaller model complexity than the others.
The trade-off between its model complexity and goodness-of-fit to the data finally leads to the largest Bayesian evidence.
On the other hand, the constant SFH has the largest Bayesian evidence as shown in the right panel of Figure \ref{fig:BF_1SED_101_sp} for the SFG ULTRAVISTA99938.
Although it has the largest Occam factor as shown in the right panel of Figure \ref{fig:ML-OF-BE_1SED_101_sp}, its goodness-of-fit to the data is much smaller than the Exp-dec SFH which actually provides the best fit to the data.
Interestingly, the trade-off between its model complexity and goodness-of-fit to the data still leads to the largest Bayesian evidence.
These results suggest that a simple definition of Occam factor similar to that in Equation \ref{eq:Omega_total_1SED_011} can provide results that are basically consistent with our intuition about the complexity of a model.
However, it seems meaningless to talk about the absolute complexity of a model in the sense of this definition without mention a particular object.

The above results are obtained for a particular PEG and SFG.
They are not necessarily representative of a whole population of galaxies.
In Figure \ref{fig:BF_NSED_DUD_101_sp}, we have compared the universality of different forms of SFH for the sample of PEGs and SFGs, respectively.
Since the results are obtained without assuming a universal SSP and a universal DAL, they are very robust against the choice of SSP and DAL for different galaxies.
Interestingly, the results show that the  Exp-dec SFH, which is the most widely used form of SFH in the literature, has the best universality for both PEGs and SFGs in our sample.
Besides, the max-likelihoods in Figure \ref{fig:ML-OF-BE_NSED_DUD_101_sp} show that the Exp-dec SFH also provides generally the best goodness-of-fit to the observational data of both PEGs and SFGs, although it has the smallest Bayes factor and consequently the largest model complexity.
These results show that the Exp-dec SFH is the most successful at explaining the multi-wavelength photometric observations of a relatively large sample of low-redshift galaxies.
However, since the results obtained with Bayesian model comparison are always data-dependent, as clearly shown in Equation \ref{eq:bayes_model}, the results for galaxies at higher redshifts could be very different.
We will check this in a future work.

\subsection{The universality of different forms of DAL} \label{ss:DAL}
An assumption about the effects of dusty ISM on the observed SEDs of galaxies is necessary when deriving the physical properties of galaxies.
The most widely used assumption is a uniform empirical or analytical attenuation law as a simple screen.
However, some works suggested that the dust laws are likely to be non-universal for galaxies with different types and redshifts.
For example, \cite{KriekM2013a} have utilized the stacked photometric SEDs to explore the variation of DAL in $0.5<z<2.0$ galaxies.
They found that the best-fit DAL varies with the spectral type of the galaxy, with more active galaxies having shallower DALs. 
\cite{SalmonB2016a} show that some individual galaxies at $z\sim1.5-3$ from CANDELS have strong Bayesian evidence in favor of one particular dust law.
Besides, they found that the shallower SB-like DAL is more favored by galaxies with high color excess, while the steeper SMC-like DAL is more favored by galaxies with low color excess.
With the CIGALE \citep{Noll2009a} SED-fitting code, \cite{SalimS2018a} studied the dust attenuation curves of 230,000 individual galaxies in the local universe, including PEGs and intensely SFGs.
Similar to \cite{SalmonB2016a}, they found a strong correlation between the attenuation curve slope and the optical opacity ($A_{\rm v}$), with more opaque galaxies having shallower curves.
These results are consistent with the predictions based on some radiative transfer models \citep{ChevallardJ2013a}.

An important difference between our method and that of \cite{SalmonB2016a} is that we have calculated the Bayesian evidence of different DALs with the marginalization over not only the stellar population parameters but also the different choices of the SSP model and the form of SFH.
By using this method, more robust results about the DAL can be obtained.
In Figure \ref{fig:BF_1SED_110_sp}, we have compared the performance of different DALs with the Bayesian evidence for the PEG ULTRAVISTA114558 and the SFG ULTRAVISTA99938.
The results show that the SB-like DAL is more favored by the PEG and the MW-like DAL is more favored by the SFG.
However, the ML-OF-BE diagram in Figure \ref{fig:ML-OF-BE_1SED_110_sp} show that the more favored DALs not necessarily provide much better fit to the data, although they do have relatively larger Occam factor which indicate lower model complexity for the two galaxies.

\begin{figure}[]
  \begin{center}
	\includegraphics[scale=0.7]{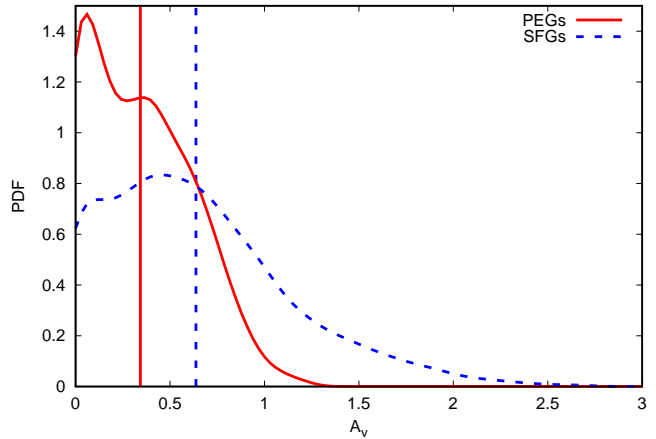}
  \end{center}
  \caption{The distribution of optical opacity $A_{\rm v}$ for the PEGs and SFGs in our sample. The vertical lines indicate the mean of the two distributions. On average, the SFGs have larger optical opacity than PEGs.}
  \label{fig:dist_Av}
\end{figure}

Another very important difference between our method and that of \cite{SalmonB2016a} is that we have defined the Bayesian evidence for the SED modeling of a sample of galaxies in addition to that for individual galaxies.
In Figure \ref{fig:BF_NSED_DDU_110_sp}, we have compared the performance of different DALs for the SED modeling of a sample of PEGs and SFGs, respectively.
By using the Bayesian evidence defined for the SED modeling of a sample of galaxies, we find that the steeper SMC-like DAL is systematically more favored by the PEGs, while the shallower SB-like DAL is systematically more favored by the SFGs.
Besides, the ML-OF-BE diagram in Figure \ref{fig:ML-OF-BE_NSED_DDU_110_sp} show that the SMC-like DAL also provides the best fit to the observational data of PEGs, while the SB-like DAL also provides the best fit to the observational data of SFGs.
Since these results are obtained without assuming a universal SSP and SFH, they should be more robust.
As shown in Figure \ref{fig:dist_Av}, for the sample used in this work, the SFGs have a mean value of  optical opacity larger than that of PEGs.
So, basically, our results are consistent with the findings of \cite{SalmonB2016a} and \cite{SalimS2018a}, and the prediction of \cite{ChevallardJ2013a} based on radiative transfer models.
However, our results are based on the assumption of a universal DAL.
We have tried to find out which DAL is the better if it is assumed to be universal.
So, the results may highly depend on the used sample if the attenuation curve is actually object-dependent.
\cite{SalimS2018a} have used a much larger sample than us.
They show that the average attenuation curve of local star-forming galaxies in their sample is almost as steep as that of SMC.
With a parameterized form of DAL, a more detailed investigation of the variation of DAL in different galaxies and its possible evolution  with redshift will be the subject of a future work.

\section{Summary and conclusions}\label{sec:sum}
In this work, we have proposed a new method to define the Bayesian evidence for the SED modeling of an individual galaxy and a sample of galaxies, respectively.
With the application of the newly defined Bayesian evidences and the new version of our BayeSED code to a Ks-selected, low redshift ($z\lesssim 1$) sample in the COSMOS/UltraVISTA field, we have demonstrated a comprehensive Bayesian discrimination of the different assumptions about SSP, SFH and DAL in the SED modeling of galaxies.

We summarize our main results as follows:

$\bullet$ The more ``TP-AGB heavy'' SSP model of \cb\ and \ma\ are not systematically more favored by both PEGs and SFGs in our sample, although it could be favored by some individual galaxies.

$\bullet$ A reasonable consideration of binaries is important for the SED modeling of both PEGs and SFGs. For the two publicly available SSP models with the consideration of binaries, the \ynII\ model is more favored than the \bpass\ model by both the PEGs and SFGs in our sample.

$\bullet$ For both the PEGs and SFGs in our sample, the Kroupa01 IMF is systematically more favored than that of Salpeter55.

$\bullet$ A simple but reasonable consideration of nebular emission lines, such as that implemented in the \galev\ SSP model, can significantly improve the performance of the SED modeling of SFGs.

$\bullet$ The widely used Exp-dec SFH is the one best supported by the multi-wavelength photometric data of both PEGs and SFGs in our sample, although it is not necessarily more physically reasonable than others.

$\bullet$ For the galaxies in our sample, the SMC-like DAL is systematically more favored by the PEGs, while the SB-like DAL is systematically more favored by the SFGs.

The above results are either obvious or understandable in the context of galaxy physics.
So, we conclude that the Bayesian evidence, which is determined by the trade-off between the complexity of a model (quantified by the Occam factor) and its goodness-of-fit to the data (quantified by the max-likelihood), is very useful for discriminating the different assumptions in the SED modeling of galaxies.
By using the Bayesian evidence marginalized over not only the normal parameters but also the different choices of all the irrelevant and uncertain components, it is possible to obtain much more robust conclusions.
Especially, the Bayesian evidence defined for the SED modeling of a sample of galaxies allows us to compare the universality of any assumption made in the modeling procedure.
This opens the door for many interesting investigations.
Based on a simple procedure and widely used SSPs, SFHs and DALs to model the SEDs of galaxies, we have demonstrated the usefulness of Bayesian model comparison method, evaluated its effectiveness, and built a reference for the future works.
In the future, with a more flexible and sophisticated SED modeling procedure, we will apply the Bayesian method developed in this work to a larger sample of galaxies covering a much larger redshift range.
\acknowledgments
We thank an anonymous referee for his/her very specific and constructive comments that greatly helped us to improve the paper.
The authors gratefully acknowledge the computing time granted by the Yunnan Observatories, and provided on the facilities at the Yunnan Observatories Supercomputing Platform.
We thank Professor X. Kong for helpful discussions about the classification of galaxies and TP-AGB star issue in stellar population synthesis models.
This work is supported by the National Natural Science Foundation of China (NSFC, Grant Nos. 11773063, 11521303, 11733008), Chinese Academy of Sciences (CAS, no. KJZD-EW-M06-01) and Yunnan Province (Grant No. 2017FB007, 2017HC018).





\bibliography{ms.bbl}

\appendix 

\section{The sensitivity of results to assumptions of the priors}\label{sec:sensitivity}
The choice of priors is indispensable in any Bayesian data analysis.
In principle, the priors should be chosen to best represent our state of knowledge before the analysis of the data.
In the main body of this paper, we have assumed a uniform prior truncated at the allowed range for all free parameters of the SED model.
This just reflects the fact that, except for the allowed range, the SED model itself tell us nothing about the detailed physical distribution of its free parameters.
However, to make us notice the possible variation of the results with the assumed priors, we present here an analysis of the sensitivity of results to assumptions of the priors in the parameter space.
As in \S \ref{ss:1SED_par}, we demonstrate this with the analysis of the multi-wavelength SEDs of the PEG ULTRAVISTA114558 and the SFG ULTRAVISTA99938 by assuming the commonly used BC03 SSP model with a \cite{Chabrier2003a} IMF (bc03\_ch), an Exp-dec SFH, and the \cite{Calzetti2000a} DAL.
In addition to the truncated uniform prior, we also considered three Gaussian priors with $\sigma=\frac{b-a}{2}$ while centered at $a$, $(a+b)/2$ and $b$, respectively, for all free parameters of the SED model.
Here, $a$ and $b$ represent the lower and upper limit of the parameters set by the used SED model.
The Gaussian prior are not necessarily more physically reasonable than the uniform prior.
However, it provides a way to test the sensitivity of the results to the assumptions of priors.

\begin{figure}[]
  \begin{center}
	\includegraphics[scale=0.88]{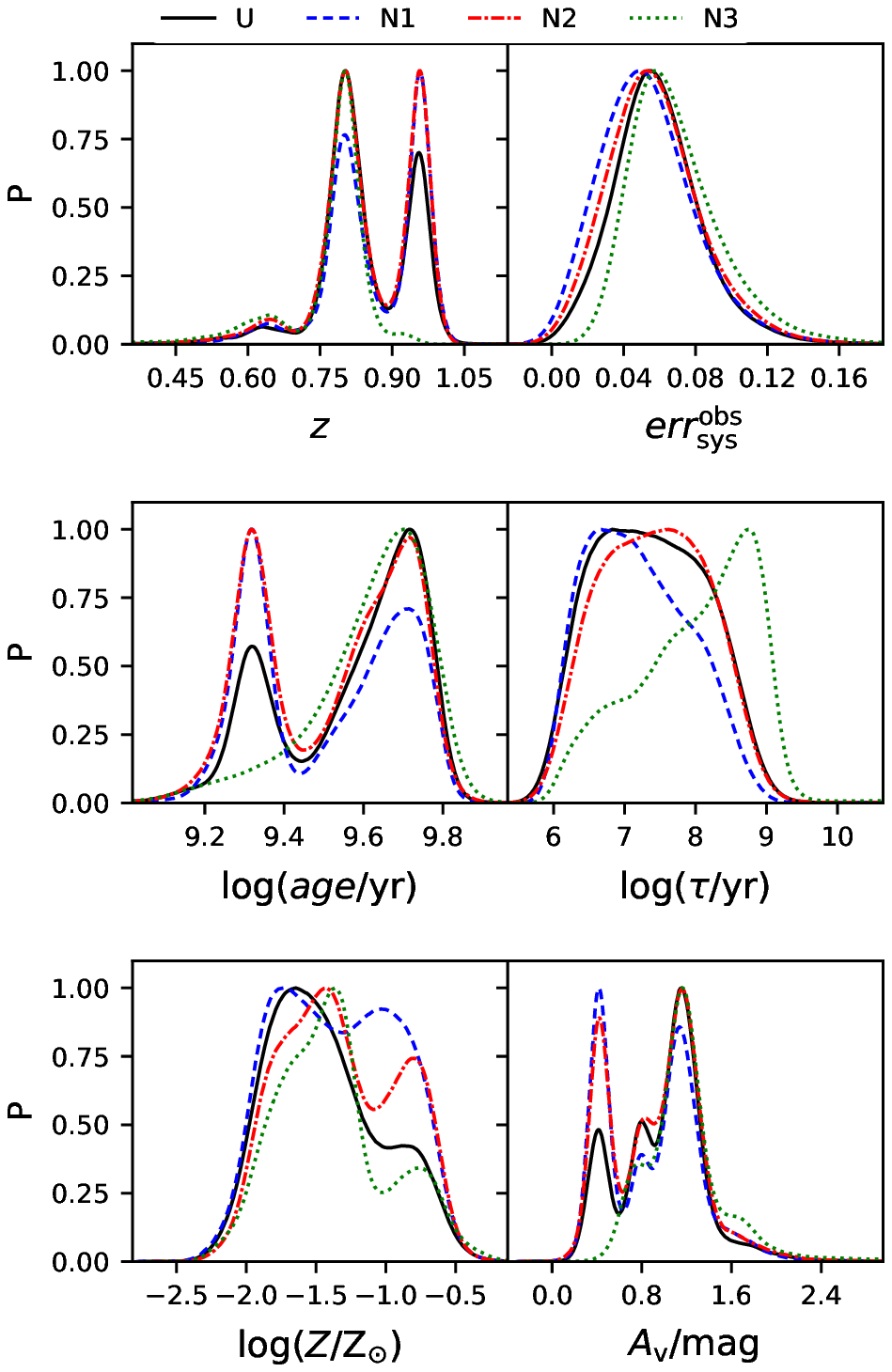}
	\includegraphics[scale=0.88]{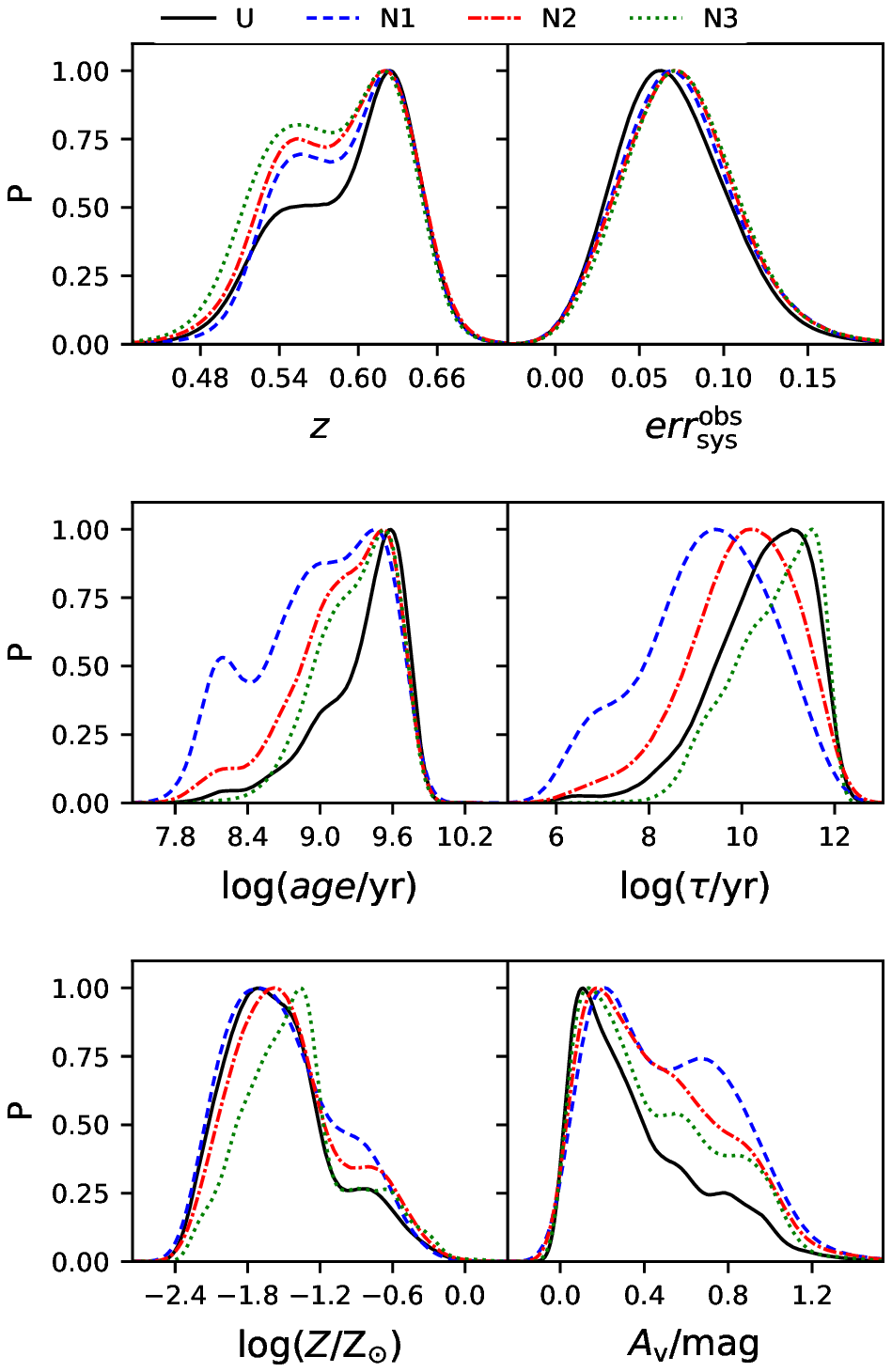}
  \end{center}
  \caption{The 1-D posterior PDFs of free parameters assuming the U:Uniform(a,b), N1:Normal$(\mu=a,\sigma=\frac{b-a}{2})$, N2:Normal$(\mu=\frac{b+a}{2},\sigma=\frac{b-a}{2})$, N3:Normal$(\mu=b,\sigma=\frac{b-a}{2})$ priors, respectively, for the PEG ULTRAVISTA114558 (left) and the SFG ULTRAVISTA99938 (right).}
  \label{fig:plotP}
\end{figure}
In Figure \ref{fig:plotP}, we present the 1-D posterior PDFs of free parameters assuming  four different types of priors for the PEG ULTRAVISTA114558 and the SFG ULTRAVISTA99938.
As shown in the figure, different types of priors can lead to somewhat different shapes of posterior PDFs.
However, the results for different parameters have different sensitivities to the assumptions of priors.
For both galaxies, the posterior PDFs of $z$ and $\sigma_{\rm sys}$ are very insensitive to the assumptions of priors, while  that of ${\rm log}(\tau/\rm{yr})$ and ${\rm log}(Z/{\rm Z_{\odot}})$ are very sensitive to the assumptions of priors.
Meanwhile, the situation for other free parameters seems object-dependent for the two galaxies.
Although the shape of posterior PDFs assuming different priors could be obviously different, the parameter estimation with the median of the posterior PDFs are actually more similar, as shown in Tables \ref{tab:ULTRAVISTA114558_All} and \ref{tab:ULTRAVISTA99938_All}.
Generally, the sensitivity of a parameter to the assumptions of priors is consistent with the estimated uncertainty of the parameter.
Some parameters, such as ${\rm log}(\tau/\rm{yr})$ and ${\rm log}(Z/{\rm Z_{\odot}})$, cannot be well constrained by the data and are therefore more sensitive to the assumed priors.
The derived parameters, such as stellar mass and luminosity can be well constrained by the data and are therefore very insensitive to the assumptions of priors.

On the other hand, in Tables \ref{tab:ULTRAVISTA114558_All} and \ref{tab:ULTRAVISTA99938_All}, we also presented the Bayesian evidences obtained with different assumptions of priors in the parameter space.
As shown in the two tables, the Bayesian evidence of the model could be very sensitive to the assumptions of priors in the parameter space.
In this paper (See \S \ref{s:apply}), to make the results about model comparison with Bayesian evidence to be more robust, we have considered different assumptions about the SSP,SFH and DAL which represent different priors in the model space, and our conclusions are obtained with the comparison of the different cases.
Furthermore, we believe that the sensitivity of the Bayesian evidence of a model to the assumptions of priors in the parameter space is actually a benefit of the method.
Since the physically more reasonable and informative priors of the parameters can be provided by a model for the distribution and evolution of the physical parameters of galaxies, it should be considered as a part of the model.
In this way, the Bayesian model comparison/selection method developed in this paper has the potential to be used as a method for the comparison/selection of the combined model of the SED and the formation and evolution of galaxies.
The results of Bayesian model comparison with the uniform priors for the physical parameters in this work could be used as a reference for this kind of investigation in the future.

\begin{table*}[htb]
	\centering
	\begin{tabular}{@{}ccccc@{}}
		\toprule
		Parameter & U & N1 & N2 & N3 \\
		\midrule
		{\boldmath$z              $} & $0.82^{+0.13}_{-0.05}$ & $0.85^{+0.12}_{-0.07}$ & $0.83^{+0.13}_{-0.06}$ & $0.79^{+0.03}_{-0.11}$ \\
		{\boldmath$\sigma_{\rm sys}      $} & $0.06^{+0.02}_{-0.02}$ & $0.05^{+0.03}_{-0.02}$ & $0.05^{+0.03}_{-0.02}$ & $0.06^{+0.03}_{-0.02}$ \\
		{\boldmath${\rm log}(age/\rm{yr})$} & $9.63^{+0.12}_{-0.30}$ & $9.54^{+0.18}_{-0.24}$ & $9.57^{+0.16}_{-0.27}$ & $9.65^{+0.11}_{-0.17}$ \\
		{\boldmath${\rm log}(\tau/\rm{yr})$} & $7.33^{+0.90}_{-0.87}$ & $7.12^{+0.90}_{-0.72}$ & $7.42^{+0.83}_{-0.86}$ & $8.15^{+0.73}_{-1.13}$ \\
		{\boldmath${\rm log}(Z/{\rm Z_{\odot}})$} & $-1.49^{+0.54}_{-0.36}$ & $-1.35^{+0.50}_{-0.48}$ & $-1.36^{+0.55}_{-0.42}$ & $-1.41^{+0.50}_{-0.37}$ \\
		{\boldmath$A_{\rm v}/{\rm mag}$} & $1.05^{+0.23}_{-0.56}$ & $0.93^{+0.33}_{-0.53}$ & $0.98^{+0.31}_{-0.55}$ & $1.15^{+0.31}_{-0.32}$ \\
		$z_{form}                     $ & $2.65^{+3.86}_{-1.03}$ & $1.98^{+3.14}_{-0.41}$ & $2.16^{+3.00}_{-0.59}$ & $2.82^{+2.95}_{-1.20}$ \\
		${\rm log}(SFR/[\rm M_{\odot}/\rm{yr}])$ & $-67.72^{+61.42}_{-931.28}$ & $-99.41^{+89.06}_{-899.59}$ & $-48.00^{+42.50}_{-951.00}$ & $-9.58^{+8.99}_{-159.99}$ \\
		$log(M_*/\rm M_{\odot})    $ & $10.78^{+0.10}_{-0.15}$ & $10.72^{+0.14}_{-0.09}$ & $10.74^{+0.12}_{-0.12}$ & $10.75^{+0.09}_{-0.19}$ \\
		${\rm log}(L_{\rm bol}/[\rm erg/s])$ & $44.60^{+0.06}_{-0.05}$ & $44.60^{+0.05}_{-0.05}$ & $44.60^{+0.05}_{-0.06}$ & $44.57^{+0.06}_{-0.11}$ \\
		\midrule
		ln(Evidence) & $-14.36^{+0.18}_{-0.18}$ & $-13.30^{+0.17}_{-0.17}$ & $-14.93^{+0.18}_{-0.18}$ & $-17.91^{+0.20}_{-0.20}$ \\
		\bottomrule
	\end{tabular}\caption{The parameter estimation and Bayesian evidence with different priors for the PEG ULTRAVISTA114558.}
	\label{tab:ULTRAVISTA114558_All}
\end{table*}

\begin{table*}[htb]
	\centering
	\begin{tabular}{@{}ccccc@{}}
		\toprule
		Parameter & U & N1 & N2 & N3 \\
		\midrule
		{\boldmath$z              $} & $0.60^{+0.03}_{-0.07}$ & $0.60^{+0.04}_{-0.06}$ & $0.59^{+0.04}_{-0.06}$ & $0.58^{+0.05}_{-0.06}$ \\
		{\boldmath$\sigma_{\rm sys}      $} & $0.07^{+0.04}_{-0.03}$ & $0.07^{+0.03}_{-0.03}$ & $0.07^{+0.03}_{-0.03}$ & $0.07^{+0.03}_{-0.03}$ \\
		{\boldmath${\rm log}(age/\rm{yr})$} & $9.45^{+0.24}_{-0.47}$ & $9.03^{+0.51}_{-0.70}$ & $9.25^{+0.36}_{-0.49}$ & $9.35^{+0.29}_{-0.37}$ \\
		{\boldmath${\rm log}(\tau/\rm{yr})$} & $10.58^{+0.94}_{-1.21}$ & $9.33^{+1.37}_{-1.60}$ & $10.10^{+1.14}_{-1.26}$ & $10.85^{+0.80}_{-1.14}$ \\
		{\boldmath${\rm log}(Z/{\rm Z_{\odot}})$} & $-1.60^{+0.49}_{-0.38}$ & $-1.57^{+0.62}_{-0.42}$ & $-1.52^{+0.60}_{-0.40}$ & $-1.42^{+0.53}_{-0.41}$ \\
		{\boldmath$A_{\rm v}/{\rm mag}$} & $0.28^{+0.42}_{-0.20}$ & $0.46^{+0.39}_{-0.32}$ & $0.40^{+0.42}_{-0.28}$ & $0.36^{+0.45}_{-0.25}$ \\
		$z_{form}                     $ & $1.22^{+1.04}_{-0.49}$ & $0.78^{+0.71}_{-0.15}$ & $0.92^{+0.85}_{-0.25}$ & $1.02^{+0.88}_{-0.31}$ \\
		${\rm log}(SFR/[\rm M_{\odot}/\rm{yr}])$ & $0.01^{+0.26}_{-0.15}$ & $-0.02^{+0.35}_{-0.42}$ & $0.05^{+0.31}_{-0.20}$ & $0.07^{+0.30}_{-0.17}$ \\
		$log(M_*/\rm M_{\odot})    $ & $9.54^{+0.10}_{-0.21}$ & $9.44^{+0.16}_{-0.20}$ & $9.49^{+0.13}_{-0.22}$ & $9.50^{+0.12}_{-0.21}$ \\
		${\rm log}(L_{\rm bol}/[\rm erg/s])$ & $43.97^{+0.24}_{-0.12}$ & $44.05^{+0.27}_{-0.16}$ & $44.02^{+0.28}_{-0.15}$ & $44.00^{+0.29}_{-0.15}$ \\
		\midrule
		ln(Evidence) & $-25.56^{+0.17}_{-0.17}$ & $-24.81^{+0.16}_{-0.16}$ & $-25.94^{+0.17}_{-0.17}$ & $-28.55^{+0.18}_{-0.18}$ \\
		\bottomrule
	\end{tabular}\caption{As Table \ref{tab:ULTRAVISTA114558_All}, but for the SFG ULTRAVISTA99938.}
	\label{tab:ULTRAVISTA99938_All}
\end{table*}
\end{document}